\documentclass[preprint ]{aastex63}

\received{}
\revised{}
\accepted{\today}
\submitjournal{ApJ}

\shorttitle{}
\shortauthors{Parenti et al.}
\graphicspath{{./}{figures/}}
\usepackage{xcolor}

\begin{document}

\title{
Validation of a wave heated 3D MHD  coronal-wind model using Polarized Brightness and EUV observations.}

\correspondingauthor{Susanna Parenti}
\email{susanna.parenti@ias.u-psud.fr}

\author[0000-0002-0786-7307]{Susanna Parenti}
\affiliation{Dept. d'Astrophysique/AIM, CEA/IRFU, CNRS/INSU, Universit\'e Paris et Paris-Saclay, 91191 Gif-sur-Yvette Cedex, France}
\affiliation{Université Paris-Saclay, CNRS, Institut d’Astrophysique Spatiale, 91405, Orsay, France}

\author[0000-0002-2916-3837]{Victor R\'eville}
\affiliation{IRAP, Université Toulouse III - Paul Sabatier, CNRS, CNES, Toulouse, France}

\author[0000-0002-1729-8267]{Allan Sacha Brun}
\affiliation{Dept. d'Astrophysique/AIM, CEA/IRFU, CNRS/INSU, Universit\'e Paris et Paris-Saclay, 91191 Gif-sur-Yvette Cedex, France}

\author[0000-0001-8247-7168]{Rui F. Pinto}
\affiliation{Dept. d'Astrophysique/AIM, CEA/IRFU, CNRS/INSU, Universit\'e Paris et Paris-Saclay, 91191 Gif-sur-Yvette Cedex, France}

\author[0000-0003-0972-7022]{Fr\'ed\'eric Auch\` ere}
\affiliation{Université Paris-Saclay, CNRS, Institut d’Astrophysique Spatiale, 91405, Orsay, France}

\author[0000-0003-4290-1897]{Éric Buchlin}
\affiliation{Université Paris-Saclay, CNRS, Institut d’Astrophysique Spatiale, 91405, Orsay, France}

\author[0000-0002-2137-2896]{Barbara Perri}
\affiliation{Centre for mathematical Plasma Astrophysics, KU Leuven, Celestijnenlaan 200b-box 2400, 3001 Leuven, Belgium\\
}
\affiliation{Dept. d'Astrophysique/AIM, CEA/IRFU, CNRS/INSU, Universit\'e Paris et Paris-Saclay, 91191 Gif-sur-Yvette Cedex, France}

\author[0000-0002-9630-6463]{Antoine Strugarek}
\affiliation{Dept. d'Astrophysique/AIM, CEA/IRFU, CNRS/INSU, Universit\'e Paris et Paris-Saclay, 91191 Gif-sur-Yvette Cedex, France}




\begin{abstract}

The physical properties responsible for the formation and evolution of the corona and heliosphere are still not completely understood. 3D MHD global modeling is a powerful tool to investigate all the possible candidate processes. To fully understand the role of each of them, we need a validation process where the output from the simulations is quantitatively compared to the observational data.

In this work, we present the results from our validation process applied to the wave turbulence driven 3D MHD corona-wind model WindPredict-AW. At this stage of the model development, we focus the work to the coronal regime in quiescent condition. We analyze three simulations results, which differ by the boundary values. We use the 3D distributions  of density and temperature, output from the simulations at the time of around the first Parker Solar Probe perihelion (during minimum of the solar activity), to synthesize both extreme ultraviolet (EUV) and white light polarized (WL pB) images to reproduce the observed solar corona. For these tests, we selected AIA 193\,\AA,\ 211\,\AA\ and 171\,\AA\  EUV emissions, MLSO K-Cor and LASCO C2 pB images obtained the 6 and 7 November 2018. We then make quantitative comparisons of the disk and off limb corona. We show that our model is able to produce synthetic images comparable to those of the observed corona. 

\end{abstract}


\section{Introduction}
\label{sec:intro}

Understanding the mechanism responsible for coronal heating and solar wind acceleration  \citep{parker58} remains, as of today, one of the biggest question of solar physics. Alfv\'en waves have been observed in the solar wind very early in the space exploration era and have been proposed as a potential driver of the solar wind acceleration \citep{Belcher1971}. \citet{Velli1989} proposed wave reflection as a way to create counter-propagating populations of Alfv\'en  waves and trigger non-linear interactions responsible for the turbulent cascade and subsequent heating. In recent years, an increasing focus has been set on Alfvén wave turbulent driven models, which can be explored in multiple configurations. Theoretical and numerical models have shown increasing success to power the open solar wind \citep{SuzukiInutsuka2005,VerdiniVelli2007,Verdini2009,Shoda2018a} and to heat coronal loops \citep{BuchlinVelli2007,Downs2016}.

Using photospheric magnetic field information as lower boundary, 3D MHD global models recreate the structure of the solar corona and the heliosphere. They are, as such, powerful tools to test and select physical mechanisms that may play a major role in the energy deposition and dissipation in the corona. The progresses made in the past twenty years on  Alfv\'en wave turbulence have been included in 3D global models \citep{Sokolov2013,vanderHolst2014,Mikic2018} and tested against several observables \citep[see for instance,][]{Lionello2009, Oran2015, Oran2017, Mikic2018, Sachdeva2019}. In particular, the model we are discussing in this paper, WindPredict-AW, has been shown to reproduce accurately the \emph{in-situ} observations made by the Parker Solar Probe made during its first perihelion in November 2018 \citep{Reville2020ApJS}, as well the global variation of mass loss and solar wind speed during a whole solar cycle \citep{Hazra2021}.

Because the coronal magnetic field is extremely difficult to measure, 3D MHD models are essential to connect and interpret \emph{in-situ} data with their sources in the solar atmosphere, observed through remote sensing \citep[see for instance,][]{parenti2021}.
3D models can also be used to predict future configurations and simulate 'observables' to be compared to future observations. For instance, this was recently done for solar eclipses by \cite{Mikic2018}. Other important forecast are solar eruptions \citep[e.g.][]{Leka2003, Leka2019, Georgoulis2021} with possible Earth impact depending on the connectivity, as well as the preparation of the Solar Orbiter remote sensing observations. Solar Orbiter \citep[][]{muller2020}, launched in February 2020,  has a mission profile where the remote sensing instruments (some of them with a limited field of view) are observing only during limited periods called 'windows', part of which  will cover the passage along the far side of the Sun. Precursor observations and modeling predictions of the corona will be a major support for the mission pointing strategy during those observations. 

For all these reasons, an accurate validation of the 3D MHD models through remote sensing observations is of primary importance. 'Observables' that can be directly compared to real data must consequently be computed from the model. For instance, in the corona, we can measure directly neither the large scale magnetic field nor the electron temperature of the plasma. But we can measure the radiation emission of this plasma frozen to the magnetic field, and from this infer the plasma parameters. In general, to be carefully constrained a model  needs to be compared with the larger possible number of observables. 
In this work we aim at validating the 3D MHD WindPredict-AW model of the corona by following the above recommendations. In particular, from the output of the simulations, we derive cotemporal synthetic images for both optically thin EUV corona  and polarized brightness (pB) white light (WL). These are quantitatively compared to Solar Dynamics Observatory (SDO) EUV AIA \citep[Atmospheric Imaging Assembly,][]{lemen12}, MLSO KCor \citep[COSMO K-Coronagraph,][]{Hou2013} and SOHO/LASCO \citep[Large Angle and Spectrometric Coronagraph Experiment,][]{Brueckner1995} pB data. Because of the different formation process of light in both these bands (processes having a different sensitivity to the electron temperature and density), we could infer more accurate constraints to a set of models.  

Going beyond the in situ validation of WindPredict-AW shown in \cite{Reville2020ApJS}, we here look closer to the Sun, and analyze synthetic remote sensing observables in the coronal quiescent regime. We chose a period of the minimum of the solar activity and limited the analysis to the corona. Indeed, the model is not yet able to adequately treat the low layers of the solar atmosphere, which will be addressed in future works for more active periods of the solar cycle. We have chosen to compare three realizations of the solar corona and wind solutions for synoptic magnetic maps close to the first PSP perihelion by changing key control parameters of the model. The model to observation comparison of the absolute intensity values and their modulation is made on the  solar disk and off the limb, up to about 6 R$_\odot$, and along different latitudes. For our tests, we selected observation data from the 6 and 7 November 2018, which correspond to the first Parker Solar Probe perihelion.

The paper is organized as follows. Section \ref{sec:model} presents the 3D MHD model and the emissivity models to construct the EUV and pB synthetic images. Section \ref{sec:data} illustrates the data used for the comparison with the synthetic images. Section \ref{sec:res_gen} provides global  results, while more detailed results for November 6 are given in Section  \ref{sec:res_disk}  (EUV on disk data) and  Section \ref{sec:res_corona} (off--disk corona). Section \ref{res:res_corona_7} reports the results for November 7, while we draw our conclusions in Section \ref{sec:conclusion}.

\section{The models}
\label{sec:model}

\subsection{The 3D MHD model}
\label{sec:mod_mhd}

In this work, we rely on the 3D MHD coronal model presented in \citet{Reville2020ApJS}, \citep[see also][]{Reville2020ApJL,Hazra2021}. The model uses the PLUTO code \citep{Mignone2007} and has been further developed to fully integrate the evolution of Alfvén waves packets launched from the inner boundary of the simulation, as well as their propagation and dissipation in the solar wind. The full set of equations is presented in appendix \ref{app:eq}. The code solves, in addition to the usual MHD equations, two equations related to the transport of parallel and antiparallel Alfvén waves energy densities integrated over the whole frequency spectrum \citep[see, e.g.,][]{Chandran2021}. The dissipation is obtained by a Kolmogorov phenomenology, assuming a turbulent correlation length scale $\lambda_{\perp}$ \citep{Dmitruk2002}, which evolves with distance to the Sun and solar wind expansion. The total source term in the  fluid's energy equation is 

\begin{equation}
    Q = Q_w + Q_h - Q_c - Q_r,
\end{equation}
where $Q_w$ is the wave heating term, $Q_h$ a small exponential heating term, $Q_r$ the radiative cooling and $Q_c$ the thermal conduction. For the domain of interest here, i.e. and the low and middle corona, we considered only an optically thin cooling with $Q_r = n^2 \Lambda(T)$, where $n$ is the electron/proton number density of the MHD model, and $\Lambda(T)$ comes from the derivation of \citet{Athay1986} which assumes coronal elemental fractional abundances. The thermal conduction $Q_c$, below $5R_{\odot}$, takes the form of a Spitzer-Härm collisional heat flux with $\kappa= 9 \times 10^{-7} T^{5/2}$ erg s$^{-1}$ K$^{-1}$ cm$^{-1}$. The heat flux is further transitioned to a free-streaming heat flux \citep{Hollweg1978d,Reville2020ApJS}. 

The energy flux that enters the simulation domain is the sum $F_h + F_w$, where $F_h$, is related to $Q_h$ through the relation :

\begin{equation}
    Q_h = \frac{F_h}{H} \left(\frac{R_{\odot}}{r}\right)^2 \exp \left(- \frac{r-R_{\odot}}{H} \right),
\end{equation}
with $H=1 R_{\odot}$. The wave energy flux is given by : 

\begin{equation}
    F_w = \rho_{\odot} v_{A,\odot} (\theta, \phi) \delta v_{\odot}^2,
\end{equation}
where $\delta v$ is the input velocity perturbation at the inner (coronal) boundary. $F_w$ is thus a function of the magnetic field that varies with latitude and longitude. Yet, we aim to get a total input energy flux around $10^5$ erg.cm$^{2}$s$^{-1}$, which corresponds to the kinetic energy flux in the distant solar wind \citep[see, e.g.][]{Reville2018}. We use a spherical grid of $[224 \times 96 \times 192]$ cells, with a uniform spacing in $(\theta, \phi)$ and a stretched grid in radius. The maximum resolution is at the base of the domain with $dr = 0.01 R_{\odot}$.

\begin{figure}
    \centering
    \includegraphics[width=0.5\textwidth]{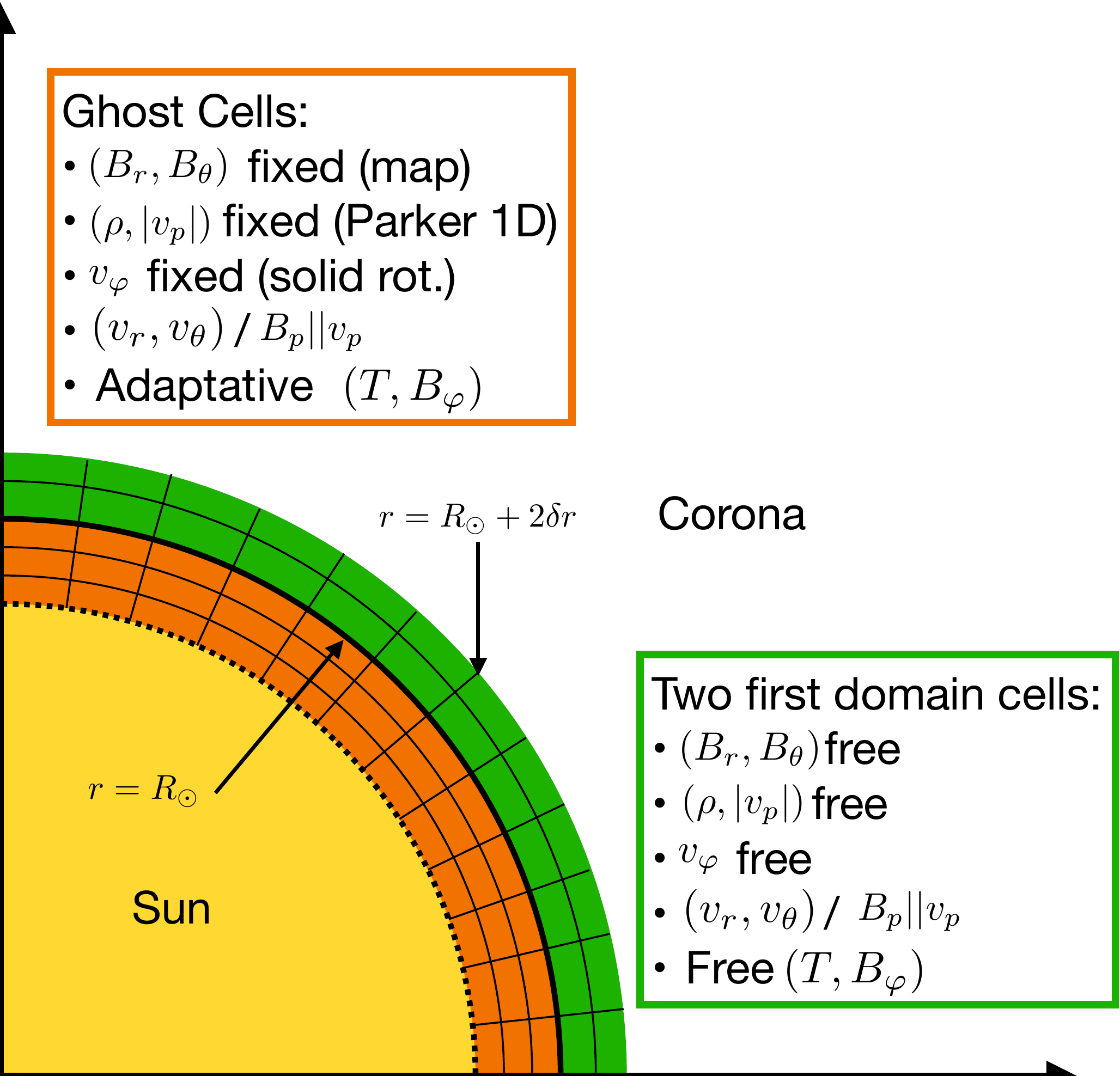}
    \caption{Schematics of the boundary conditions used in the model. The inner boundary condition is made of two zones. In the ghost cells (in orange), the equations are not advanced, but some quantities are dynamically modified. In the two-cell thick "buffer" zone in the domain (green), the azimuthal electric field remain constrained while other variables are free. More details can be found in the text.}
    \label{fig:boundary_conditions}
\end{figure}

Figure \ref{fig:boundary_conditions} describe the structure of the boundary conditions of the MHD model. It is inspired and adapted from \citet{Reville2015a}. As PLUTO uses Godunov type Riemann solvers, the inner boundary size is dependent on the reconstruction order of the solver. Here we use a parabolic reconstruction that requires three ghost cells below the numerical domain. These cells are used to set boundary conditions. In the ghost cells, we fix three quantities, the poloidal components of the magnetic field $(B_r, B_\theta)$ using a multipolar expansion of the magnetic map and the density using a 1D Parker wind model computed for a coronal temperature of 1 MK. As in \citet{Reville2015a}, we also maintain the poloidal magnetic and velocity fields parallel to each other, using the amplitude of the poloidal velocity field $v_p$. While developing the model, we found that fixing the sign of $v_p$ or changing it dynamically in the ghost cells depending on the presence of inflows or outflows in the domain, did not change significantly the solution. We thus keep it positive, the total amplitude corresponding to the (small) value of an initial 1D Parker wind. There are consequently two quantities evolved dynamically in the ghosts cells, the azimuthal magnetic field and the pressure (or temperature). These quantities take information from the solution in the domain to adapt and improve the behavior of the solution. For instance, $B_\varphi$ is modified such that $\partial B_{\varphi}/\partial r$ = cste, using the value at the first domain cell and propagating it into the boundary. This ensures a good conservation of the angular momentum carried along open field lines \citep[see, e.g.,][]{Reville2015a}. The pressure inside the ghost cells is updated using the following equation:
\begin{equation}
    P_{i,j,k} = P_{i+1,j,k} + \delta r \rho \mathbf{g}_{i+1/2,j,k} + \delta P_w,
\end{equation}
from $i=-1$ to $i=-3$, $i$ being the index of the radial direction (and $i=0$ the first domain cell). We include in this balance the wave pressure and its variation $\delta P_w$ across two adjacent cells separated by $\delta r$. This makes the thermal profile of the solution in the ghost cells close to a radial hydrostatic equilibrium. We add a "buffer" layer of two domain cells, where only the poloidal velocity field is constrained to be parallel to the poloidal magnetic field. All others variables are free to evolve. Moreover, Riemann solvers are, by design, able to handle discontinuities. At the boundaries, the solver sometimes favor the solution coming from the outer domain and allow, for instance, inflows in the first few cells of the computational domain, to balance thermal conduction, radiation losses and heating.

\begin{figure}
    \centering
    \includegraphics[width=0.8\textwidth]{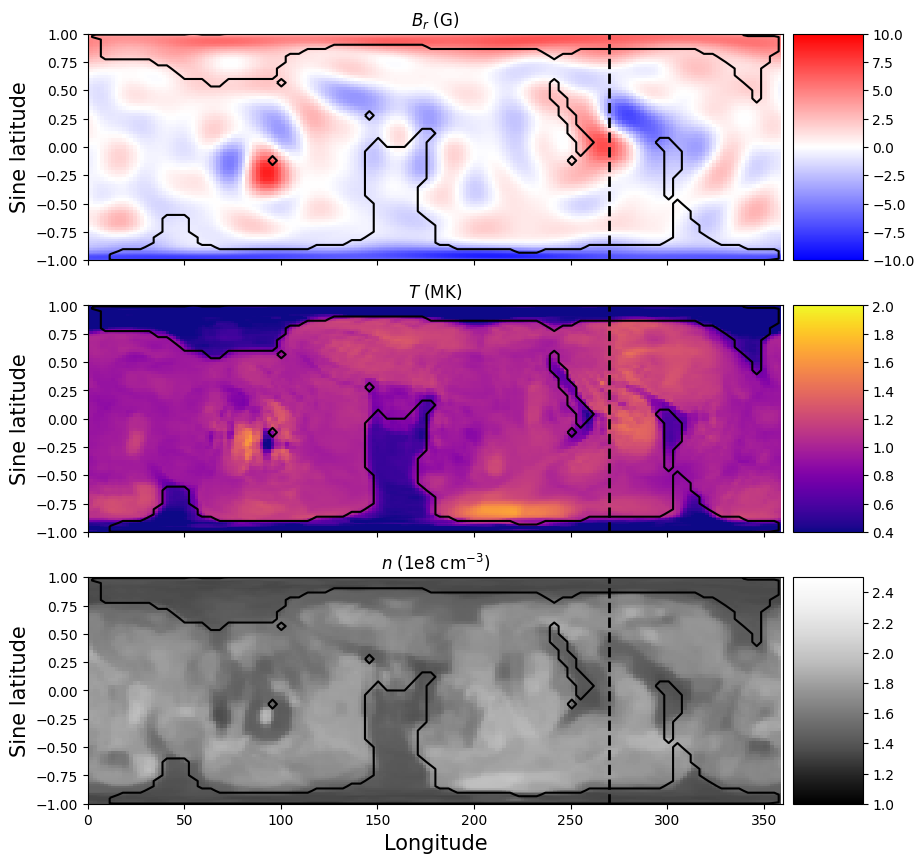}
    \caption{Inner boundary of the reference model (model 2) on the first cell of the simulation. The top panel shows the radial magnetic field at the base of the corona. The middle and bottom panel shows the temperature and density, where we can see contrast between open and close regions. The black line marks the open/closed field line boundary. The black dashed vertical line indicate the position of the west limb, discussed in details further in the paper.}
    \label{fig:bc_temp}
\end{figure}

In Figure \ref{fig:bc_temp}, we show the structure of the first cell in the computational domain of one of our simulation (model 2, see Table \ref{table:MHDparams}). The top panel shows the ADAPT map of November 6th 2018 at 12:00 UTC projected on 15 spherical harmonics and used as a fixed inner boundary condition for the poloidal magnetic field. The middle and bottom panel show the temperature and density of the inner boundary. There is a strong contrast in temperature, enabled by our adaptive boundary condition, and the inner temperature varies between 0.4 and 1.5 MK, with lower temperatures in coronal holes and higher inside closed field regions as expected. Our model is thus only meant to describe what is above the transition region, which is consistent with the location of the first cell center at 3500 km above the photosphere. The density contrast is less important but still follows the structure of open and closed field regions.

\begin{table}
\caption{Input parameters of the MHD models}
\label{table:MHDparams}
\center
\setlength{\tabcolsep}{3pt}
\begin{tabular}{cccc}
\hline\hline
Parameter & Model 1 & Model 2 & Model 3 \\
\hline
$\delta v_{\odot}$ (km/s) & 48 & 36 & 36 \\
$\rho_{\odot}$ ($10^8 m_p$ cm$^{-3}$)  & 1 & 2 & 3 \\
$ \langle B_r(R_{\odot}) \rangle$ (G) &  1.8 & 1.8  & 1.8 \\
$F_h$ ($10^5$ erg.cm$^{-2}$.s$^{-1}$) &  0.2 & 0.2 & 0.2 \\
$\langle F_w \rangle$ ($10^5$ erg.cm$^{-2}$.s$^{-1}$) & 1.5 & 1.2 & 1.5\\
\hline
\end{tabular}
\end{table}

In this work, we analyze three different simulations, to better understand the possible range of predicted white light and EUV emission and compare them with observations. The three simulations are identical except for the base density and the base velocity perturbations. Table \ref{table:MHDparams} sums up the different parameters and the energy flux entering the simulations. Model 2 is our reference model with a total input energy flux of $F=F_w + F_h = 1.4 \times 10^5$ erg.cm$^{-2}$s$^{-1}$. Model 1 and 3 use a reduced and increased base density respectively compared to the reference case. The total energy input at the inner boundary is increased by 25\% in model 1 and 3 with respect to model 2. This increase is however achieved through varying different parameters. In model 1, we decrease the base density and increase the base velocity perturbations, while in model 3 we keep the same $\delta v_{\odot} = 36$ km/s, but the base density is increased. These different choices have significant consequences on the emissions predicted by the models, for reasons we describe in the following paragraph. 

\begin{figure}
    \centering
    \includegraphics[width=1.0\textwidth]{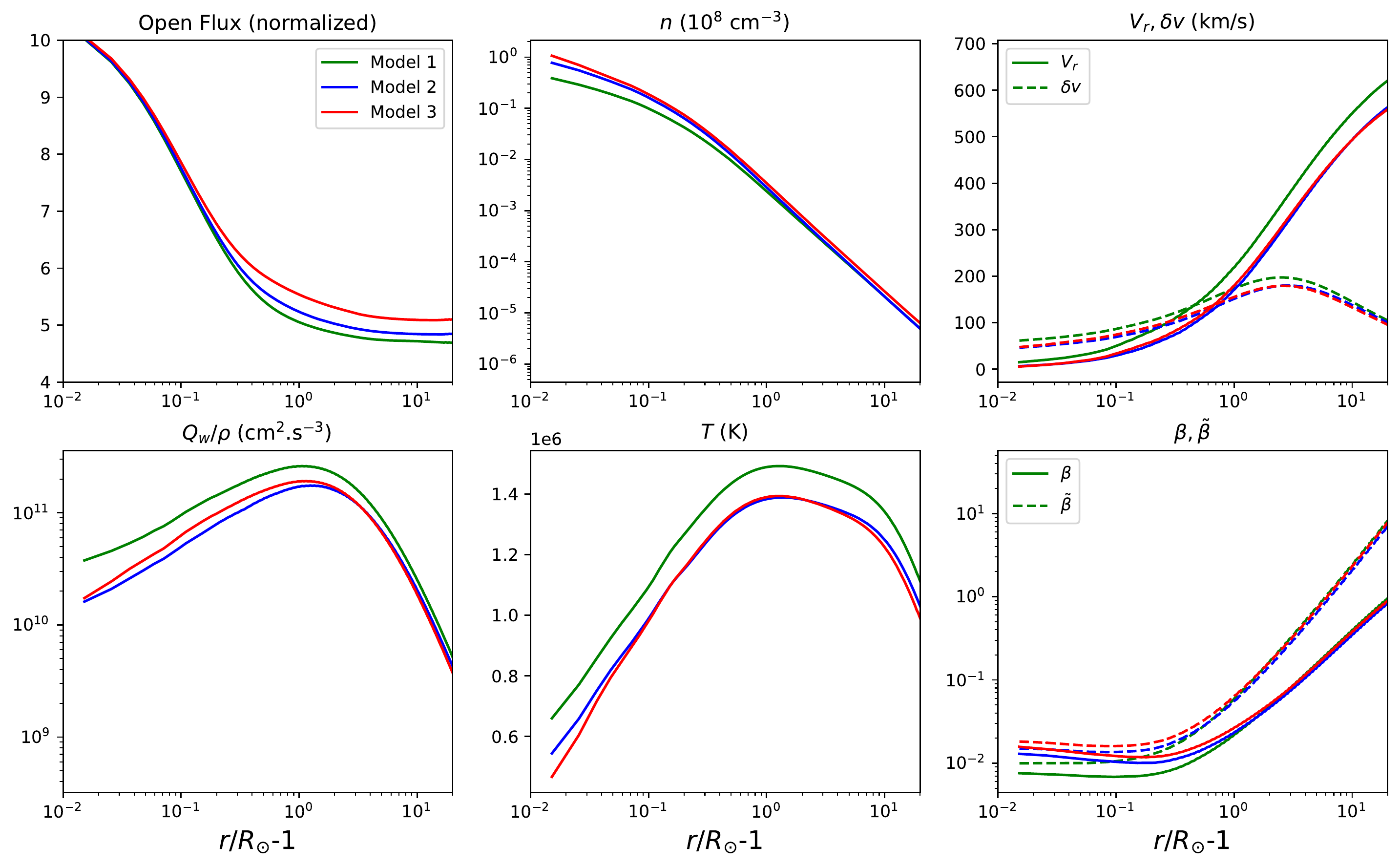}
    \caption{Top left panel: the global open flux integrated on expanding spheres. In the other panels, we show average profiles of characteristics quantities on hundreds of open field lines in the three different models. In the top panel, we show the number density, radial velocity and velocity perturbations (dashed). In the bottom row, the wave heating per unit mass, the temperature, and the plasma $\beta$ and $\tilde{\beta}$ (including the ram pressure, see text) are shown.}
    \label{fig:profiles}
\end{figure}
In the top left panel of Figure \ref{fig:profiles}, we show the open flux, i.e. the unsigned magnetic flux integrated on concentric spheres of radius $r$, as a function of $r$. In any coronal solution, the open flux decreases from its surface value until reaching a plateau, when all field lines are open by the solar wind. We see a clear increase of the open flux, with the base density going from 1 to 3. The next panels show the averaged profiles of characteristic quantities of the runs on open field lines. These profiles are averaged on several hundreds of field lines. Nonetheless, clear tendencies can be uncovered. The density profiles follow the hierarchy of the base densities. Note that the initial value of the density is slightly lower than the one imposed in the boundary condition, as we are dealing with open field only (see also Figure \ref{fig:bc_temp}). Model 3 has thus the denser wind, associated with lower velocities, and lower temperatures. In the top right panel, we plot the average velocity perturbations $\delta v(r)$ in the domain, which are logically higher in model 1, because of the larger inner boundary value. Model 2 and 3 average velocity perturbations remain close, although slightly higher in model 2 beyond several solar radii. It is interesting to note that, even though model 1 and 3 have the same input energy, the denser model (model 3) will have more intense EUV coronal emissions, because of their scaling as $n^2$, (see section \ref{sec:mod_ima}). Finally, in the bottom right panel, we show the different profiles of the $\beta= P_{\mathrm{th}} / P_{\mathrm{mag}}$ and $\tilde{\beta}= (P_{\mathrm{th}} + P_{\mathrm{ram}}) / P_{\mathrm{mag}}$, where $P_{\mathrm{ram}}=\rho v^2 /2$ is the ram pressure. Here again, the base density seems to set the hierarchy of the profiles, and higher $\beta$ for model 3 can explain more flux opening in this case.

\begin{figure}
    \centering
    \includegraphics[width=0.8\textwidth]{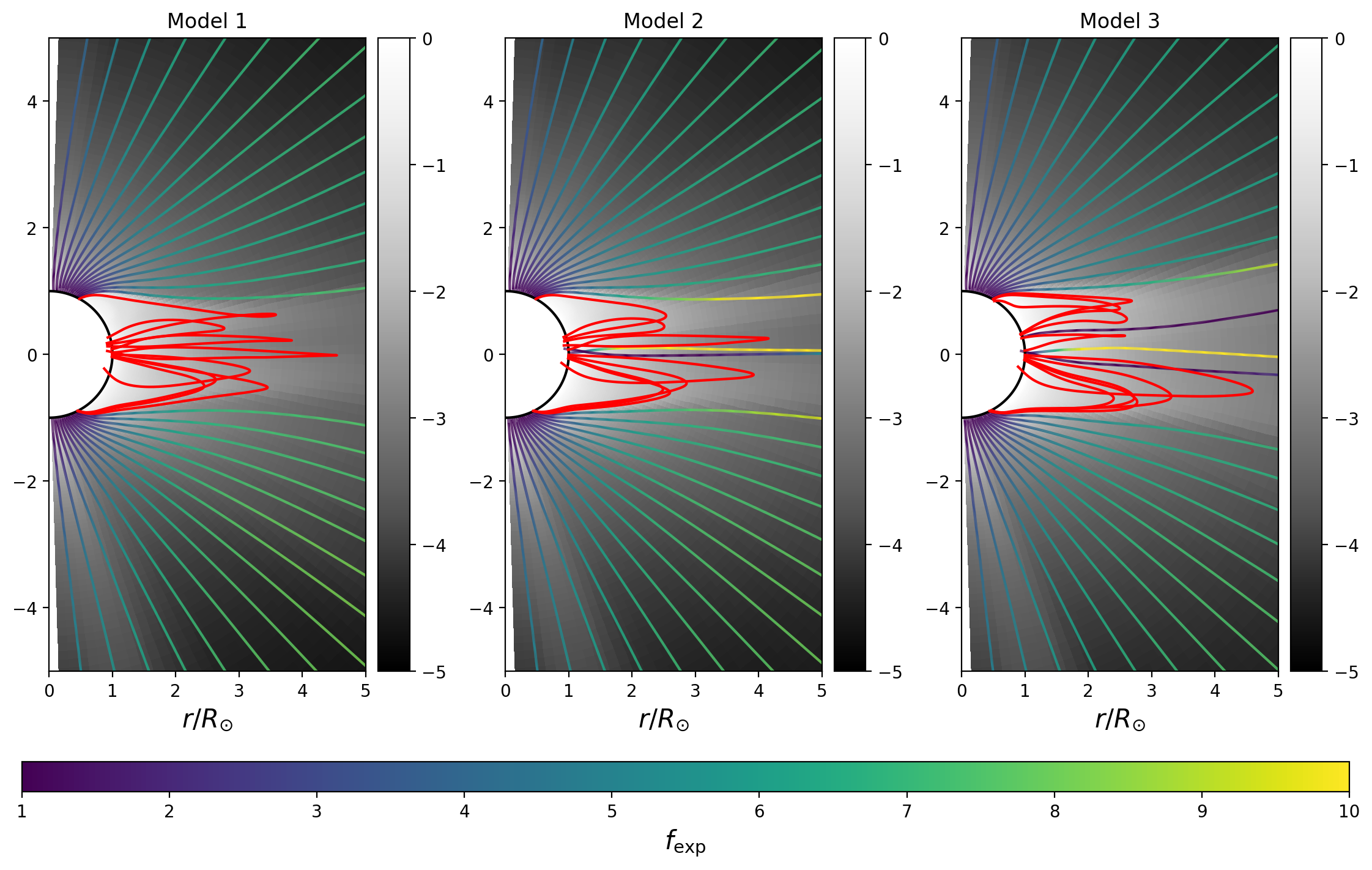}
    \caption{Structure of the coronal magnetic field for the three models on the west limb. The background shows the density logarithm (in units of $10^8$ cm$^{-3}$) in gray scales. Field lines are traced from $1.5 R_{\odot}$, and projected in the plane of the west limb. Closed field lines are colored in red, while open field lines are colored by their local expansion factor in a scale from 1 to 10.}
    \label{fig:Expansion}
\end{figure}

As a consequence, the three models display different coronal structures and extension of coronal streamers. Preparing for following white light analyses, we show in Figure \ref{fig:Expansion}, the coronal structure on the west limb of the simulations. The background color shows the log density in gray scales. 3D field lines are computed from source points on the west limb plane, at $2.3 R_{\odot}$ and projected back on the plane. Closed field lines are shown in red, while open field lines are colored with their local expansion factor

\begin{equation}
    f_{\mathrm{exp}}(s) = \frac{B_0}{B(s)} \left( \frac{r_0}{s} \right)^2,
\end{equation}
where $s$ is the curvilinear abscissa along the field line. The west limb cut is shown in Figure \ref{fig:bc_temp} with the dashed lines and is close to the coronal hole visible at 260 degrees of longitude. This structure appears in Figure \ref{fig:Expansion} with two main streamers extending up to 2 to 3 solar radii in the corona. Differences can be noted in the orientation and angular extent of the streamer structure. For model 2 and 3, several open field lines with large expansion factors (up to 10) are coming from this equatorial coronal hole. This could be the reason of the larger latitudinal angular extent of the streamer belt in model 2 and 3, as these open flux tubes push the streamers on both side of the equator. This illustrates again the importance of the  base density and location of the energy deposition in the coronal properties, as model 1 and 3 differ significantly despite having the same input wave energy flux. In the following sections, we compare all these features with multi-instruments observations. 

\subsection{Models for EUV and WL pB synthetic images}
\label{sec:mod_ima}

The core model used to create the synthetic images is {\it TomograPy\footnote{https://github.com/nbarbey/TomograPy}} \citep{barbey2013}, an open-source Python package. The model is composed by two main blocks: the first one  calculates the emission produced by the Sun and the corona; the second one calculates how much of such emission falls into the detector of a given instrument. The latter is done by the projection along a given line of sight of the cube containing the Sun and the heliospheric emissions.

The electron density and temperature cubes ($r$, $\theta$, $\phi$), output of the MHD simulation, are used to build the physical model, that is the emission $dI_j$ from the plasma within each voxel $j$ of the cube.  For the  SDO/AIA  EUV coronal bands (see details in the next section \ref{sec:data}) we assumed that excitation by collision of electrons and ions followed by spontaneous emission are the dominant process for the observed emission: 
\begin{equation}
\label{eq:i}
    dI_j = A_f G(T_j, n_j) n_j^2 dV
\end{equation}

\noindent
where $A_f$ is the spectral response function of the instrument we want to simulate, $G(T, n)$ contains all the atomic physics involved in the process of the spectral line formation and is a function of the local electron density ($n$, number electron density $\rho_{\odot}/m_p$) and temperature ($T$). $A_f$ is provided by the SDO/AIA instrument team (we used the version 10 distributed in SolarSoft library), $G(T, n)$ is calculated using the CHIANTI atomic physics database version 9.0 \citep{Dere1997, Dere2019}, assuming ionization equilibrium and coronal composition, $n_j$ and $T_j$ are provided from the output of the MHD model.

The AIA images are simulated using  128 $\times$ 128 pixels format and unit of Dn/s, which is the unit of the calibrated data. The cube representing the corona has 810 voxels side representing a length of 7 R$_\odot$.

We convolved our synthetic images with the Point Spread Function provided by the instrument team  to test the modification of intensity distribution in the neighboring pixels and found  that the effect on our binned images is not very important. This is shown in Appendix \ref{app:aia_PSF}.
In the plots in this work showing the synthetic intensities we also plot the uncertainty of 35\% \citep{guennou2012}, originates from propagating the uncertainties in the atomic physics (25\%) and instrument radiometric calibration (25\%). 

The WL pB solar emission is formed under a different physical process, that is, by Thomson scattering: the scattering of photospheric radiation by free coronal electrons. The model for calculating this contribution follows \cite{Billings1966} and it is summarized in \cite{barbey2013} as part of the {\it TomograPy} package. In this case, the emission $dI_j$ is a function of the photospheric disk intensity, the geometry between the incoming photon and the scattering electron, the Thomson-scattering cross-section, the solar distance, and the local density. Thus, the model uses as input the density obtained from the  the cube provided by the output of the MHD simulation. 

The second block in the model integrates $dI_j$ from each voxel along the line of sight corresponding to each pixel of the simulated instrument detector. For the K-Cor synthetic images, we have chosen a larger format of $256\times$256 pixels, while the original simulation cube representing the Sun and the corona has 24 $R_\odot$ side. The LASCO C2 images were built in $512\times512$ pixels format, corresponding to a coronal cube side of 49.6 $R_\odot$.

\section{Solar corona observations}
\label{sec:data}

The observations of reference used for this study were selected during the Parker Solar Probe \citep[PSP, ][]{Fox2016} first perihelion passage on November 6th 2018. 
For our study, we selected both EUV and WL pb images in order to be able to compare our simulation results with the observations both on disc and off--limb covering several solar radii. The different formation process for the two spectral bands also allows investigating temperature and density effects on the intensity variation, as it will be discussed later. 

We compare the EUV and pB WL synthesized images with the EUV SDO/AIA 193\,\AA\, 171\,\AA\, and 211\,\AA\, channels, covering a radial distance up to about 1.5 R$_\odot$, the HAO COSMO K-Coronagraph pB (K-Cor,  $1.05 < R_\odot < 1.5$) and SOHO/LASCO C2 pB ($2.2 < R_\odot < 6.5$)
 
The data used for November 6 analysis  (LASCO) and November 7 (AIA and K-Cor) are shown in Figure \ref{fig:exe_m2}.

The SDO AIA observational data are level L1 format, meaning that they have been corrected for instrumental effects and radiometrically calibrated (units of Data Number, $Dn$). 

The  193\,\AA\ band was chosen for the main work because it has the response function which peaks at the \ion{Fe}{12} temperature (see Figure \ref{aia_resf}), that is about 1.5\,MK. This is the average temperature of the corona, so this band provides an excellent map of this region. We also use the 171\,\AA\ (with the main peak at \ion{Fe}{9} temperature, 0.9\,MK), and 211\,\AA\ (with the main peak at \ion{Fe}{14}, 2\,MK), channels to show temperature effects within the range of validity of our model.

The images' field of view is made of $4096\times4096$ pixels with a spatial resolution of about 1.5''. The comparison with the models is made using binned data, that is using images of $128\times 128$ pixels. We adopted the Poisson noise as the error on the data converted in unit of photons.

We selected L1.5 HAO/K-Cor pB data which are fully calibrated to B/B$_0$ in the 720--750\,nm spectral range ($B_0$ is the mean solar brightness, \citet{Lamy2014}). These data are processed for instrumental effect and calibration, as well as to remove sky polarization, correction for sky transmission.
For the K-Cor data there are no available data on November 6th, that is the PSP perihelion. We selected the closest data available, which are 10\,min averaged on November 7th at 20:20\,UTC (Carrington longitude of 202$^\circ$). For this reason, we modeled EUV and pB WL data for both November 6th and 7th. The estimated background noise in the polarization brightness is $3\times 10^{-9}\,B/B_0$. The images are 1024 $\times$ 1024 pixels size. For the comparison with our simulations, the original observations were binned by 4. 

SOHO/LASCO-C2 pB data (540--640 nm, \citet{Brueckner1995}) were retrieved from LASCO-C2 legacy archive\footnote{http://idoc-lasco.ias.u-psud.fr/} \citep{lamy2020} hosted by the MEDOC\footnote{https://idoc.ias.u-psud.fr/MEDOC} data and operations center. These images have 512 $\times$ 512 pixels format and have been processed to remove instrumental effect, remove sky polarization, correction for sky transmission  and calibrated to $\mathrm{10^{-10}}$ B/B$_0$ unit. The data were taken on the November 6, 2018 at 15:01 UTC (Carrington longitude 218.3$^\circ$) and November 7 at 21:07 UTC (Carrington longitude 201.5$^\circ$). The uncertainties in the pB data are not easy to establish and are discussed in \cite{lamy2020}.  \cite{frazin2002} provides uncertainties as a function of the solar distance, while in a more recent work \citet{frazin2012} provides  an estimated 15$\%$ in the whole field of view. In this work, we assumed this later estimation for the observations' error. 

The first row of Figure \ref{fig:exe_m2} shows the corona as seen by the three instruments (November 6 for LASCO, November 7 for AIA and K-Cor). The corona 
was filled with three streamers, two on the west limb and an equatorial large one on the east limb. In this work we name SI the north-west streamer, SII the south-west streamer and SIII the east streamer (see also Figure \ref{fig:lasco_pa_str}).

\section{Global performance of the simulations}
\label{sec:res_gen}

This section is dedicated to provide the global results of the simulations and a first general comparison with the observational data. 

\begin{figure*}[ht]\centering
\includegraphics[width=0.31\textwidth, trim=0 0 10 0, clip]{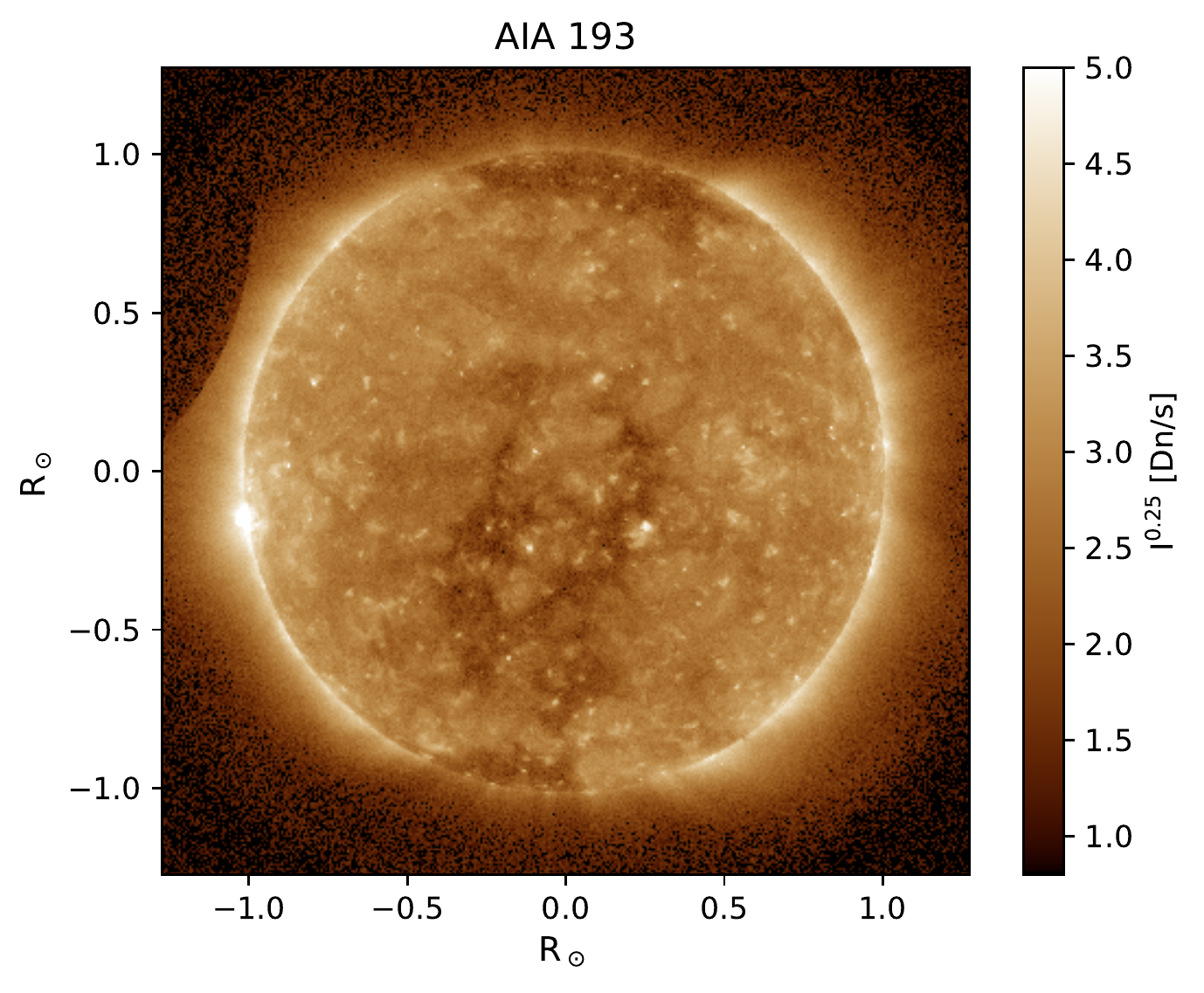}
\includegraphics[width=0.32\textwidth, trim=30 0 25 0, clip]{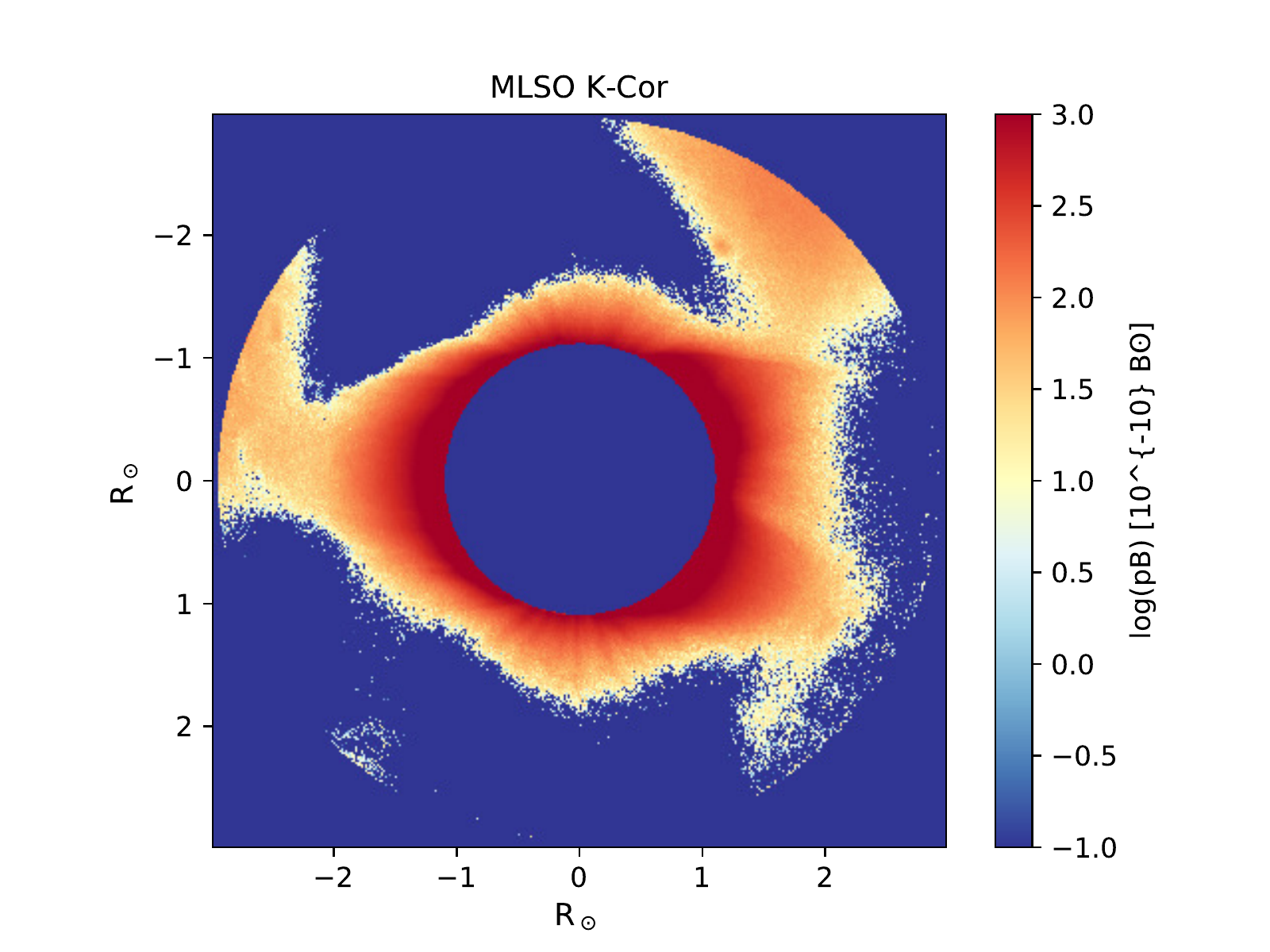}
\includegraphics[width=0.31\textwidth, trim=20 15 5 0, clip]{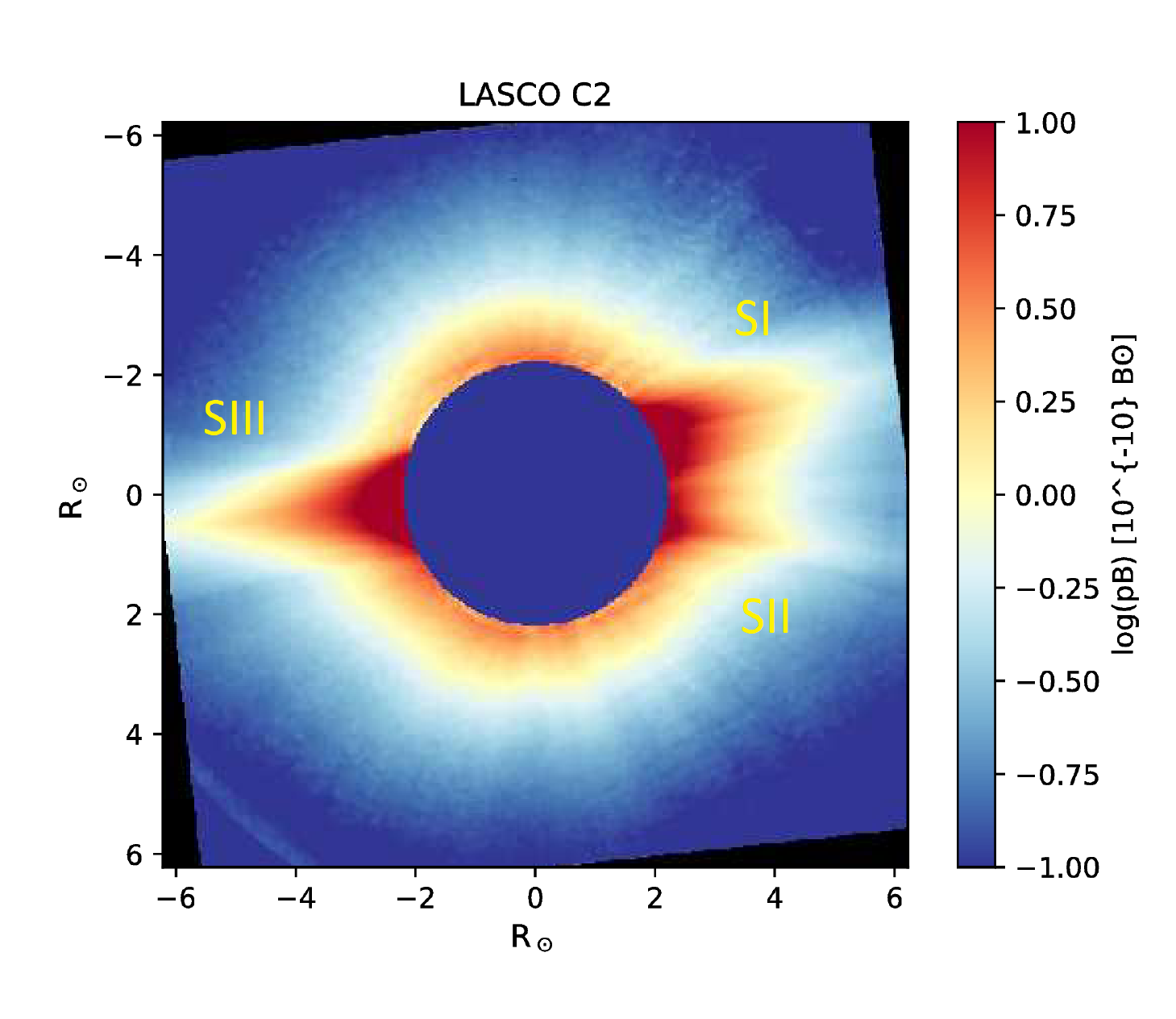}\\
\includegraphics[width=0.32\textwidth, trim= 0 0 0 0, clip]{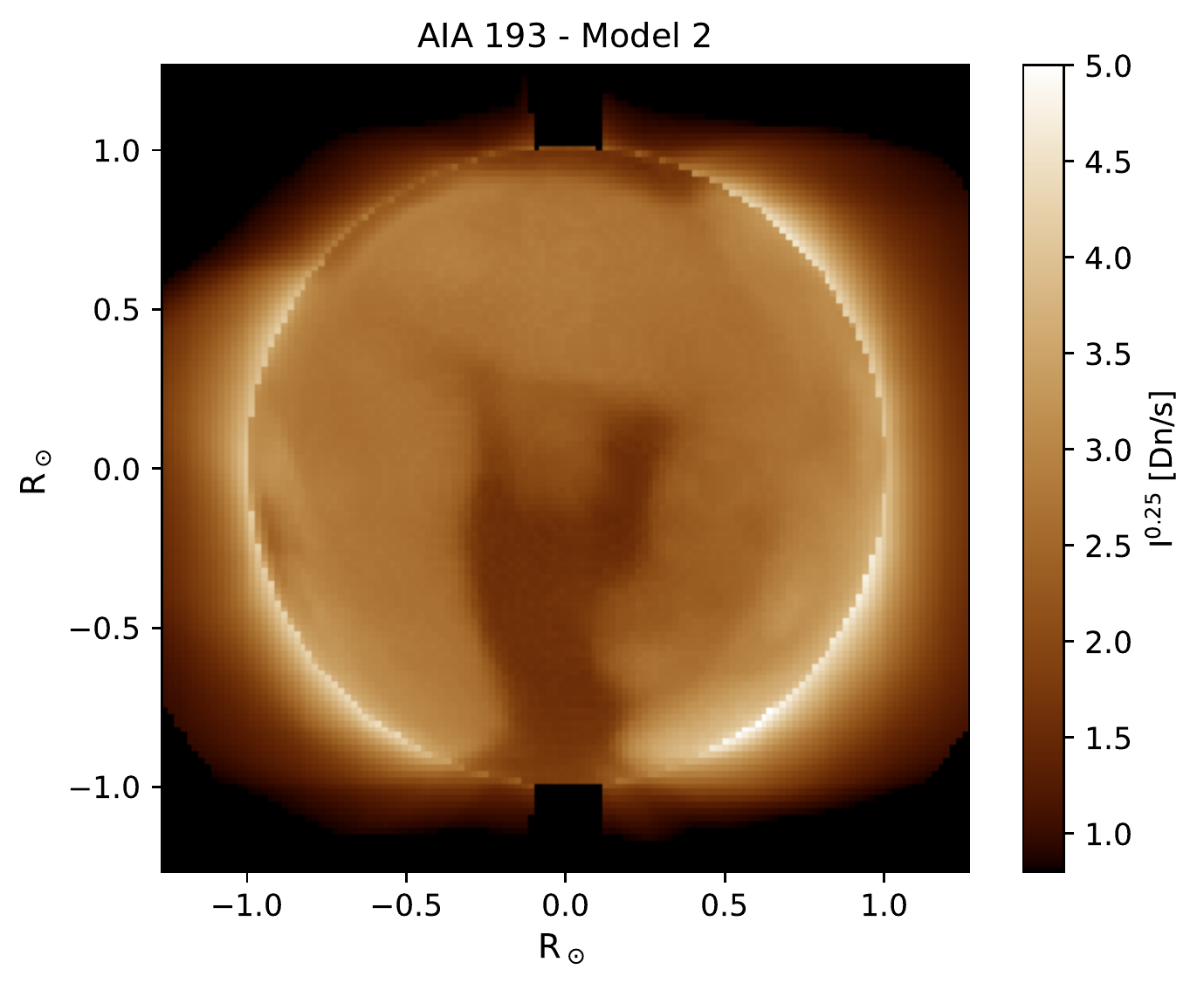}
\includegraphics[width=0.31\textwidth, trim=5 0 5 0, clip]{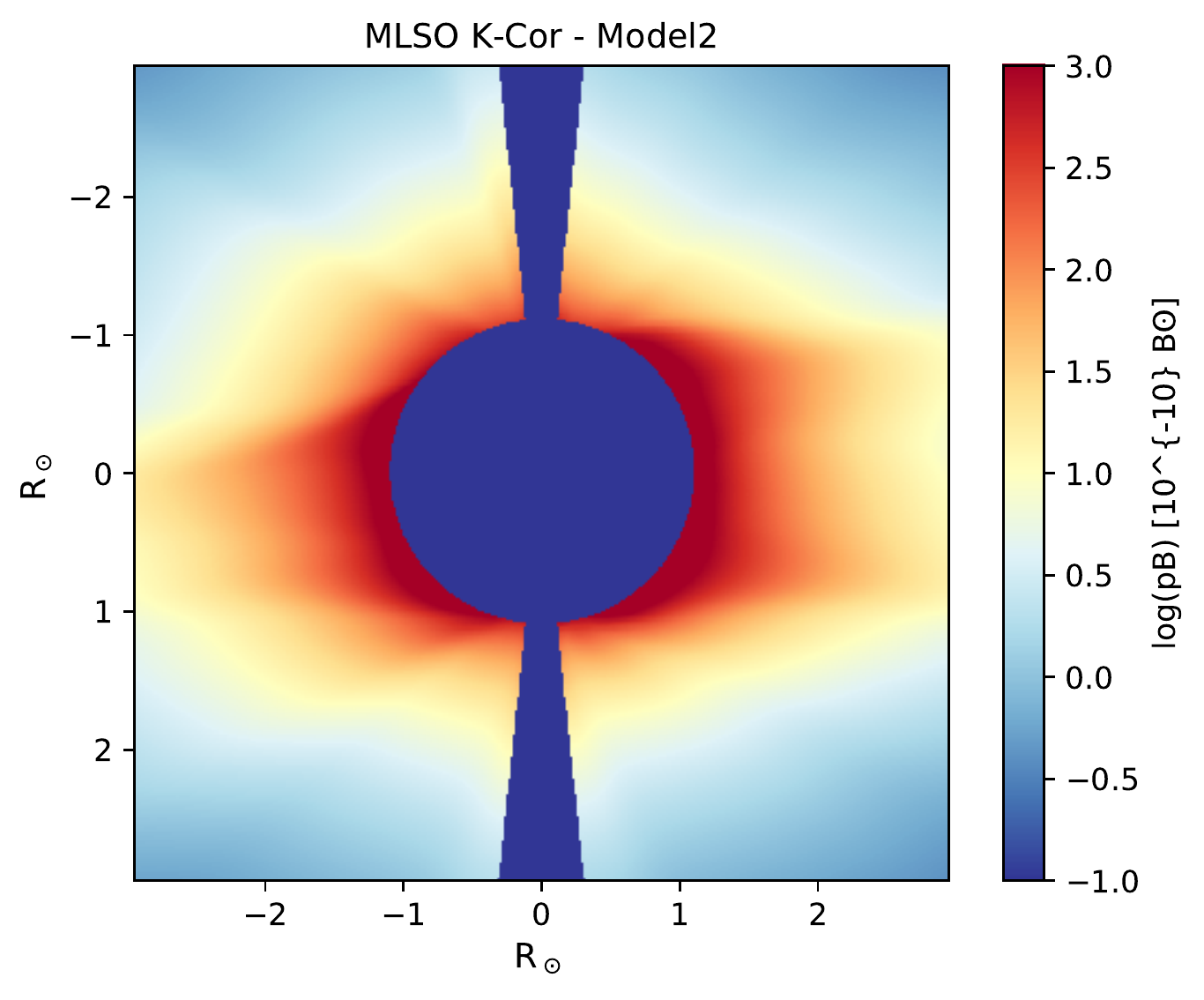}
\includegraphics[width=0.32\textwidth, trim=0 0 0 10, clip]{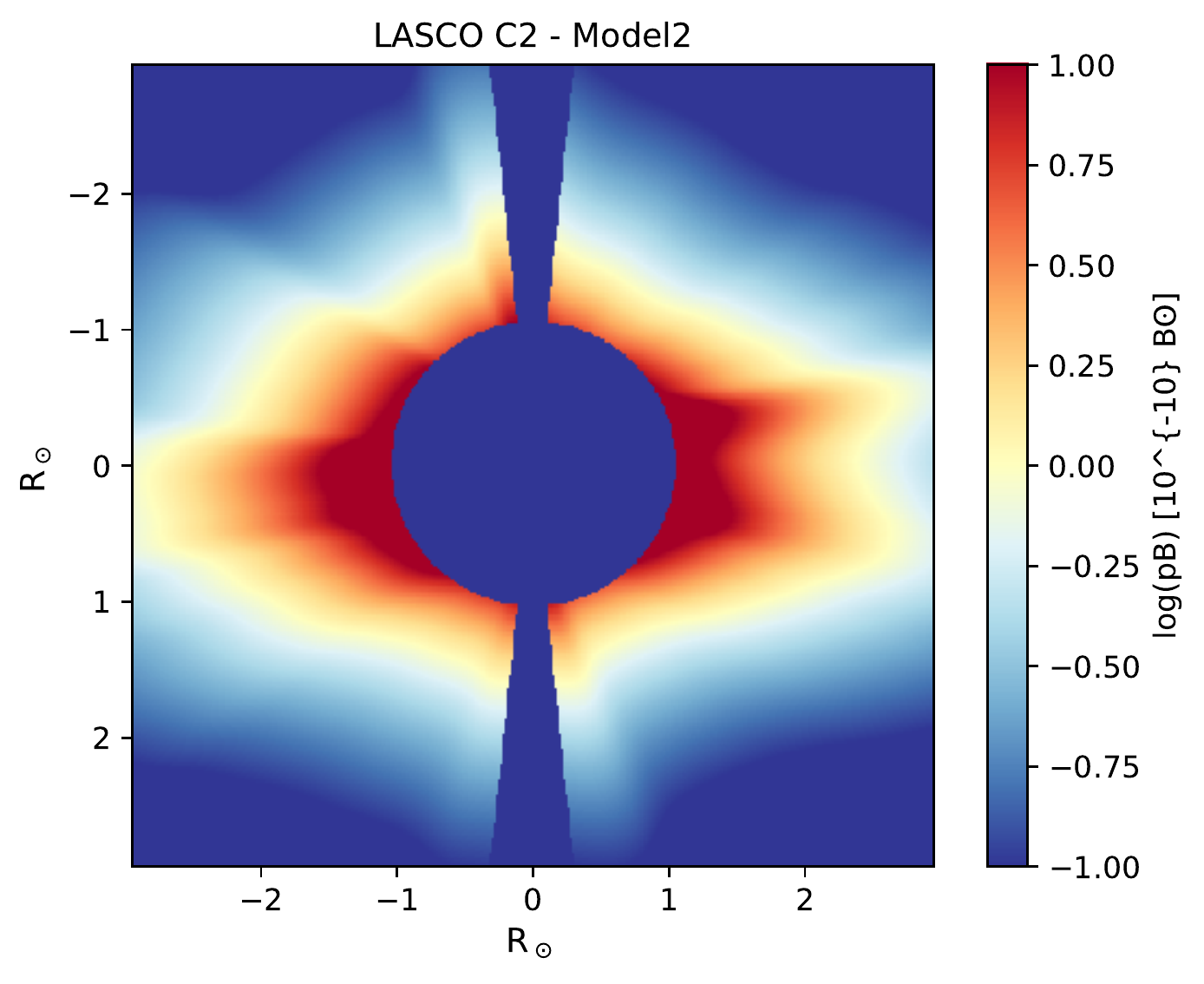}\\
\includegraphics[width=0.32\textwidth, trim=0 0 0 0, clip]{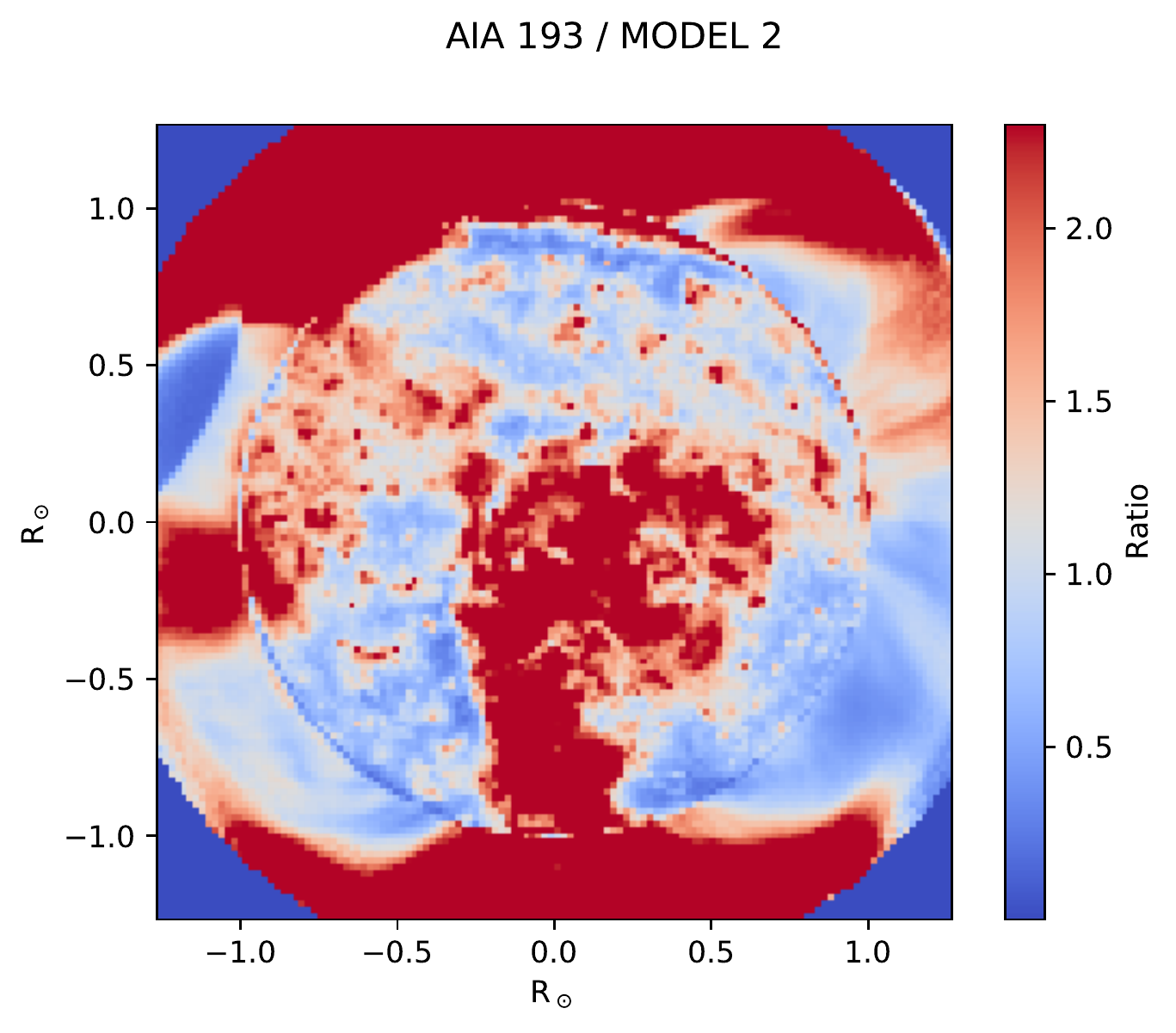}
\includegraphics[width=0.32\textwidth]{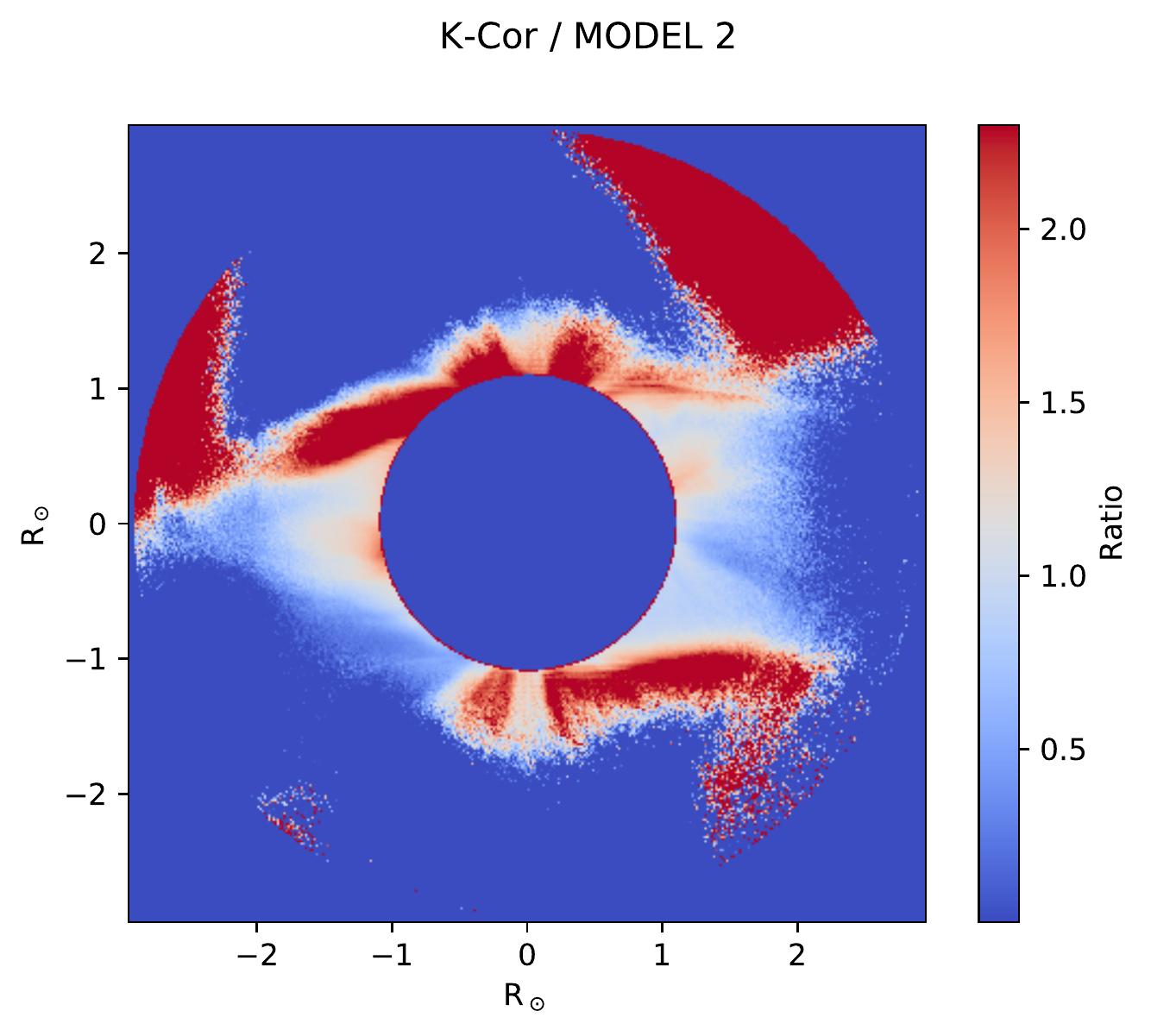}
\includegraphics[width=0.32\textwidth]{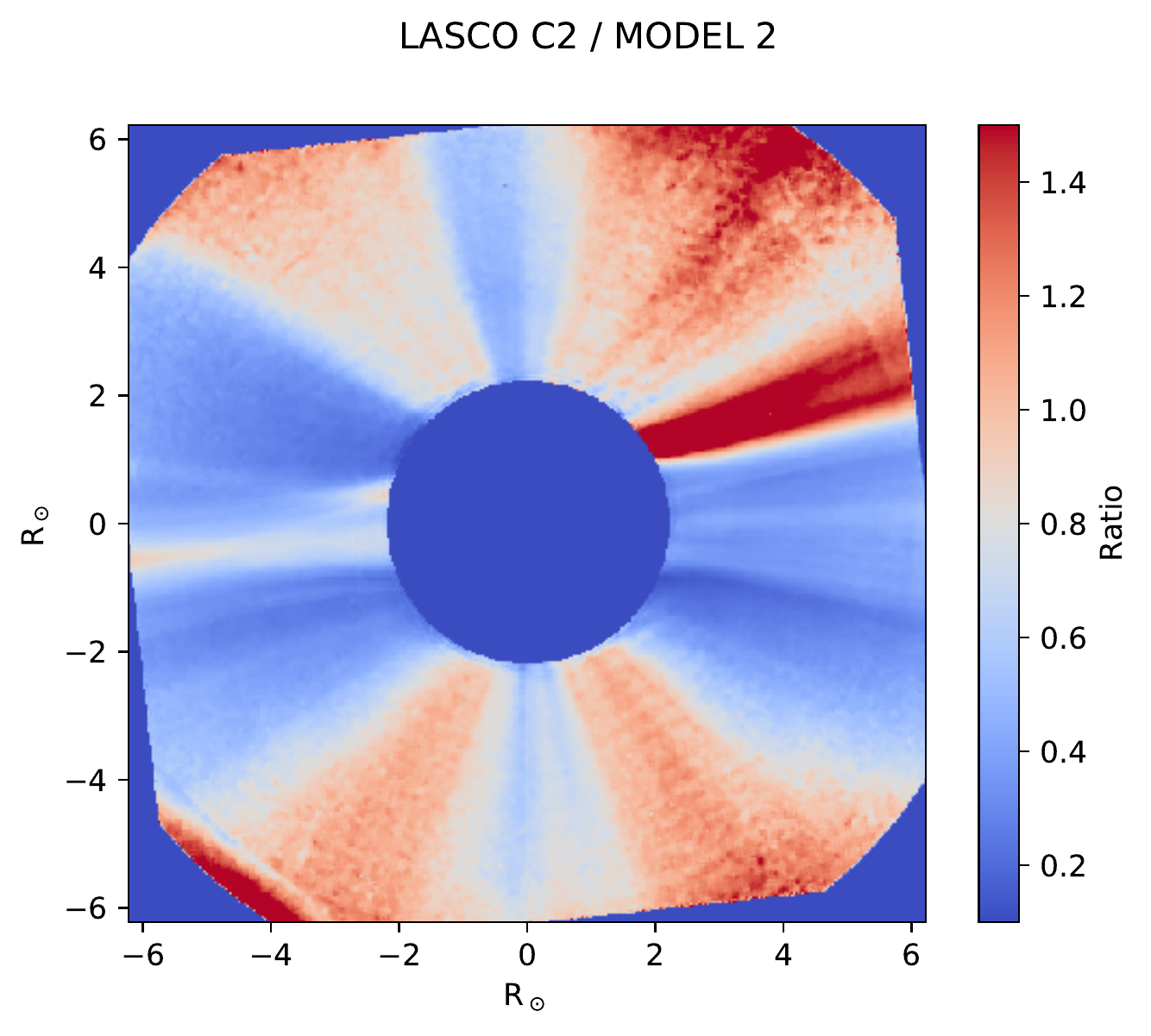}
\caption{
Top: the solar corona imaged by: EUV SDO/AIA 193 on November 7 at 20:22 UCT (left), pB COSMO K-Cor for the November 7 (middle), pB SOHO/LASCO C2 (right) for November 6. Middle: synthetic images obtained using Model 2 for the same periods. Bottom row: ratio between the observations and Model 2 outputs. The three streamers of the corona are labeled in the top right panel. }
\label{fig:exe_m2}
\end{figure*}

Figure \ref{fig:exe_m2} middle row shows the synthetic images obtained from  Model 2 which is used for a global comparison with the observations. The AIA 193 image (left panel)  represents the corona at 15:01 UTC on November 7 (the same time as the K-Cor data). A morphological comparison with the data (top row) reveals that all the large scale structures are reproduced: the polar and equatorial coronal holes (CH), the limb brightening and the base of both  west limb streamers. 
The middle and right panels show the WL pB synthetic images, which also in this case reproduce the location and size of the streamers on the west side of the corona. The east side of the WL images shows the most evident difference with the observations: the streamer is reproduced in location, but with a less compact shape. The reason for this can probably be attributed to the less accurate magnetic map at these longitudes, due to the incoming  hidden (and so not previously measured) part of the Sun.  The AIA image (top row) shows a bright spot at the East limb equator, which marks the arrival of a small sunspot. Due to its location, this small spot is indeed likely not appropriately considered by the ADAPT map.

More quantitative information are obtained by the ratio of the data to Model 2, which are shown in the third row of the figure. The quiet Sun (QS) synthetic emission in the AIA 193 is about consistent with the data, even though bright small scale spots (in red) suggest more intense emission in the observations. This is certainly due to two main elements. The temperature response of AIA 193 is also sensitive to the  transition region emission (see Appendix \ref{app:aia_tr}) while this 
is not completely reproduced by our model  setup  (the minimum temperature does not go below about 0.4 MK at the base of the corona). Secondly, the selected spatial resolution of the models (about 23 Mm at the base of the corona) is about or larger than the typical bright point diameter  \citep[see][]{Madjarska2019}.

The difference between the two images is particularly reinforced in the CHs, which are not so bright in our model than in the observations: all the small scale bright structures (plumes, bright points, jets, etc.) are blurred by the spatial resolution in the model. Furthermore, because the CH is a cool region (with an average temperature below 1 MK), it cannot be properly reproduced by our model setup. The density chosen for our simulations can also be the reason for the observed difference, this aspect will be analyzed in depth in the following sections. 

The off disk behavior of the AIA 193 ratio indicates the closed corona generally to be about or within a factor 2 more intense in the simulation than in the data. Only the north-west  streamer SI is more intense in the data. High in the corona, the K-Cor reliable data (the ones closest to the internal occulter) the ratio is close to 1, while in the LASCO case the closed corona is too bright in the model. 

These results may suggest some difference in the radial gradient within streamers between modeling and observations. Additionally, 
the different morphology of the streamers between observation and simulation can produce strong latitudinal gradient in the ratio. This is clearly the case  in the LASCO ratio for the north-west streamer. All these aspects will be further discussed in the following sections. 

It is worth nothing at this point that the ratios plotted in the figure highlight some stray light residual at the edge of the WL field of views. 

\begin{figure*}
    \centering
    \includegraphics[width=0.272\textwidth]{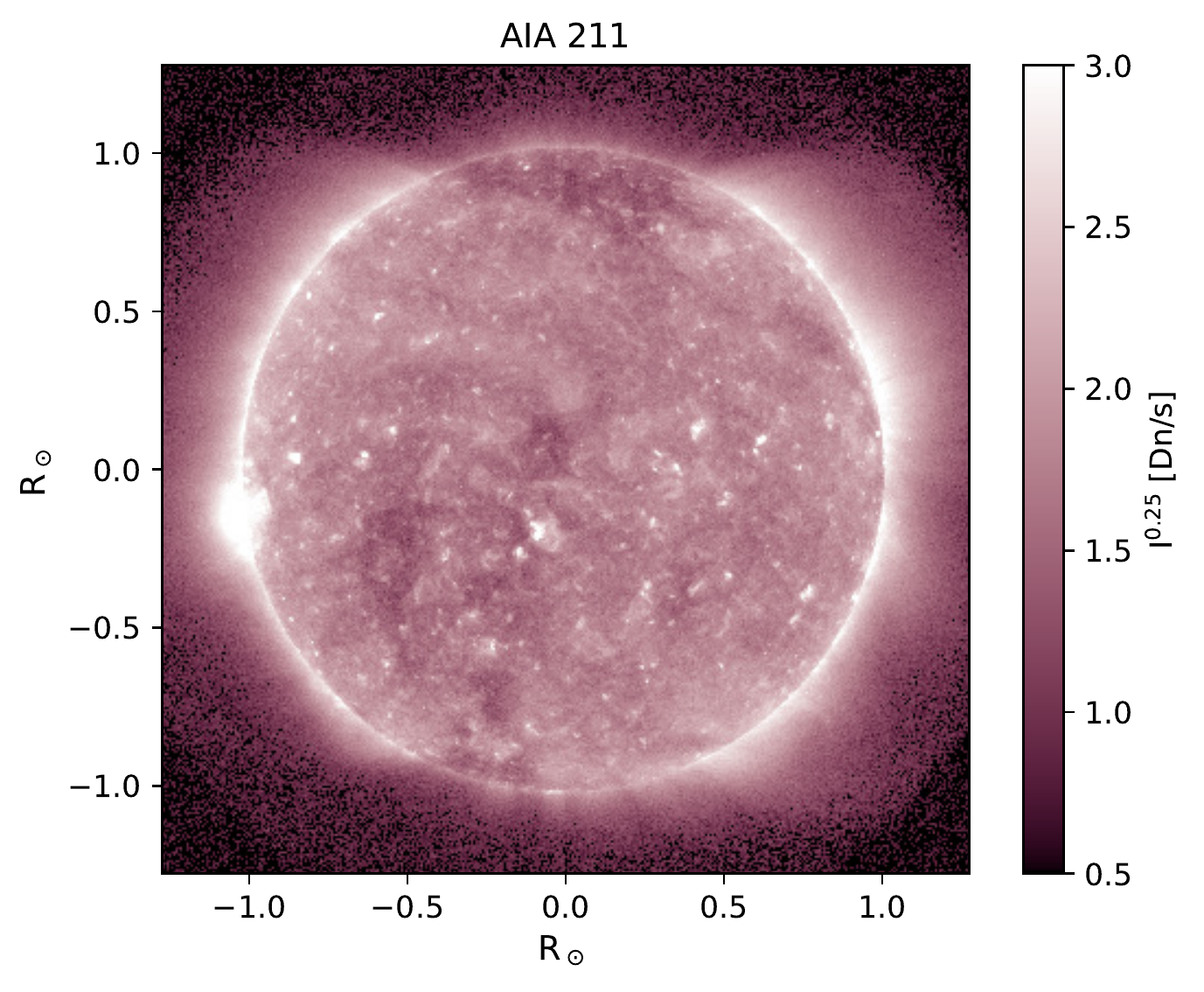}
    \includegraphics[width=0.23\textwidth]{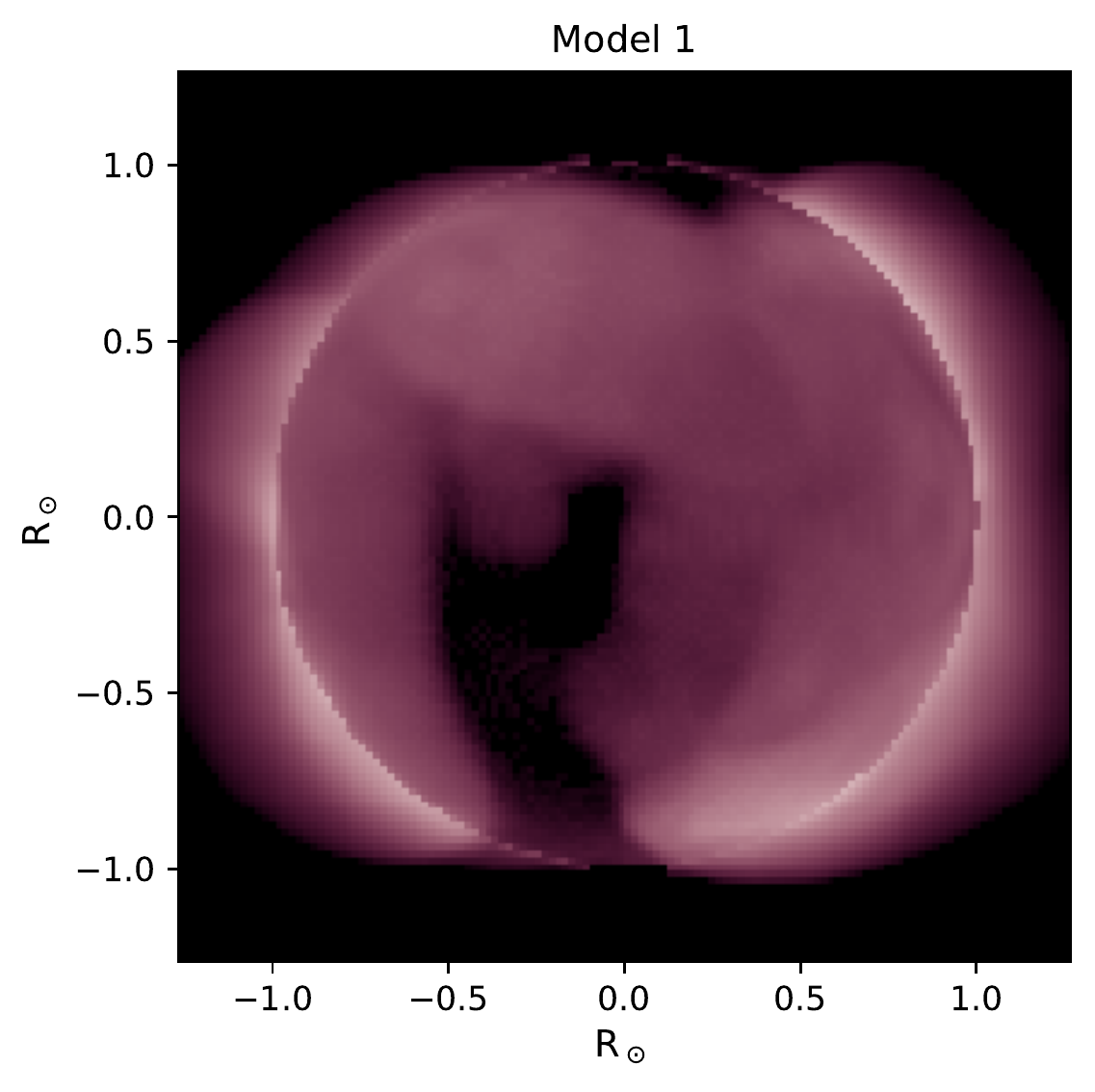}
    \includegraphics[width=0.23\textwidth]{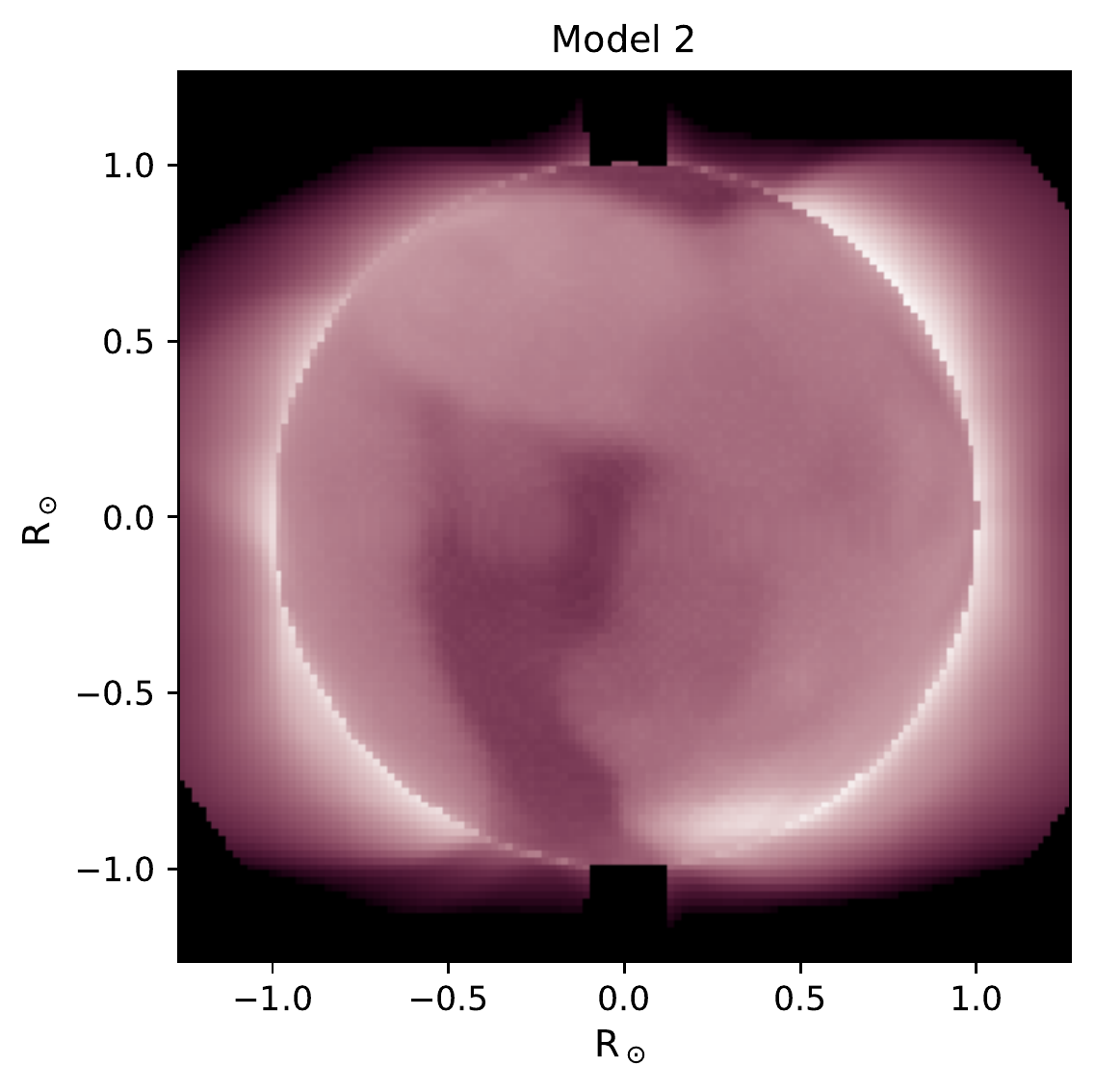}
    \includegraphics[width=0.23\textwidth]{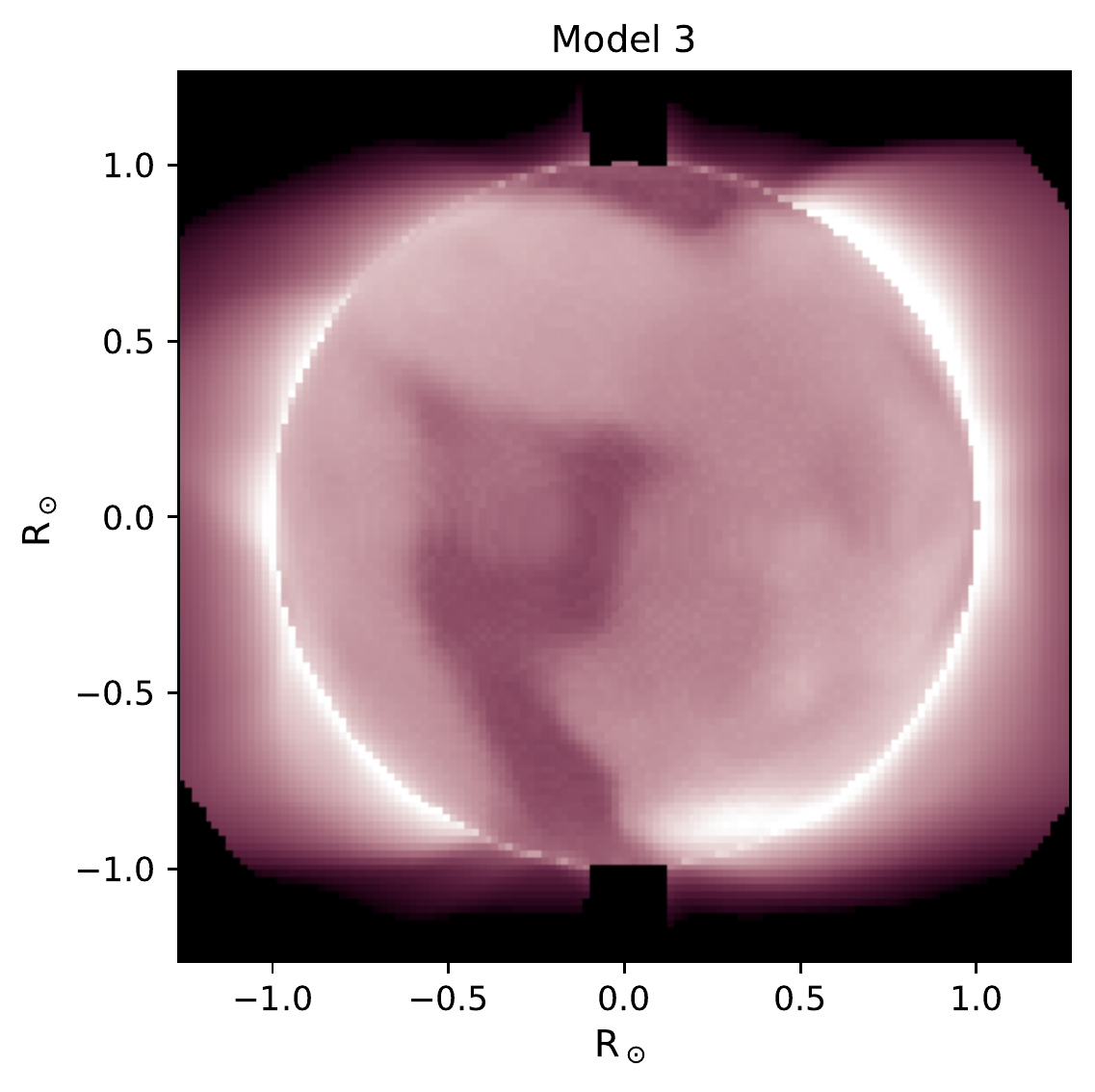}
     \includegraphics[width=0.272\textwidth, trim=0 0 10 0, clip]{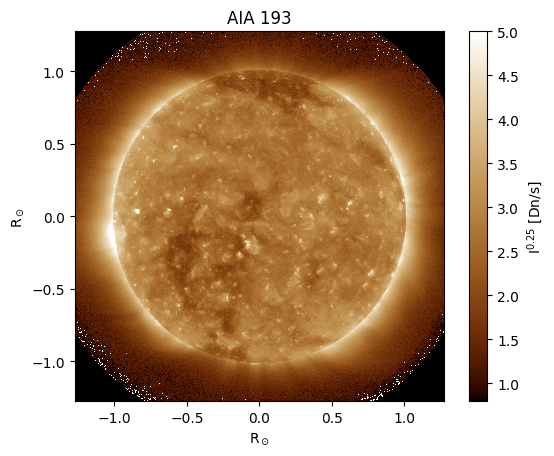}
   \includegraphics[width=0.23\textwidth]{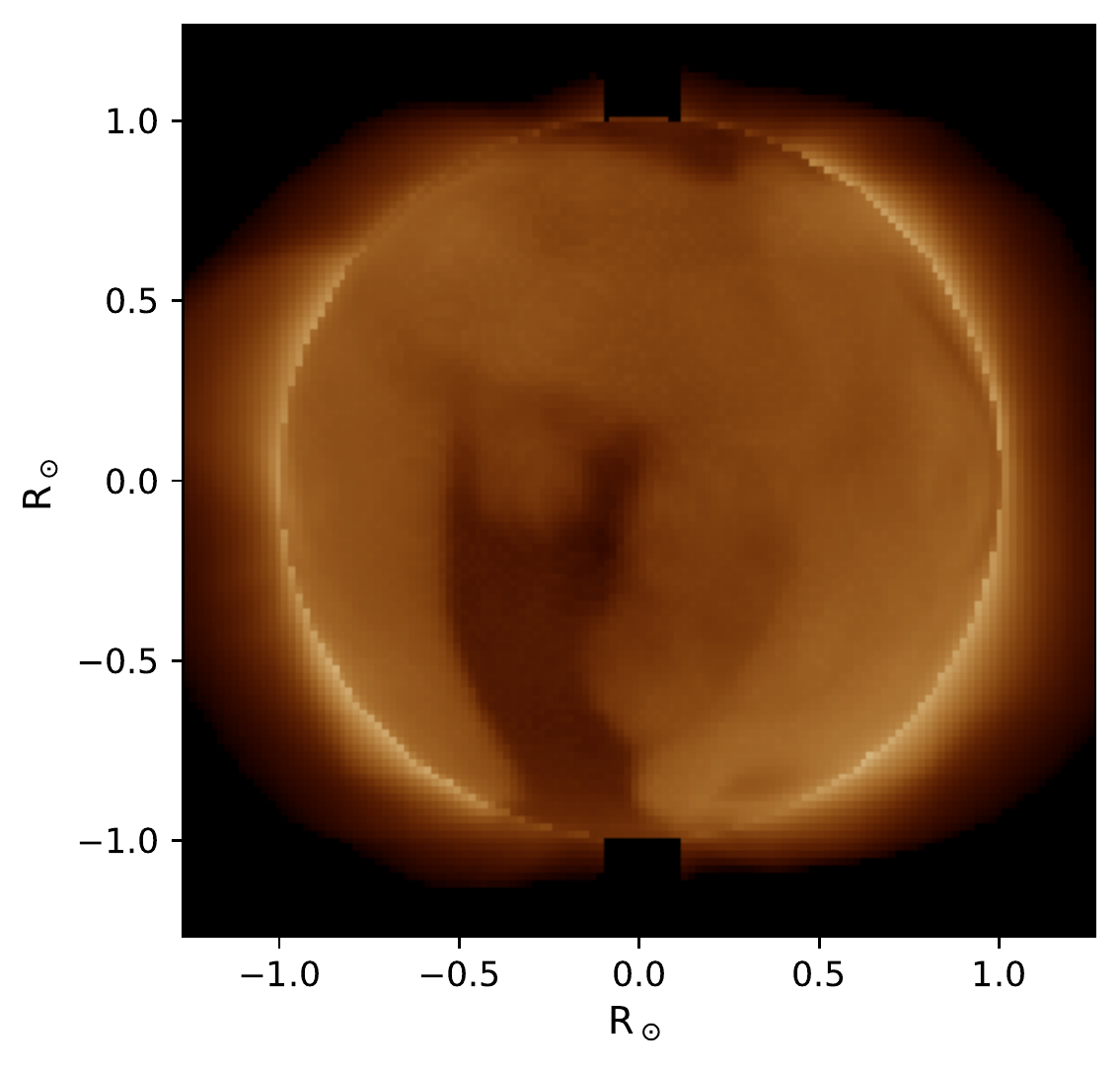}
    \includegraphics[width=0.23\textwidth]{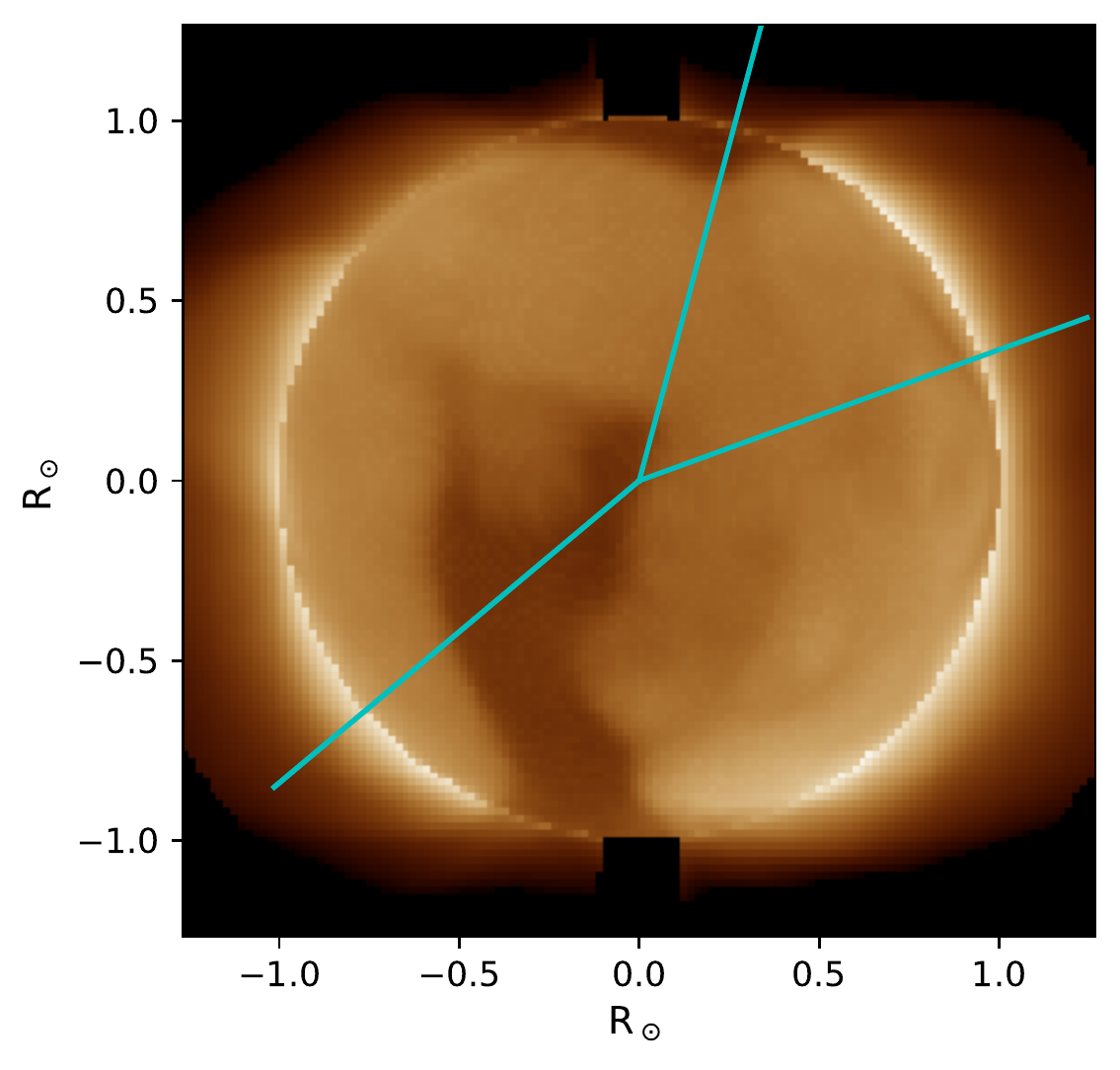}
    \includegraphics[width=0.23\textwidth]{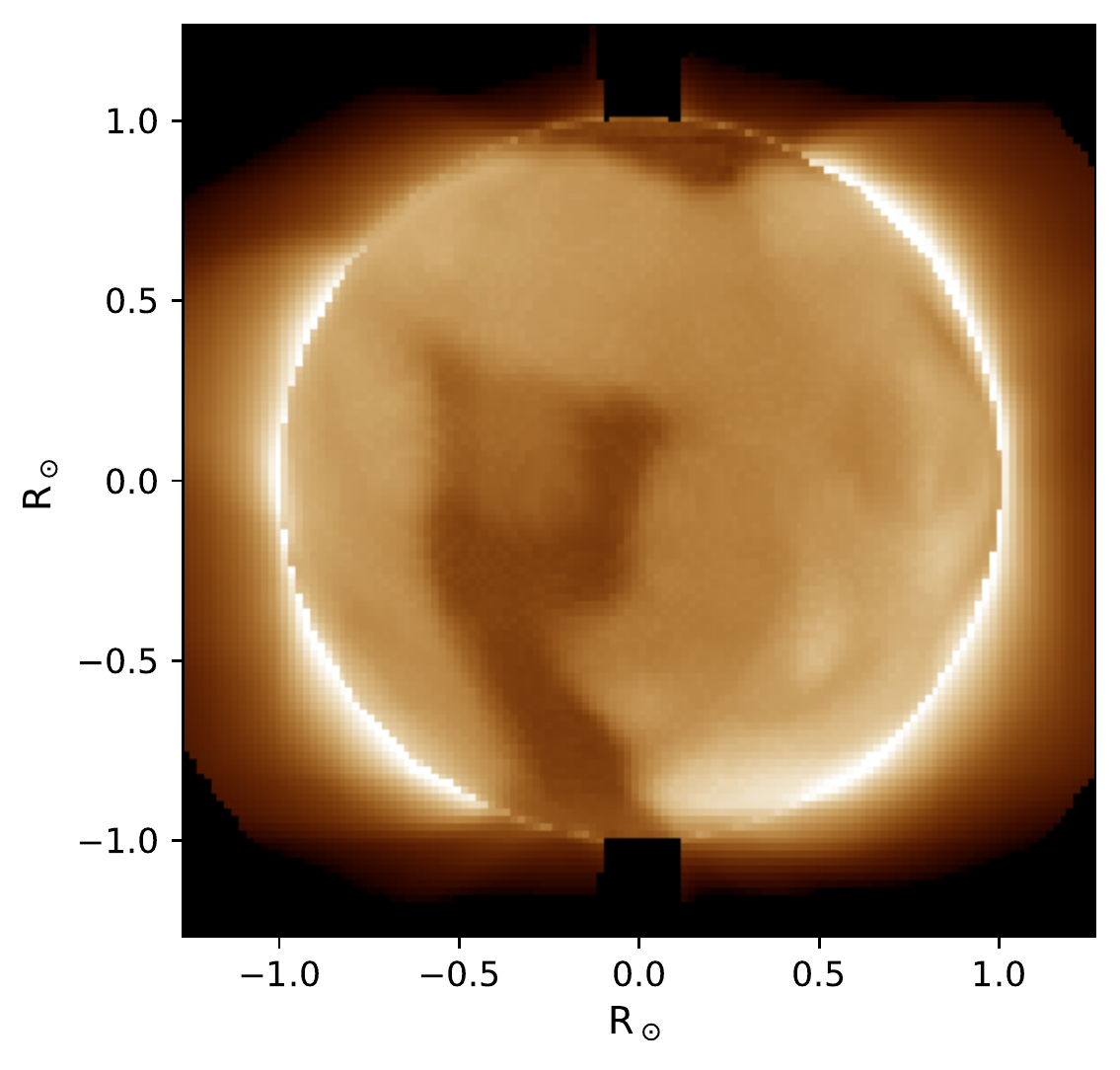}
   \includegraphics[width=0.272\textwidth]{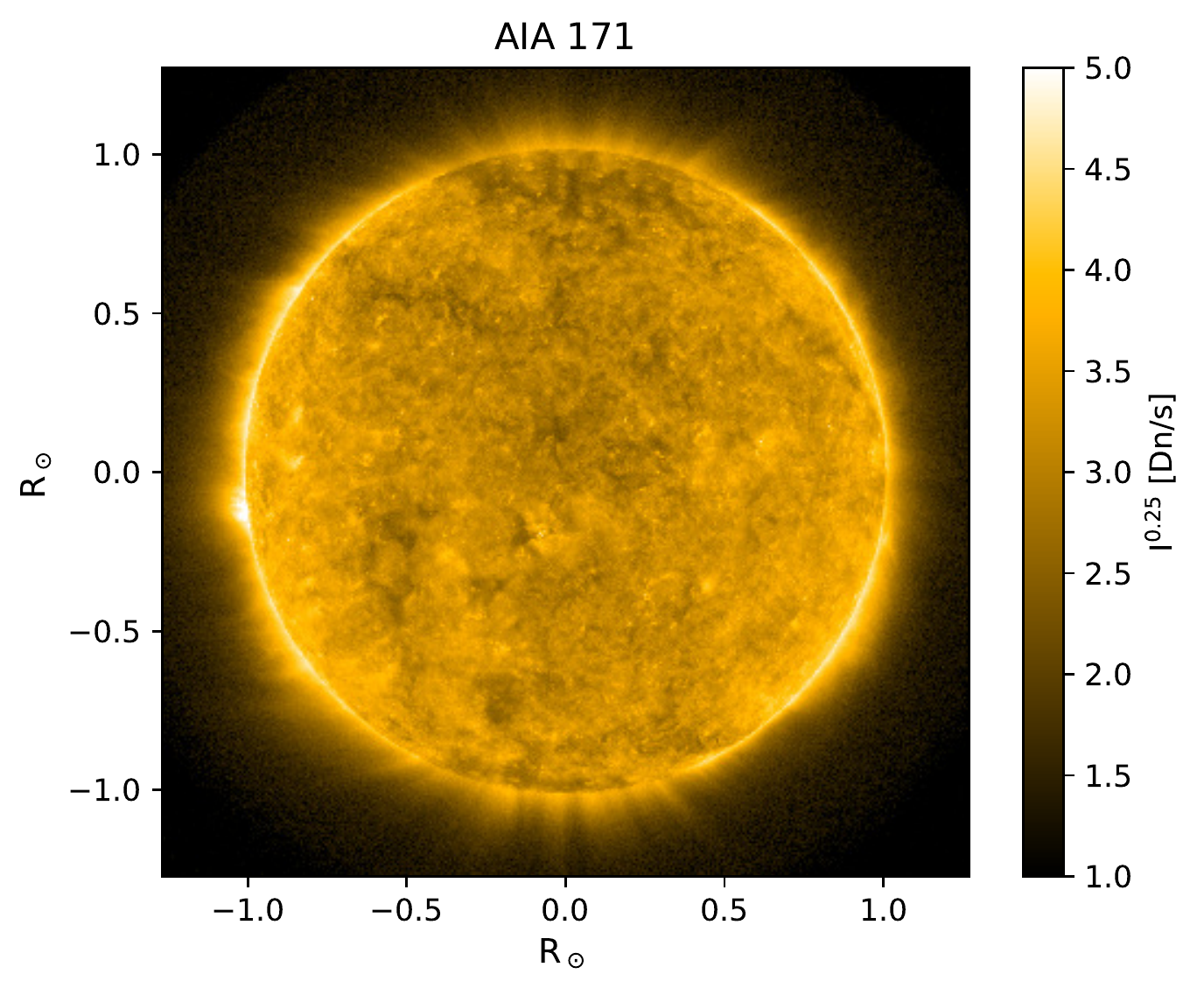}
   \includegraphics[width=0.23\textwidth]{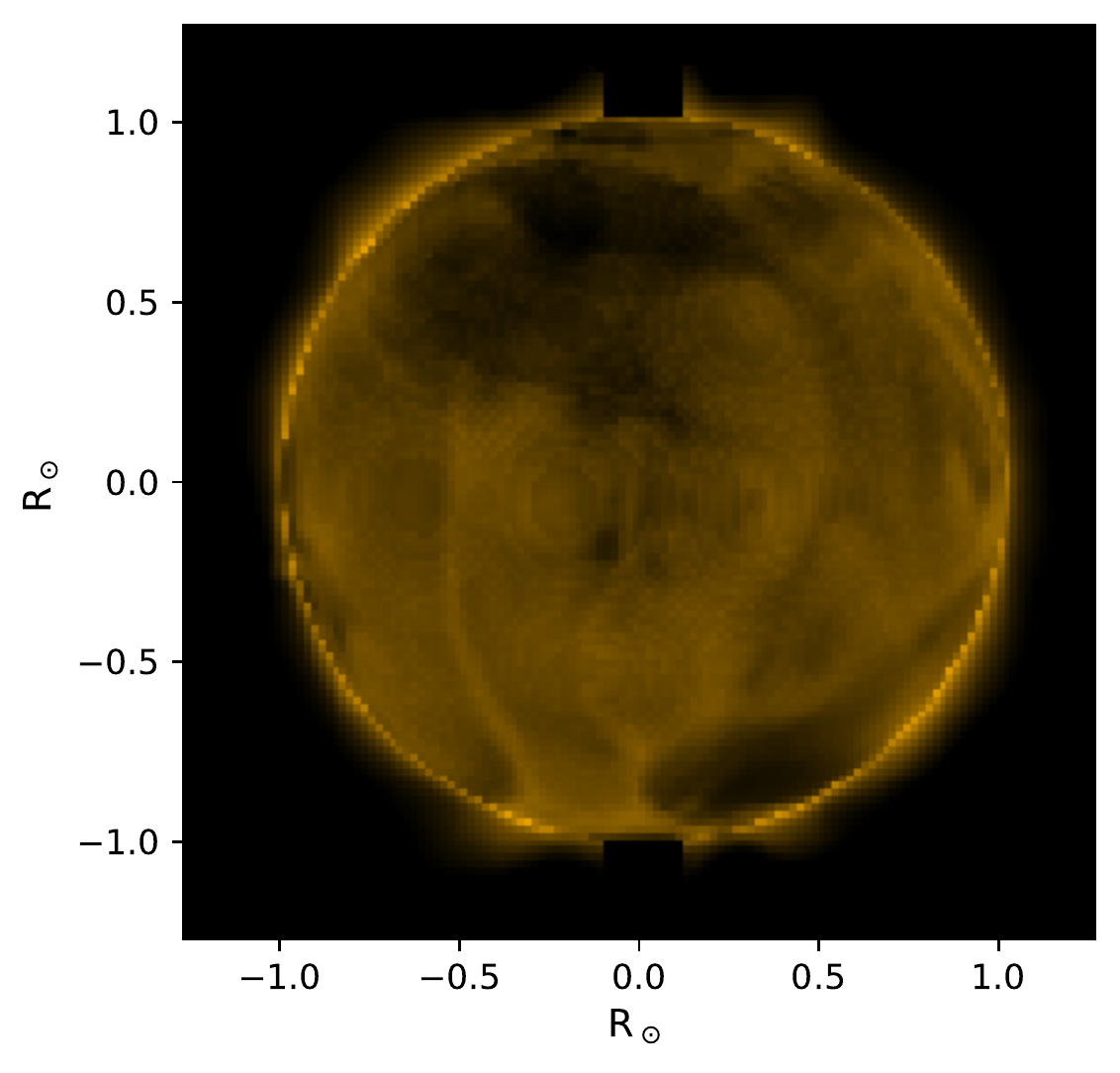}
    \includegraphics[width=0.23\textwidth]{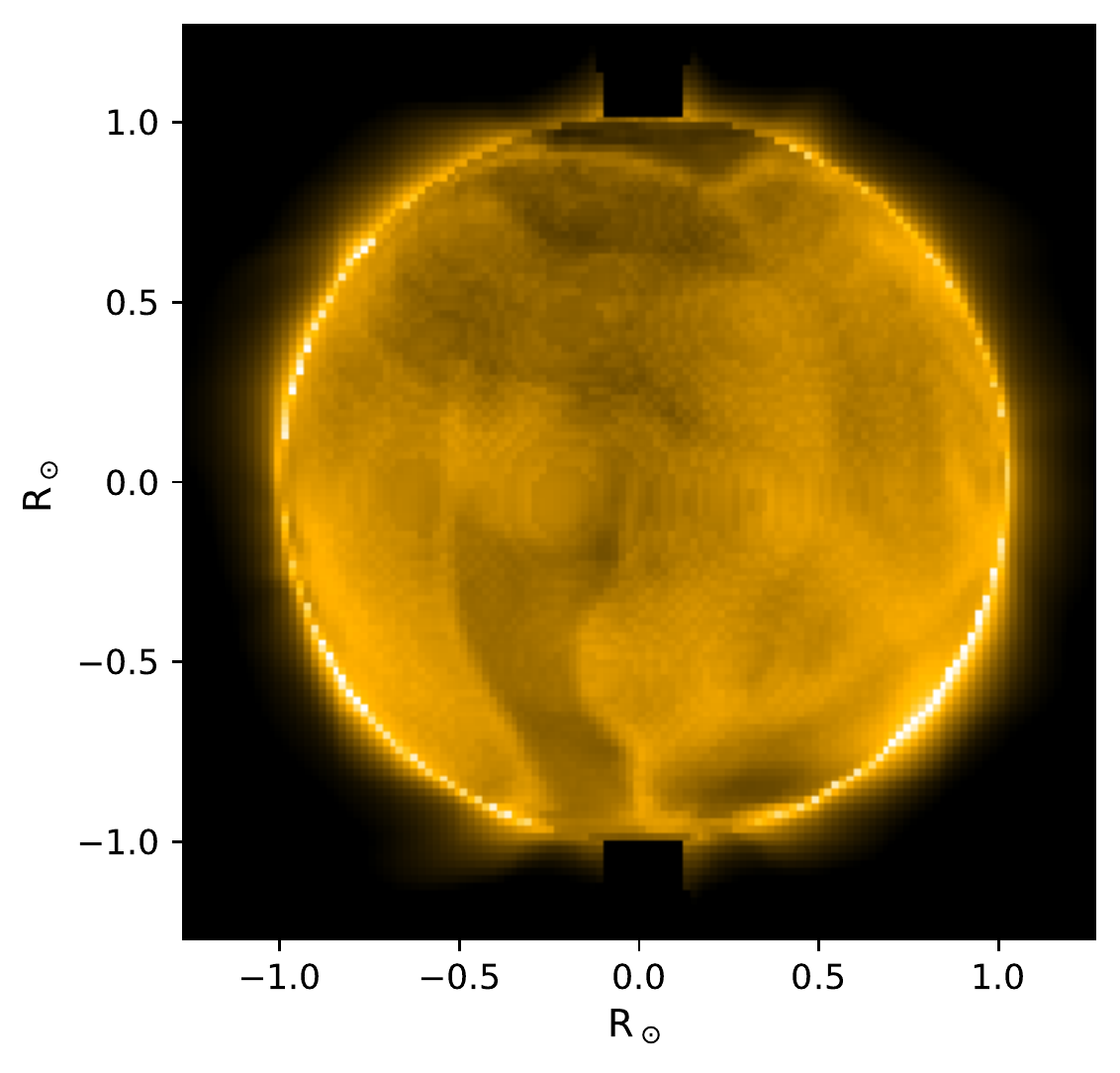}
    \includegraphics[width=0.23\textwidth]{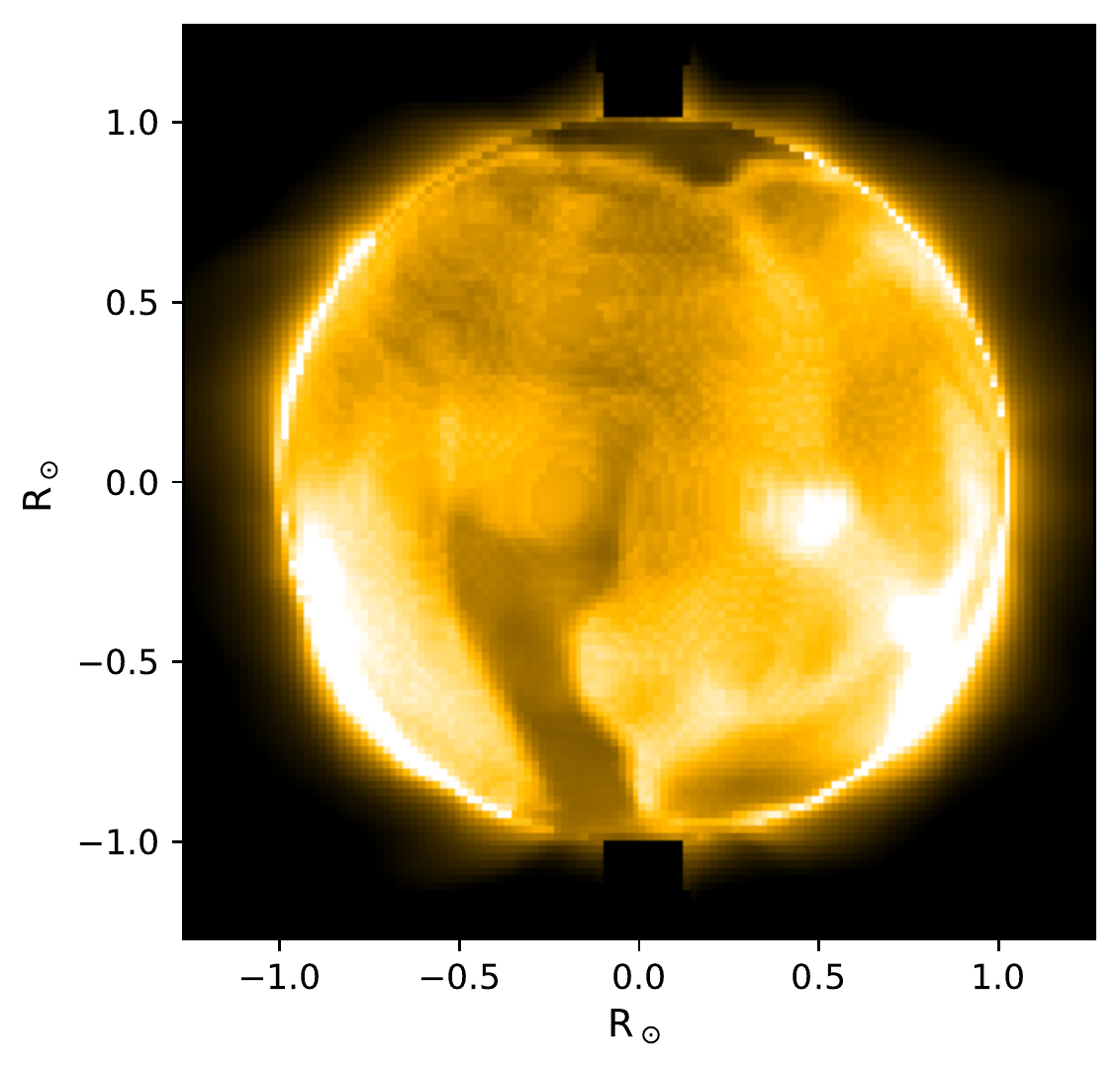}\\     
    \caption{
    Observation (first column), Model 1, 2, 3 (last three columns) for AIA 211\,\AA,\ 193\,\AA\ and 171\,\AA\ channels for November 6.}
    \label{fig:aia_all_full}
\end{figure*}

In the first column of Figure \ref{fig:aia_all_full} we display images of the Sun  for November 6 as seen through (from the hotter to the cooler channel), the  211\,\AA\, 193\,\AA\  and 171\,\AA\ bands. The other columns show the corresponding synthetic images from the three simulations. 
This figure shows that all the models are similar in reproducing the solar disk large scale structures. The 193 band images (which have a good QS/CH contrast)  show two main differences: the total intensity which increases with the increase of the density, and the extension of the CHs areas which increases from Model 1 to 3, consistent with the finding of a larger amount of open flux for Model 3 (see Section \ref{sec:mod_mhd}).
The 211\,\AA\, synthetic images are similar to the 193\,\AA\ ones, and this is quite expected as the band's main temperature sensitivity is the 2 MK corona. 
Also, the 171\,\AA\ band is well reproduced by the simulations, particularly for Model 2.

\section{Detailed EUV solar disk comparisons on November 6}  
\label{sec:res_disk}

\begin{figure*}[ht]
\includegraphics[width=0.33\textwidth, trim=15 0 40 0, clip]{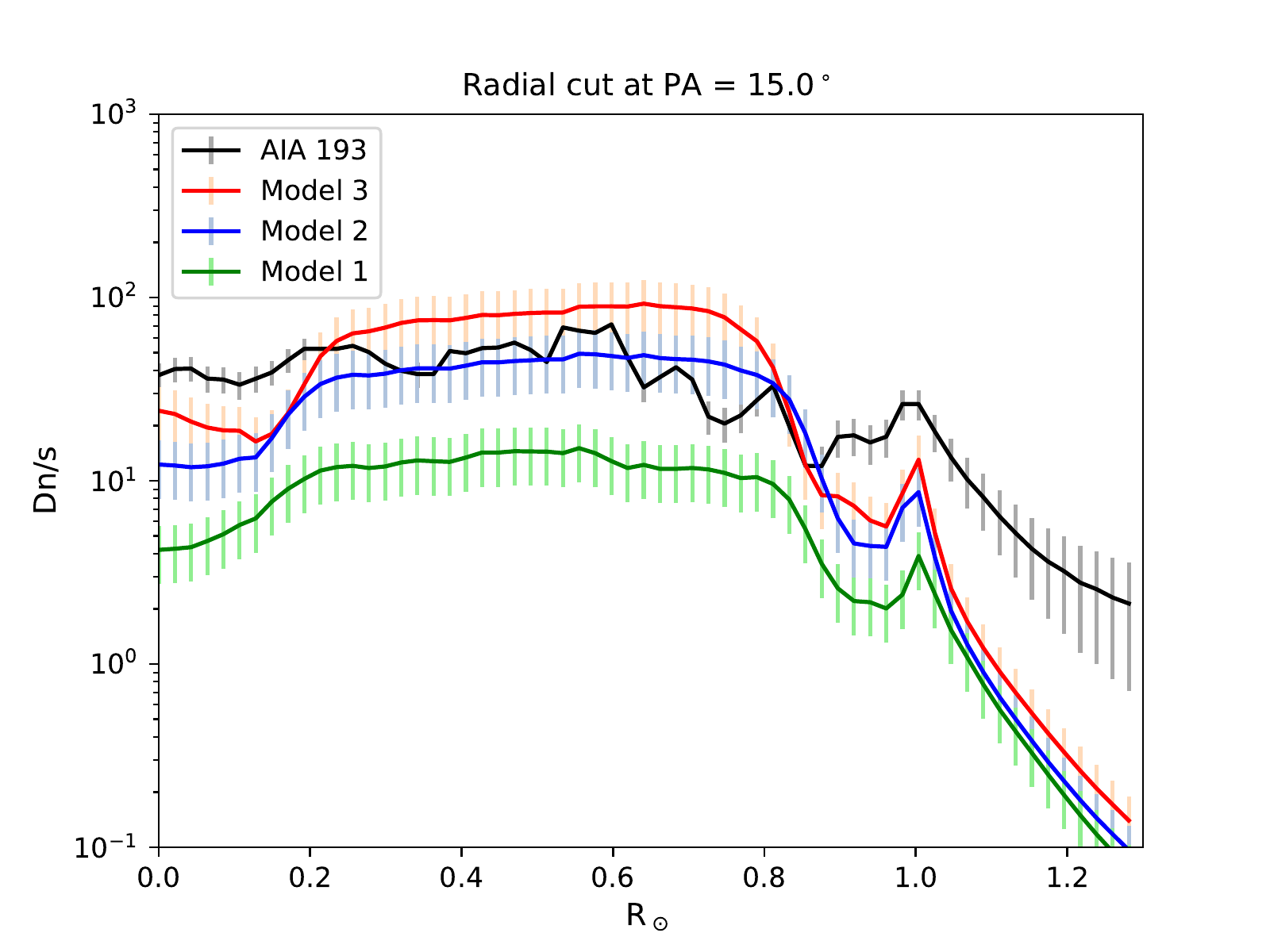}
\includegraphics[width=0.33\textwidth, trim=15 0 40 0, clip]{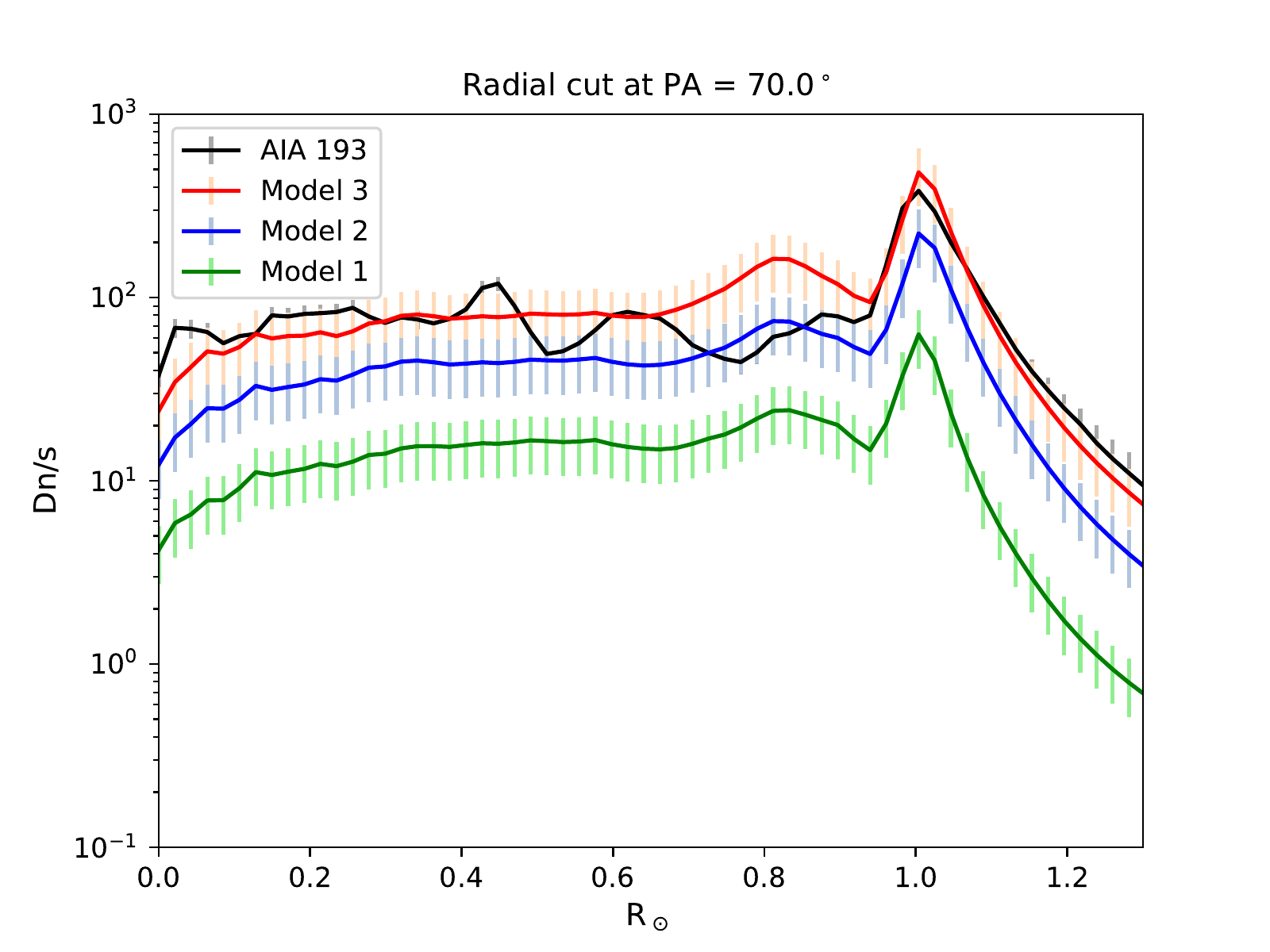}
\includegraphics[width=0.33\textwidth, trim=15 0 40 0, clip]{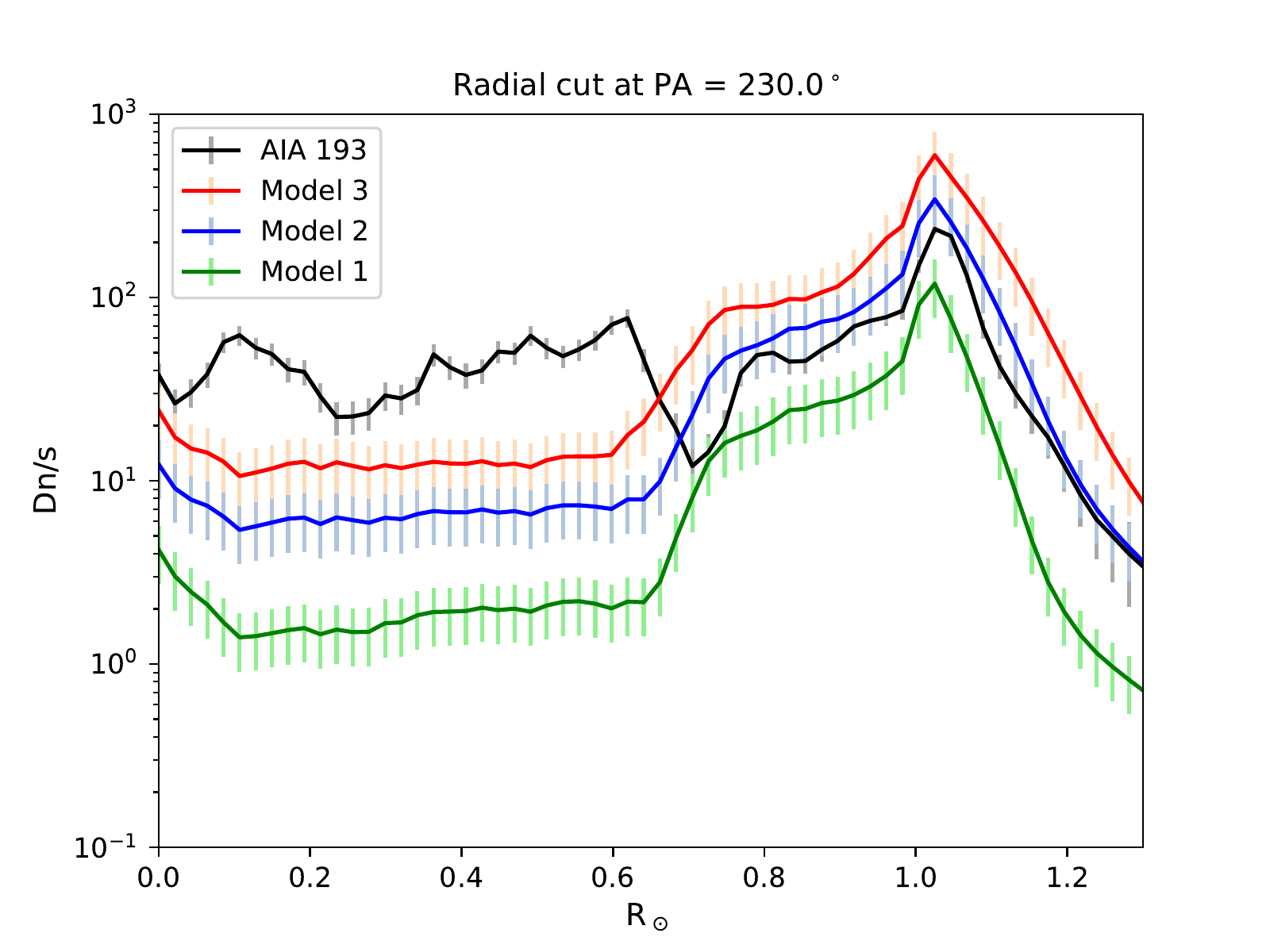}\\
\includegraphics[width=0.33\textwidth, trim=5 0 40 0, clip]{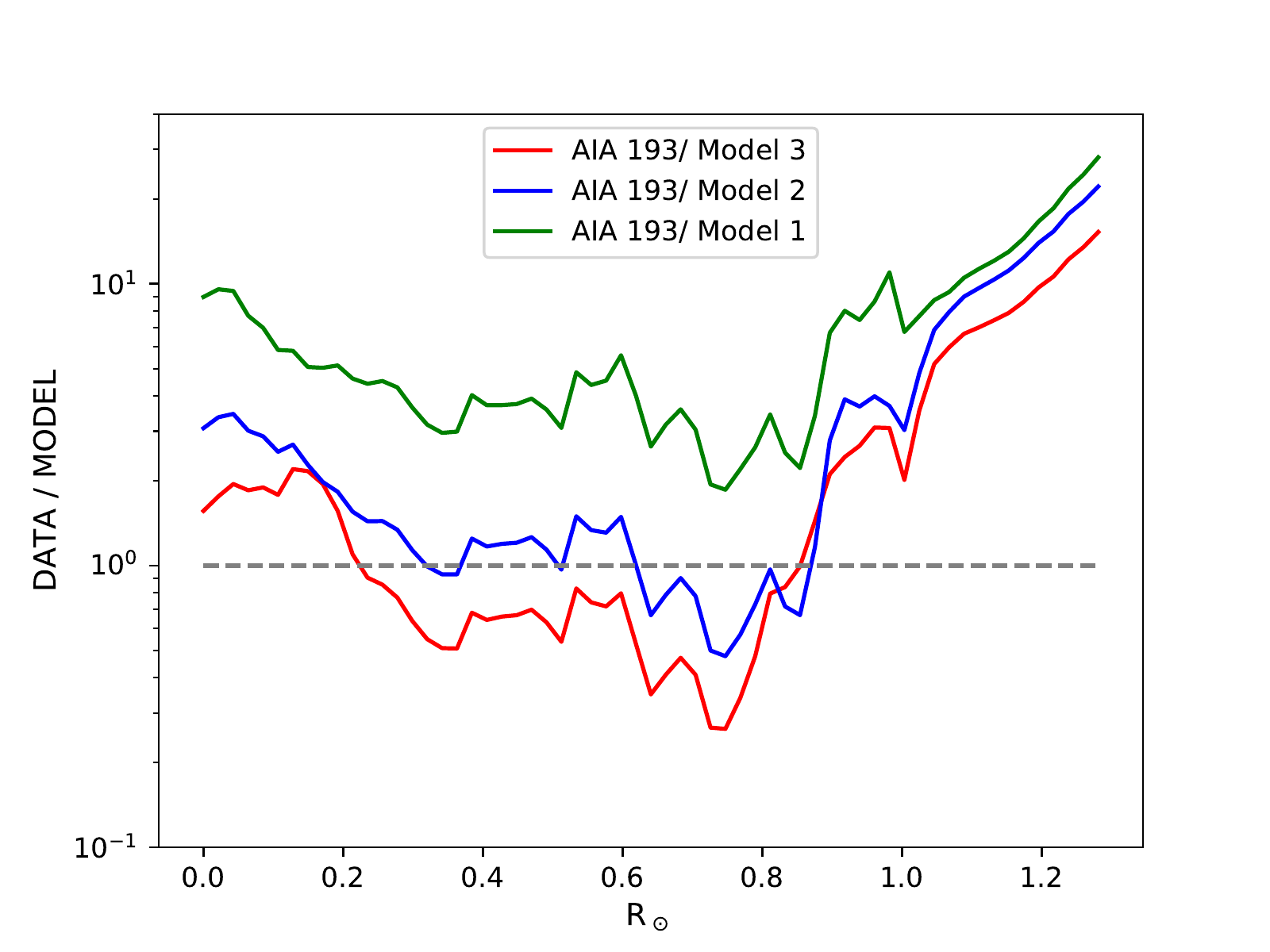}
\includegraphics[width=0.33\textwidth, trim=5 0 40 0, clip]{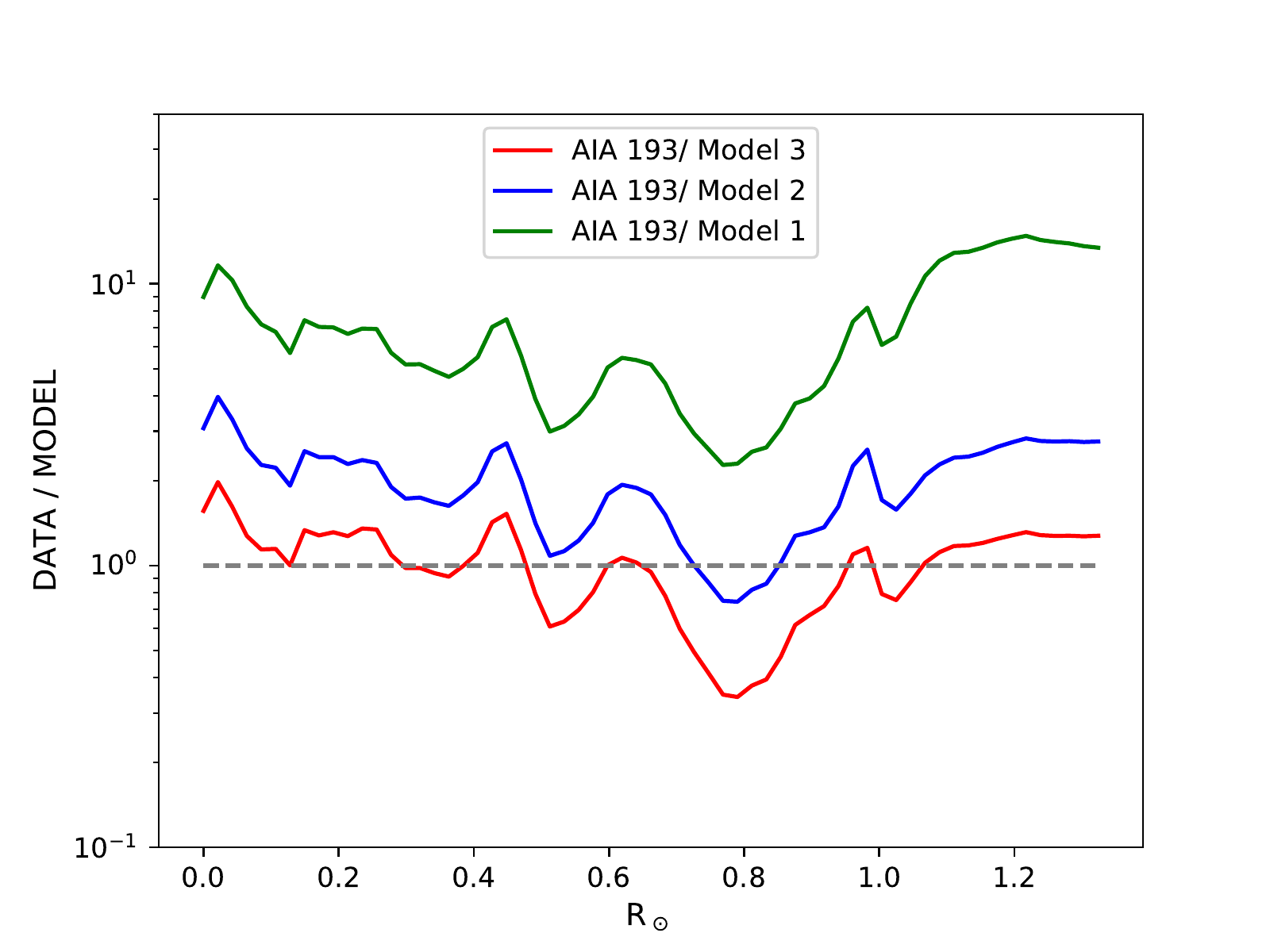}
\includegraphics[width=0.33\textwidth, trim=5 0 40 0, clip]{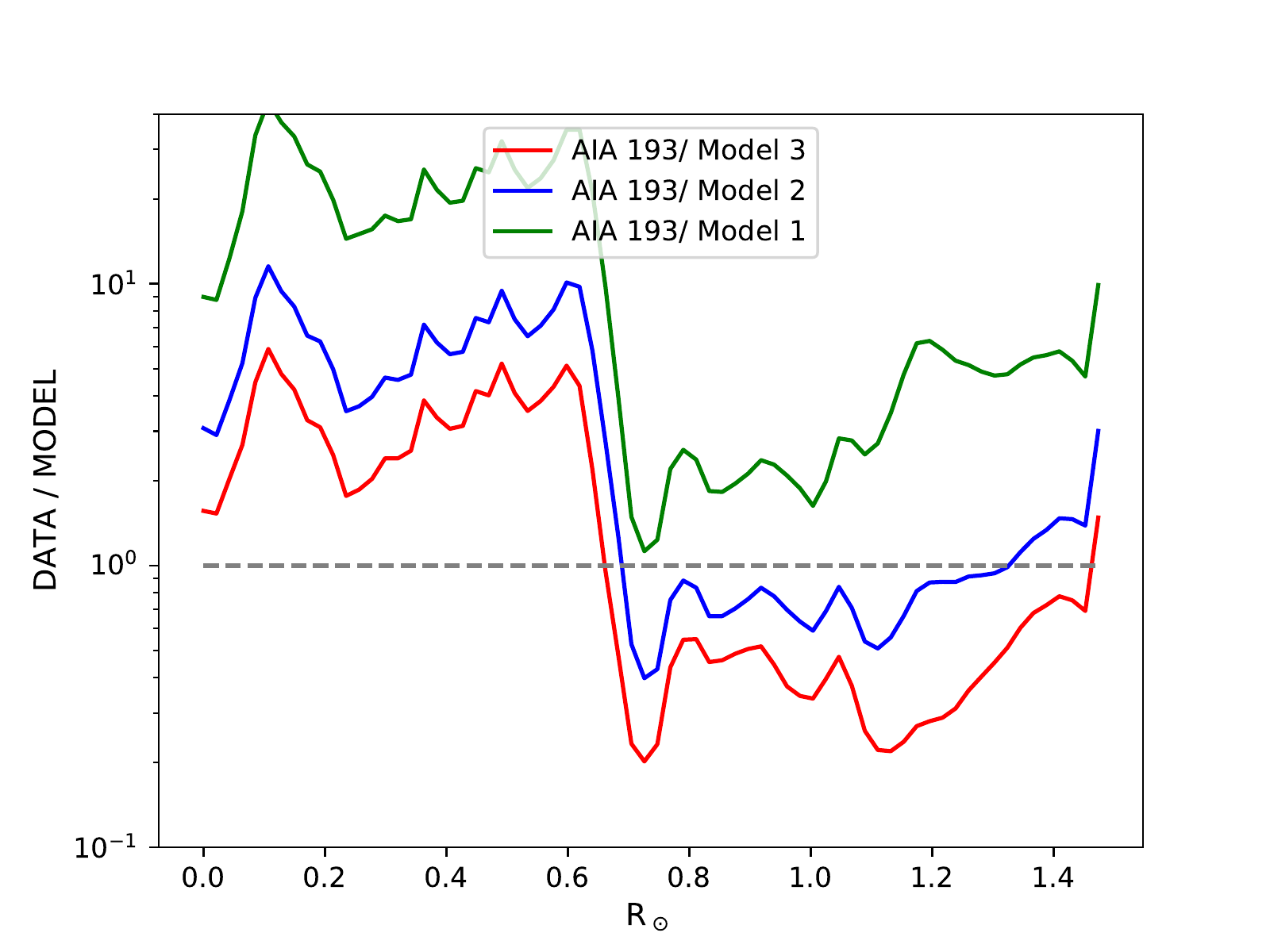}\\
\includegraphics[width=0.33\textwidth, trim=5 0 40 0, clip]{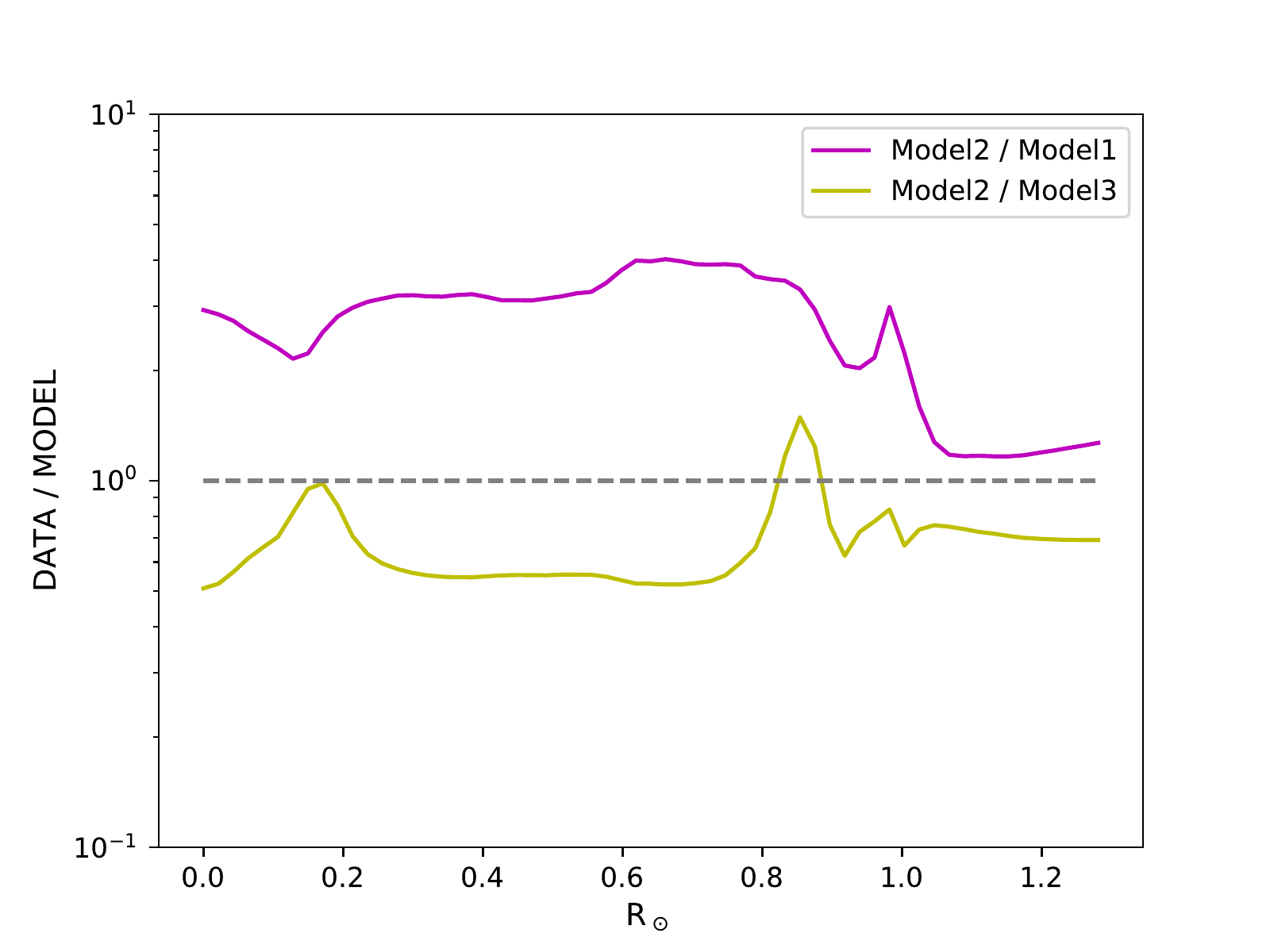}
\includegraphics[width=0.33\textwidth, trim=5 0 40 0, clip]{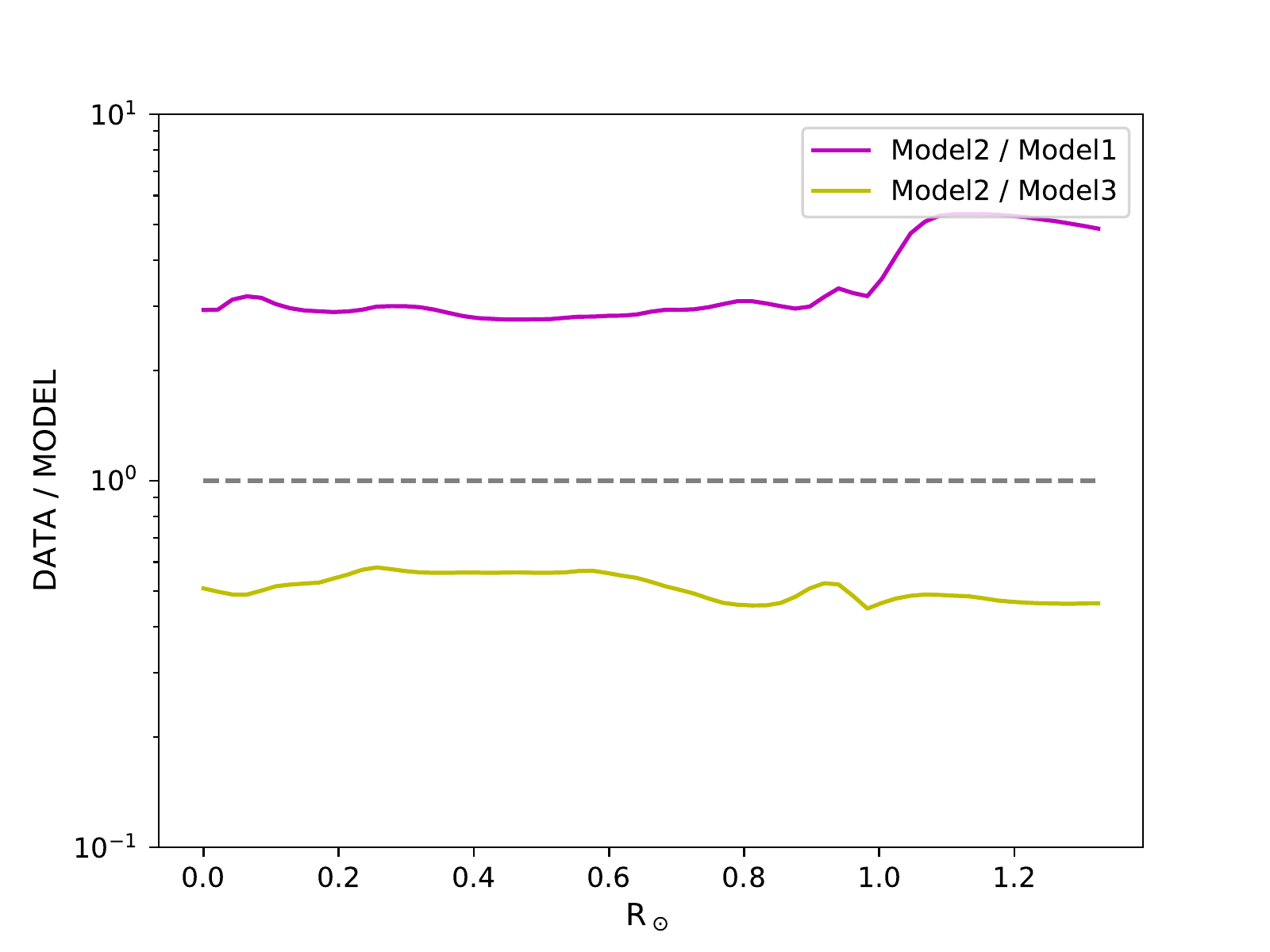}
\includegraphics[width=0.33\textwidth, trim=5 0 40 0, clip]{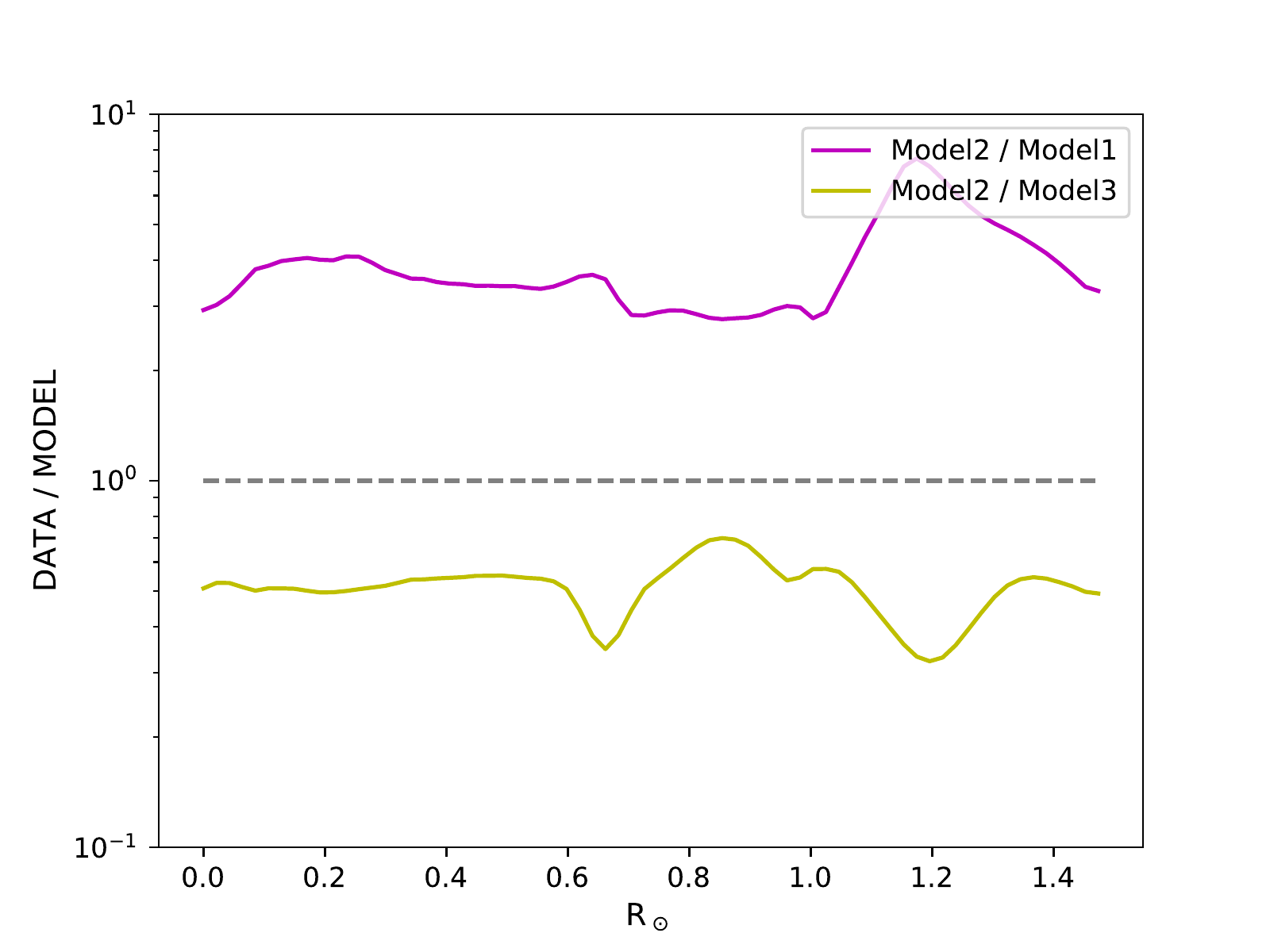}
\caption{
Top: AIA 193\,\AA\ intensities along radial cuts across the QS and north coronal hole (left), quiet Sun (middle),  equatorial CH (right). The location of these cuts are shown in Figure \ref{fig:aia_all_full} first images in the second row. Middle: ratio of the observation to model, shown on the top line. 
Bottom: ratio of Model 2 to the other models. The horizontal gray line marks the value equal 1.}
\label{fig:aia_radial}
\end{figure*}

Figure \ref{fig:aia_radial} first row shows radial cuts in the AIA 193\,\AA\ images at three co-latitudes (from now on named PA, or Position Angle), all crossing QS and CH (these cuts are labeled in the third panel of the middle row of Figure \ref{fig:aia_all_full}). The left panel (PA = 15$^\circ$) shows a profile including the central on-disk CH ($\mathrm{r} < 0.2~R_\odot$) and the north CH ($\mathrm{r} > 0.8~R_\odot$); the second one  a small equatorial CH, which is almost completely behind the west limb (PA = 70$^\circ$, $\mathrm{r} \approx 0.8~R_\odot$). The third panel 
contains the south on-disk CH  (PA = 230$^\circ$, $\mathrm{r} < 0.7~R_\odot$).  The colors are: black for the data, green for Model 1, blue for Model 2 and red for Model 3. The error bars were calculated according to what was discussed in Sections \ref{sec:mod_ima} and \ref{sec:data}. The second row reports the ratio of the observations to the models.
Figure \ref{fig:aia_radial} shows in more detail what we found from Figure \ref{fig:exe_m2}: the simulations are in general able to reproduce the modulation of the intensity profiles due to the large scale structures across the disk. Nevertheless, the CH is darker in the simulation, with a more contrasted QS to CH intensity ratio.
For instance, for PA = 15$^\circ$, in the data the QS/CH ratio (in the polar region) is about 4, while for Model 2 is about 10. For the equatorial CH (PA = 230$^\circ$) in the observation the ratio decreases getting close to 1, apart from a darker region at about 0.7 $R_\odot$. Indeed, Figure \ref{fig:exe_m2} left column shows that in the data this CH is less dark than the polar regions. 
This characteristic is also highlighted in the  second row of the figure where we plot the data to model ratio: whenever in the CH (on the disk) the ratio increases above 1.
One possible reason for a lower contrast in the observations could  be the partial absence of transition region emission in the model (already discussed in the previous section). Also,  the medium spatial resolution of the model could be one reason for the darker aspect of CHs as the simulated images lack of the emission from the small scales structures. Finally, as it will discuss in Appendix \ref{app:aia_PSF} the observation are affected by stray light which is not completely assessed.  

Comparing the three models, we can say that Model 1 has, in  general, a too low intensity. 
Model 2 is the one which best reproduces the observations, particularly on the disk.
Finally, at the bottom of Figure \ref{fig:aia_radial} we show the differences in the disk intensity variations with the models, taking Model 2 as the reference. We plot here the ratio of Model 2 to the other models. 

First, we notice that the ratios are similar in QS, CH. 
This suggests that for all the large scale structures of the corona, the density and temperature change similarly in both simulations. 
 The increase of the coronal base density for Model 3 (see Table \ref{table:MHDparams}) implies an increase of the intensity in the AIA band. Between Model 2 and Model 3 the main difference is seen at the CH boundary (around $\mathrm{r} < 0.2 ~~R_\odot$ and 0.8 $R_\odot$ for PA = 15$^\circ$, and 0.7 $R_\odot$ for PA = 230$^\circ$) due to the small change in the longitudinal extension between the two cases, as discussed before. 
The situation is slightly different when comparing  Model 2 to Model 1. The difference in the absolute intensity is more important, and the ratio
of the intensities is less modulated at the structure boundary. Stronger differences are seen off disk. From Figure \ref{fig:aia_radial} we see that this seems to be due to Model 1.

Figure \ref{fig:aia_radial} also provides the off limb coronal EUV behavior, even though the short distance covered by the images and the too low signal are not much conclusive. Within CH (PA = 15$^\circ$) the intensity fall--off is less sharp in the observations, while within closed regions the data and modelled  profiles run almost parallel for Model 1 and 2.
Further investigation of the off-limb corona is made in the next Section.

\begin{figure*}
    \includegraphics[width=0.33\textwidth, trim=5 0 40 0, clip]{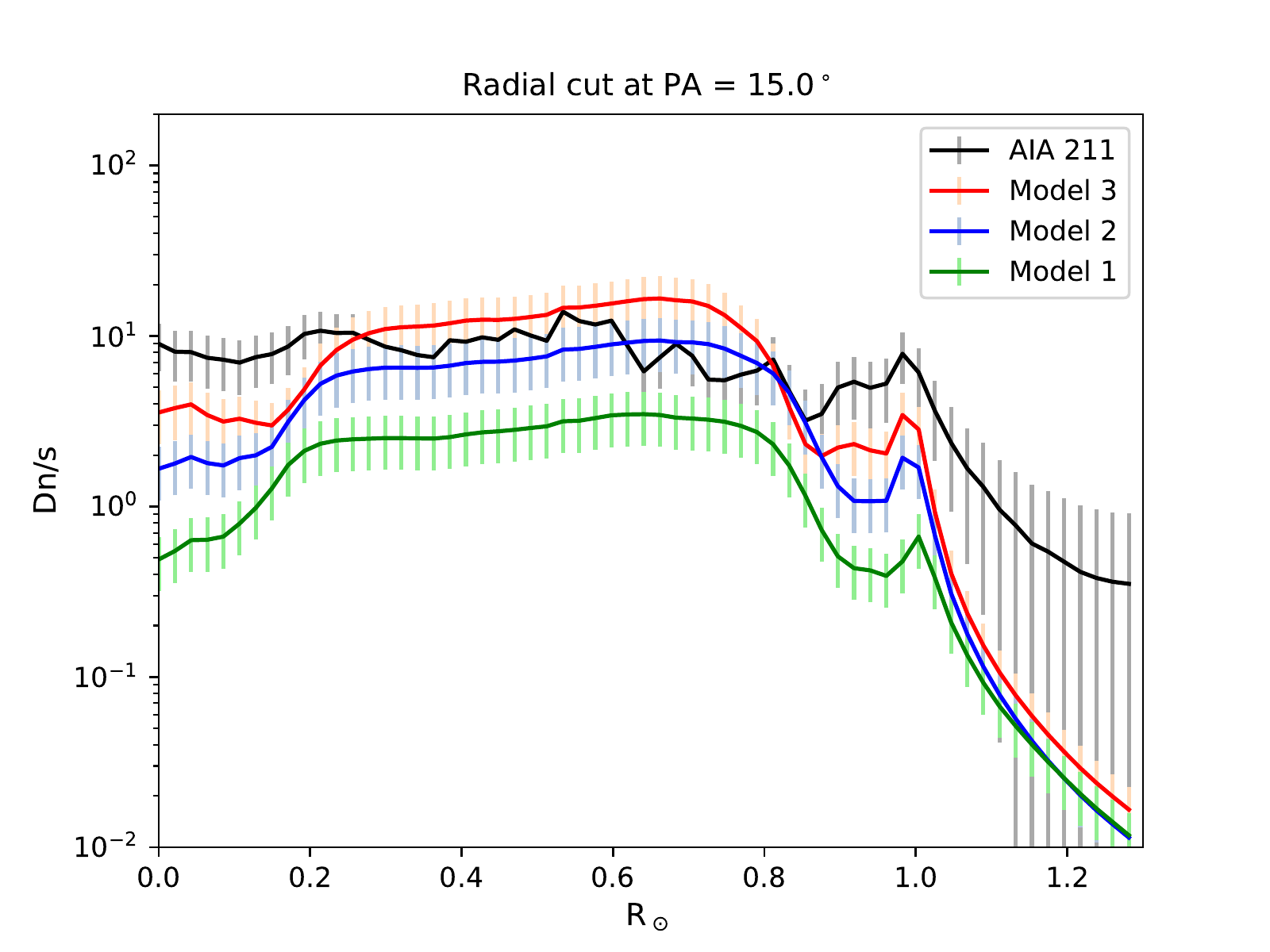}
    \includegraphics[width=0.33\textwidth, trim=5 0 40 0, clip]{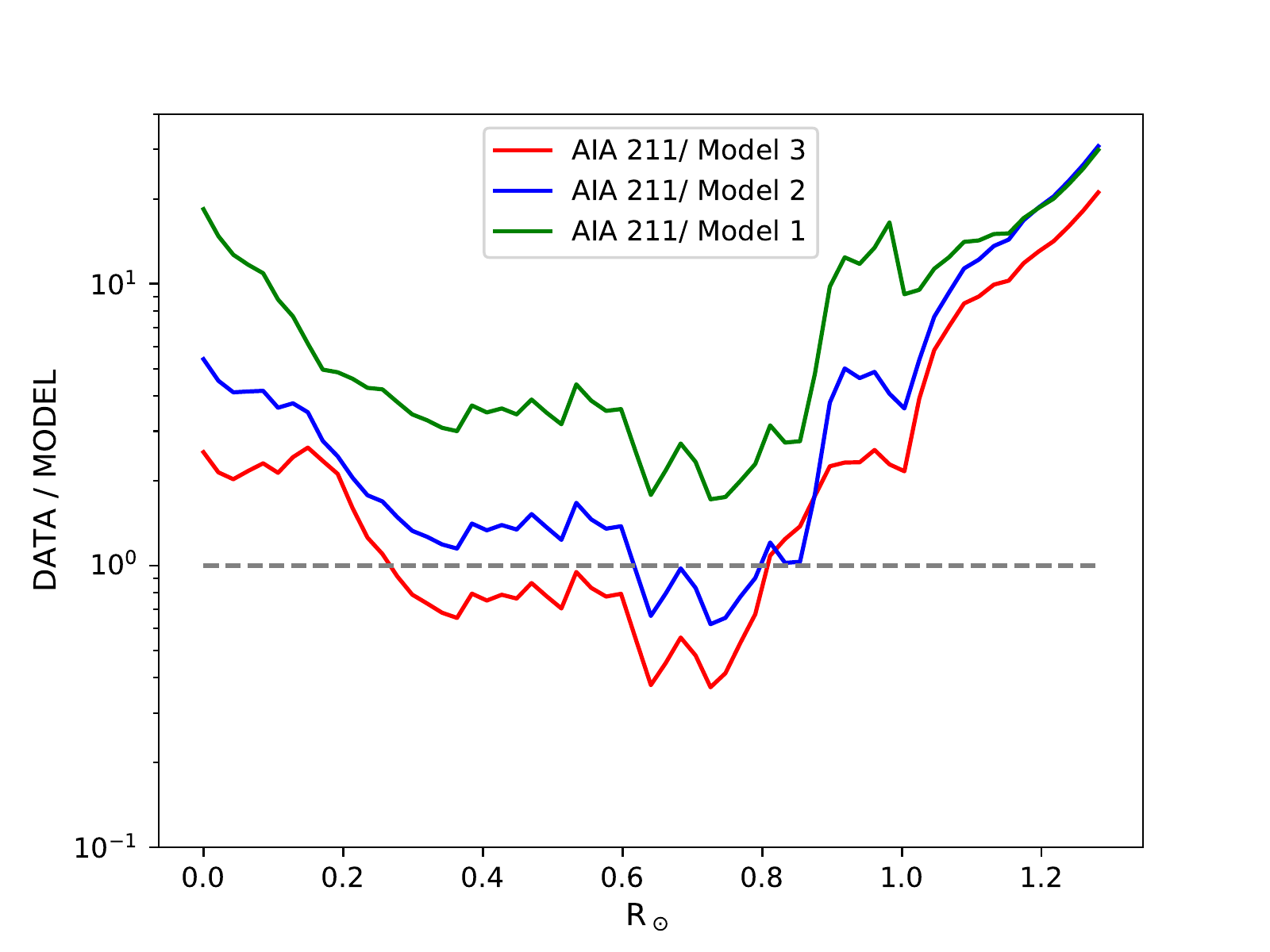}
    \includegraphics[width=0.33\textwidth, trim=5 0 40 0, clip]{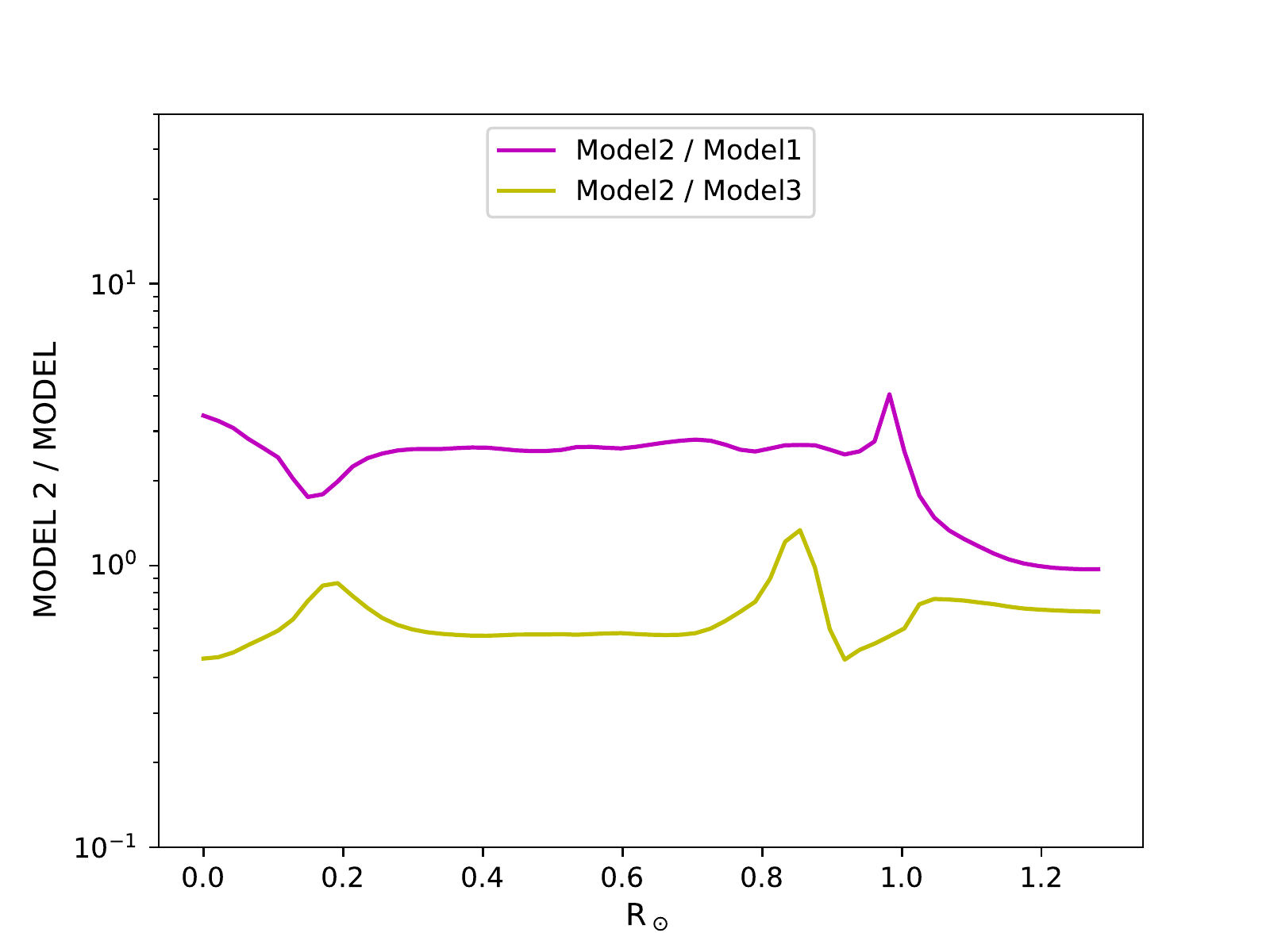}
\caption{
From  left to right: AIA 211\,\AA\ intensities along the radial at polar angle $15^\circ$, ratio between the observation and the synthetic intensities, ratio between Model 2 and the other models.  The horizontal gray line marks the ratio equal 1.} 
\label{fig:aia_211_prof}
\end{figure*}
The analysis made on the 193\,\AA\ band has permitted to mostly analyze the effect of the electron density variation with the models. As already mentioned a multi-band analysis is not useful with the present setup of the model, however we show an example of a radial cut on the 211\,\AA\ images to check for possible temperature effects at coronal regime. 
The panels in Figure \ref{fig:aia_211_prof} are the equivalent of the first column in Figure \ref{fig:aia_radial} but for the AIA 211\,\AA. 

This band is hotter than the 193\,\AA\ band, but it shows very similar behavior. Also, in this case the QS is consistent with Model 2 and Model 3, while the CHs decrease of intensity (r$<$ 0.2 and r $\simeq$ 0.9) is too large in the simulations. 

\section{Detailed off limb comparison with WL and EUV observations for November 6}
\label{sec:res_corona}

The validation of WindPredict-AW model was also performed in the off disk corona by extracting, from the intensity maps, latitudinal and radial  profiles.

\begin{figure*}[ht]
\centering
\includegraphics[width=0.95\textwidth]{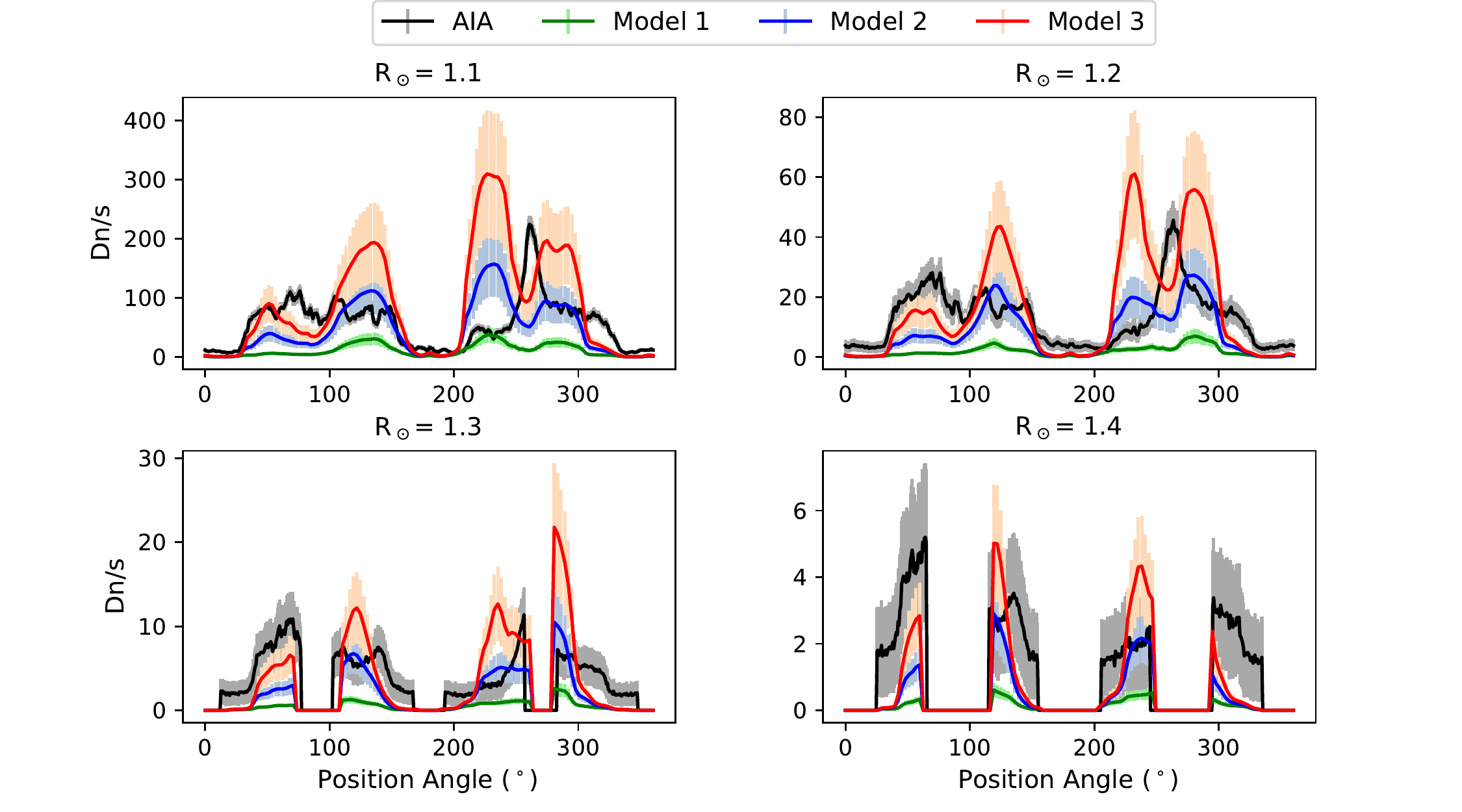}
\caption{ Intensity profiles as function of  the Position Angle for the AIA 193\,\AA\ channel at four solar distances.}
\label{fig:aia_lat}
\end{figure*}

Figure \ref{fig:aia_lat} shows the AIA 193\,\AA\ intensity as function of PA for fixed solar distances (from 1.1 to 1.4 R$_\odot$). 
 The observed intensity profiles shape the base of two west streamers, one in each quadrant (SI for the north--west, SII for the south--west) and a large one at the equator of the east limb (SIII). The three simulations reproduce the morphology of the two west structures  ($40^\circ< \textrm{PA} < 160^\circ$.), while SIII is replaced by two streamers, one for each quadrant. This difference in morphology at the east limb explains the off limb radial intensity behavior of Figure \ref{fig:aia_lat}: at PA = 230$^\circ$ the model has a bright streamer which is not seen in the data, as already previously discussed.

We now look in more details to SI and SII. Within the assumed error, SI is always less intense in the modelling than in the observation, even though Model 3 gets closer. Model 2 reproduces SII quantitatively. Above 1.3 R$_{\odot}$ the error from the instrumental noise is too high and the comparison with the models is less straightforward and probably not reliable.
We now move higher in the corona, looking at the WL LASCO data.

\subsection{Results for the streamers data}

\begin{figure*}[ht] 
\centering
\includegraphics[width=0.51\textwidth,  trim={80 0 21 0}]{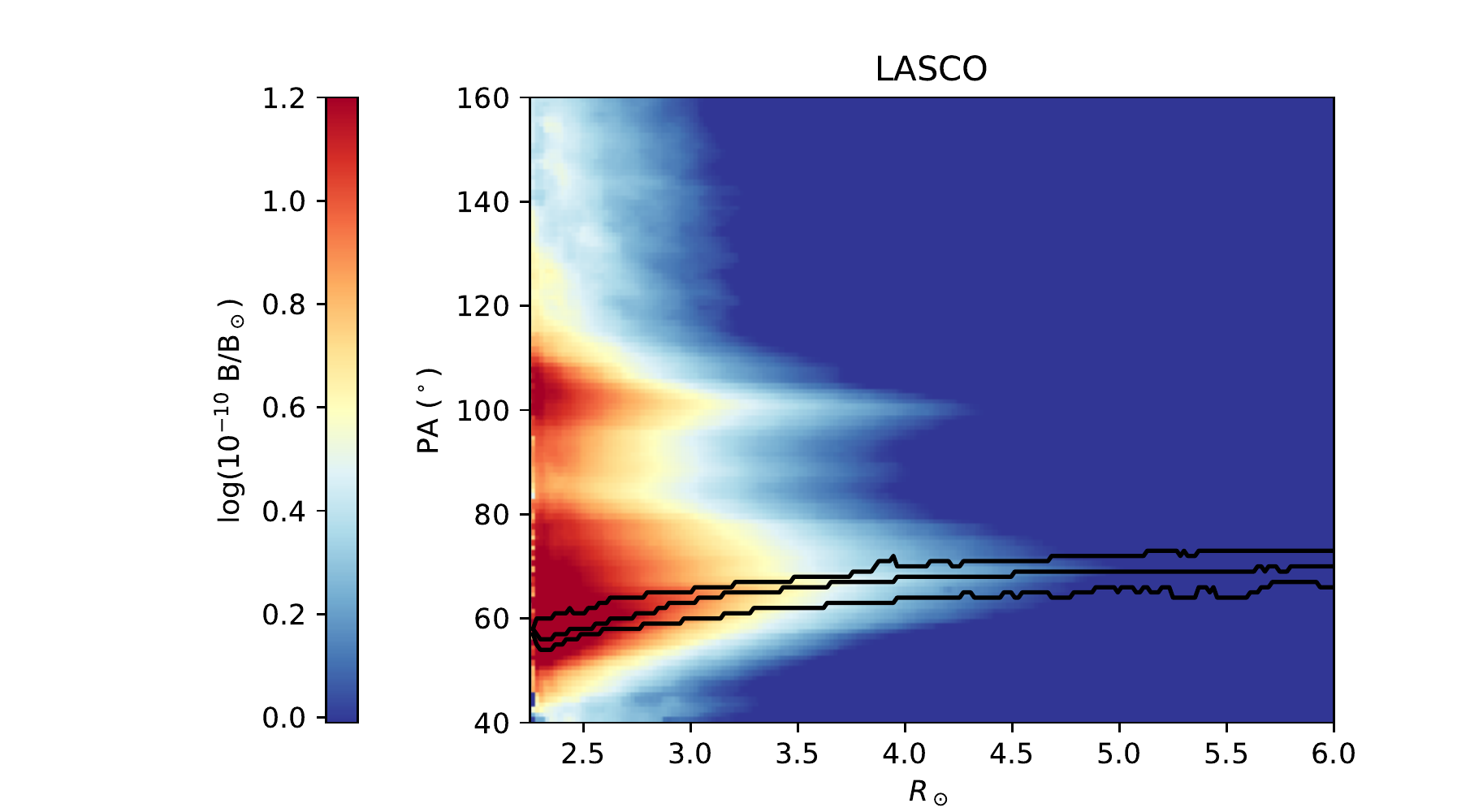}
\includegraphics[width=0.46\textwidth, trim={0 0 0 0}]{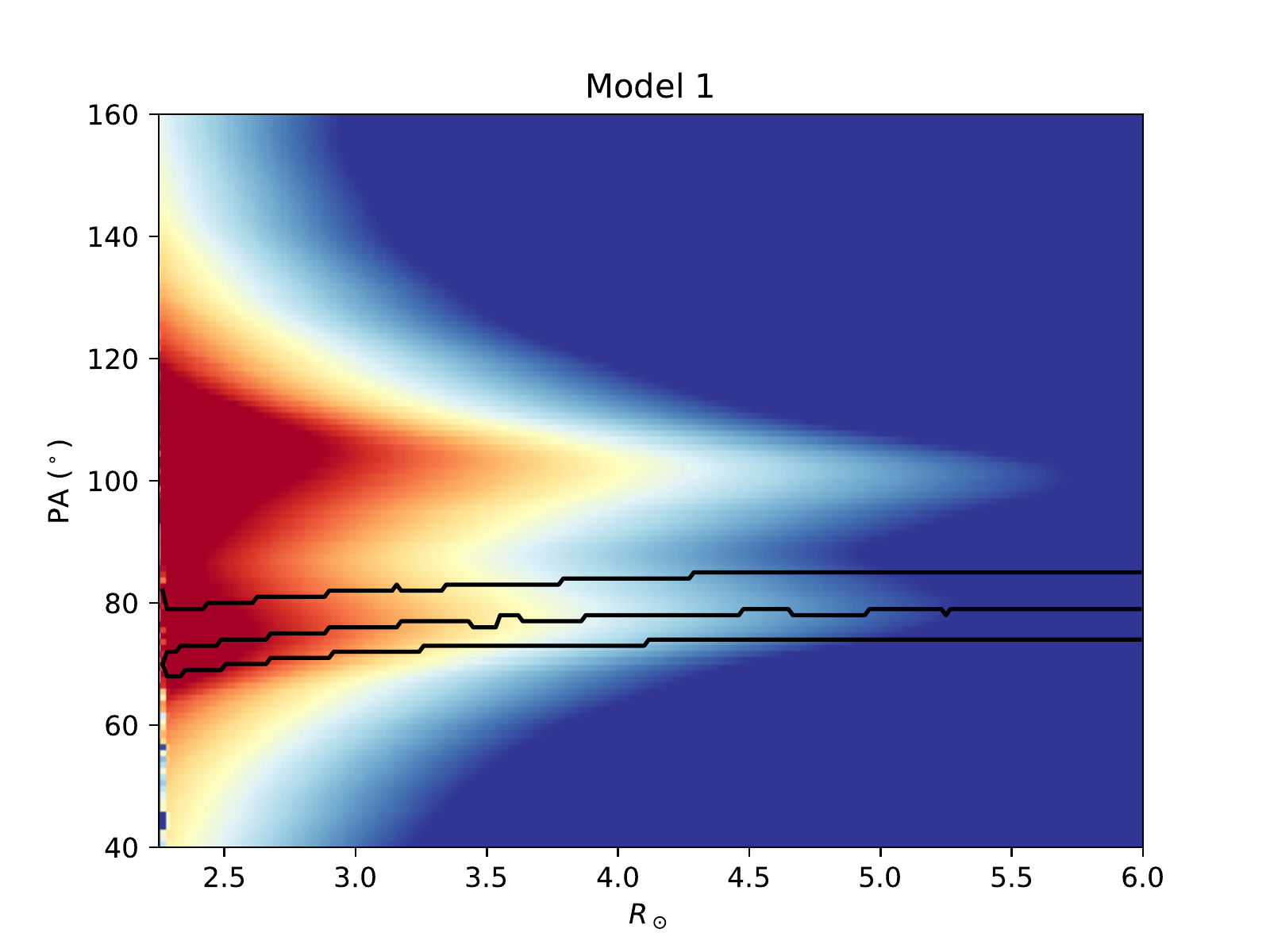}\\
\includegraphics[width=0.46\textwidth, trim={-50 0 45 0}]{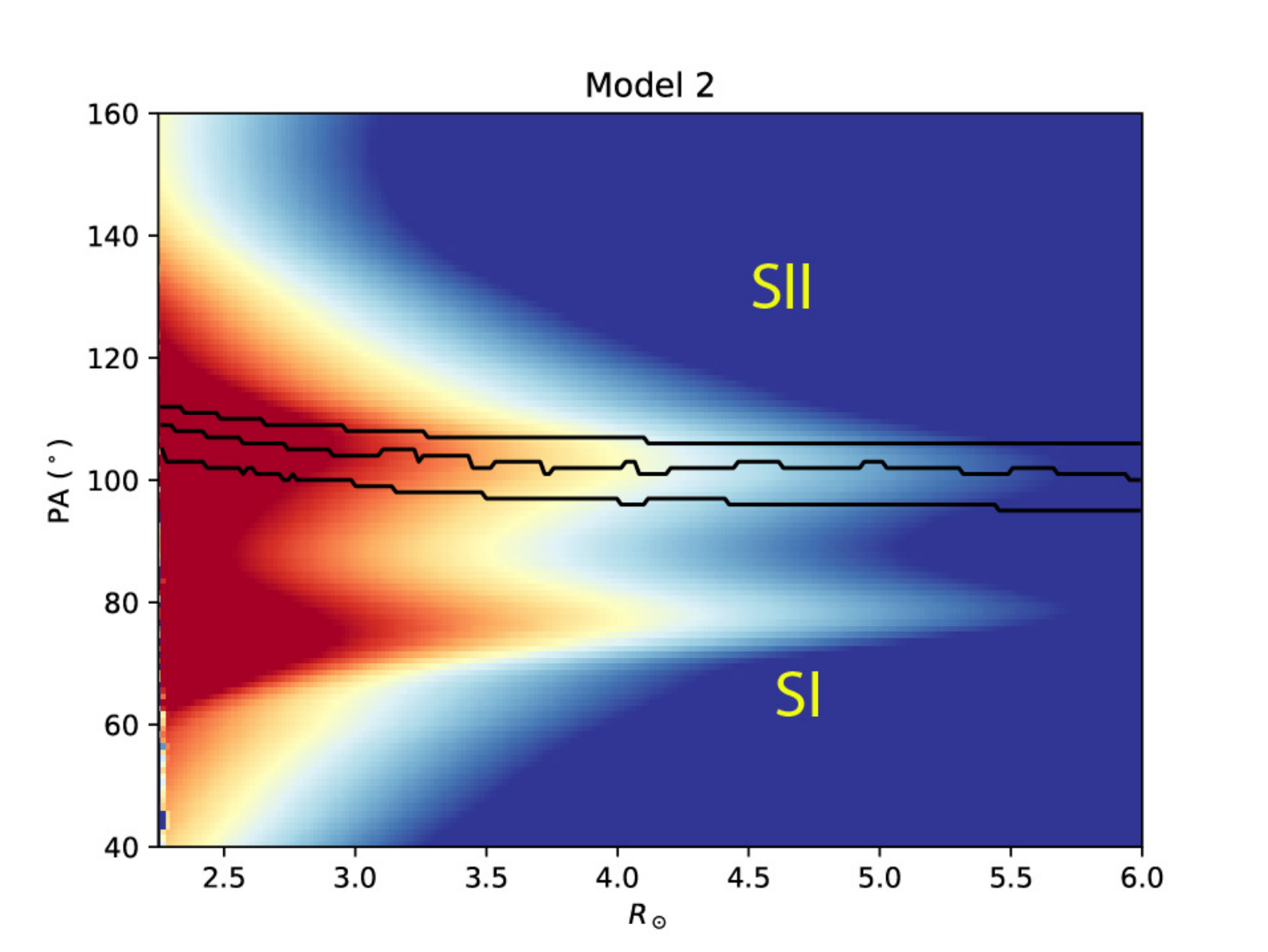}
\includegraphics[width=0.46\textwidth, trim={-18 0 30 0}]{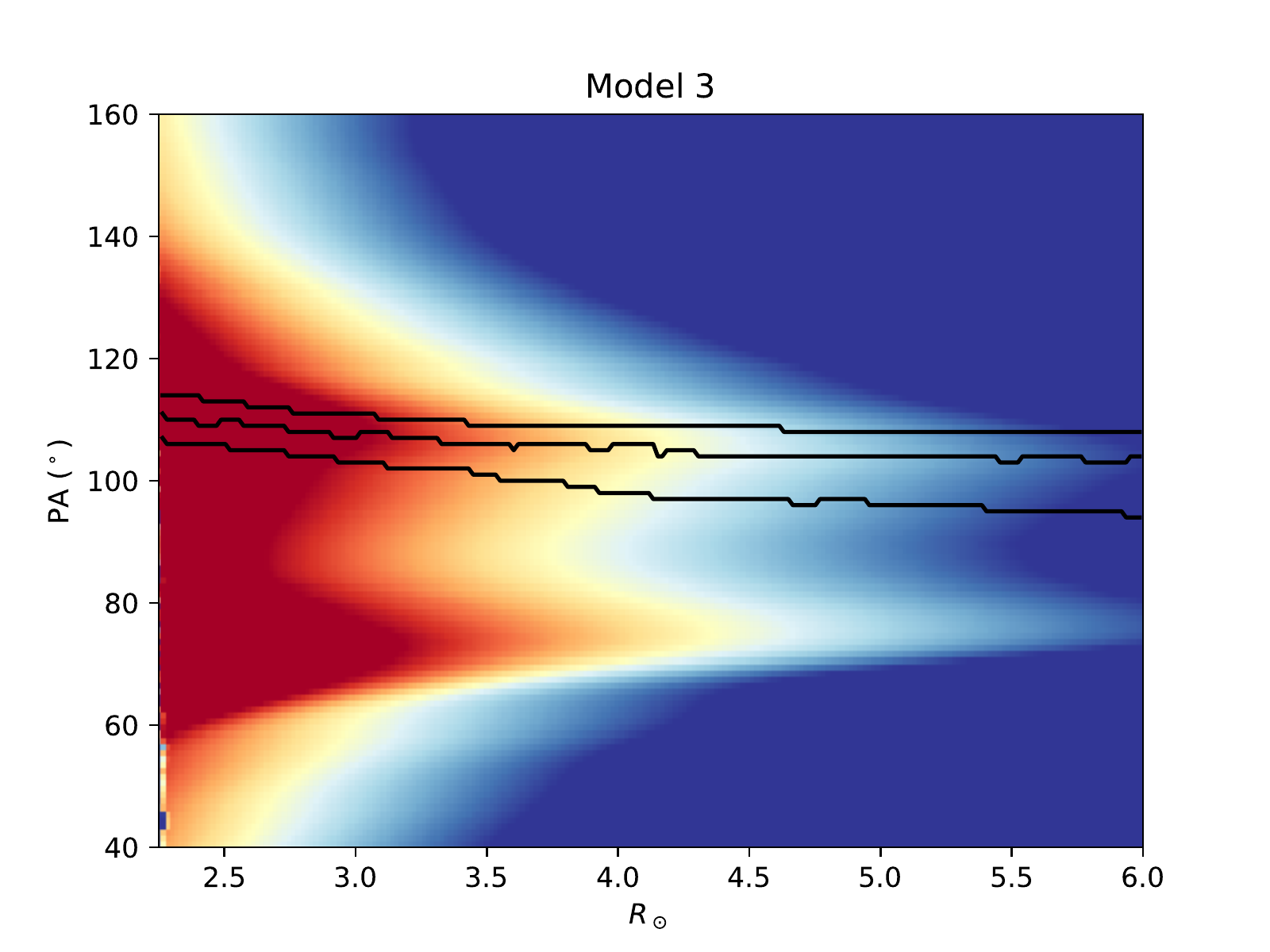}
\caption{
Images of LASCO C2 pB in polar coordinates for November 6th. From left to right and top to bottom: observation, Model 1, Model 2, and Model 3. In each panel, the central black profile marks, at each distance, the angle of the peak of intensity within the selected streamer. The other two profiles marks the interval where the pB is within 0.8 of the peak intensity value. At each distance, the intensities within this interval are averaged to build the profile along the solar height.}
\label{fig:lasco_pa_str}
\end{figure*}

\begin{figure*}[ht]
\centering
\includegraphics[width=0.9\textwidth]{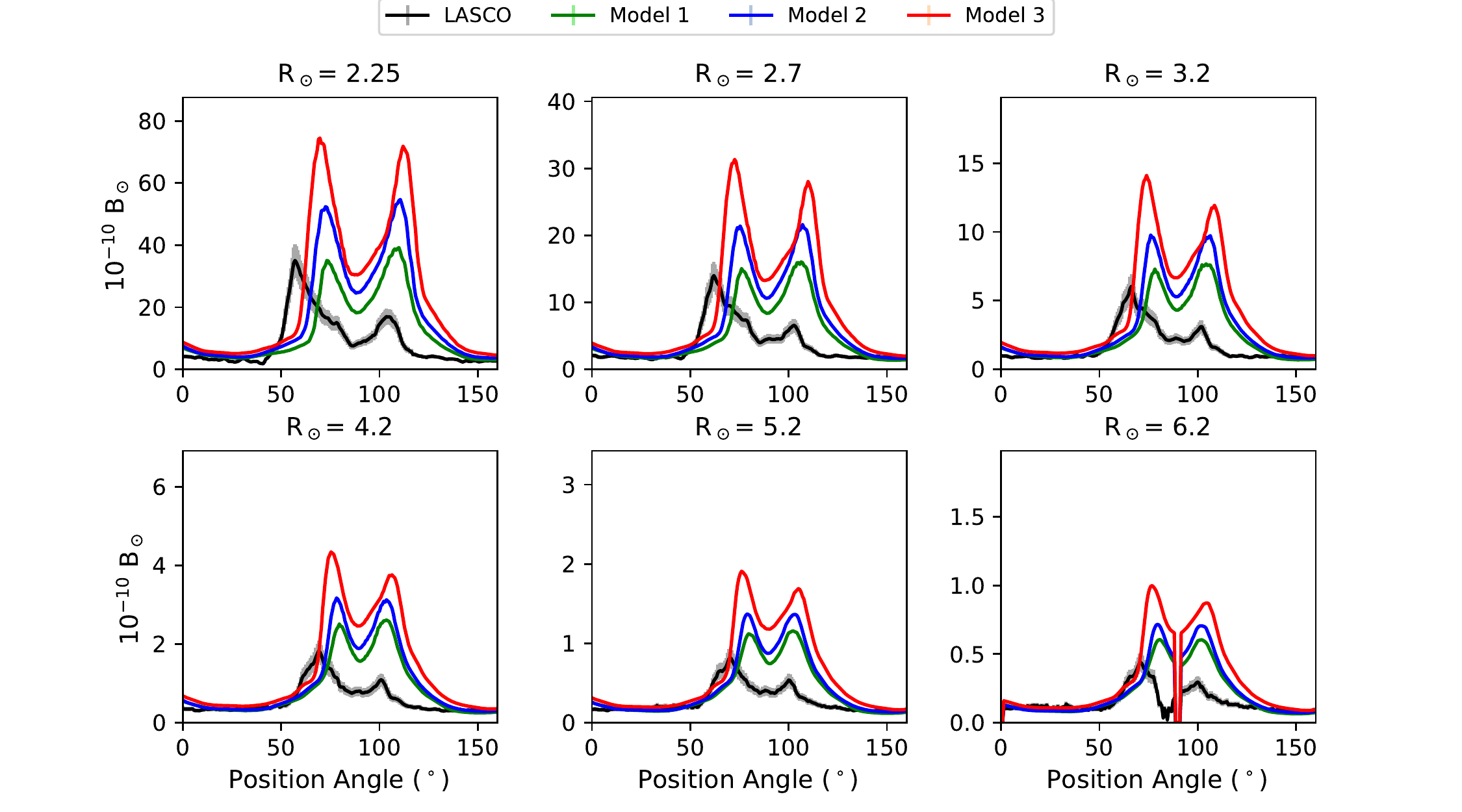}
\caption{ LASCO C2 pB latitudinal intensity profiles for IS ($\approx 70^\circ$) and SII ($\approx 110^\circ$, see Figure \ref{fig:lasco_pa_str}) at different solar distances.}
\label{fig:lasco218_pa}
\end{figure*}

\begin{figure*}[ht] 
\centering
\includegraphics[width=0.45\textwidth]{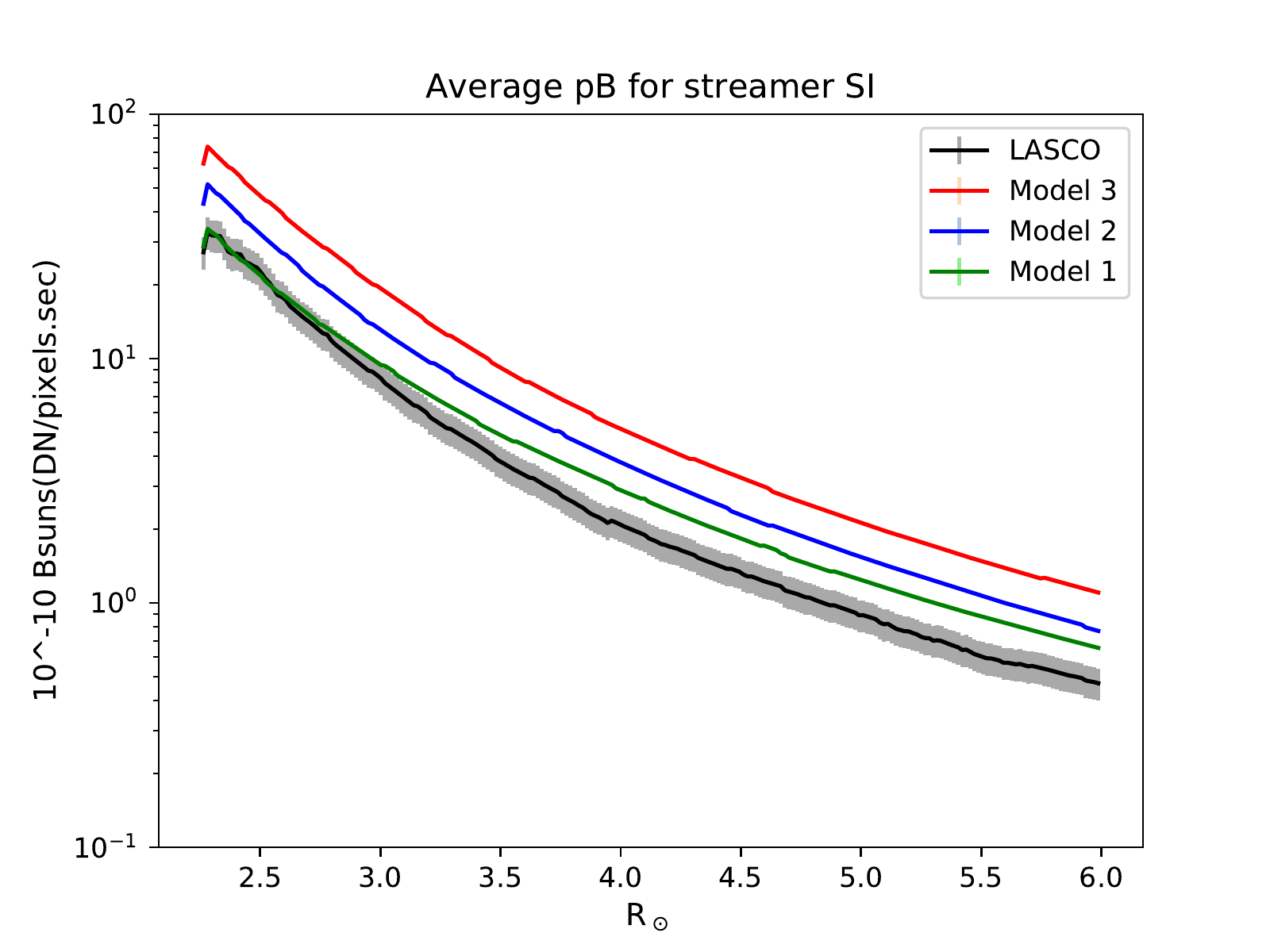}
\includegraphics[width=0.45\textwidth]{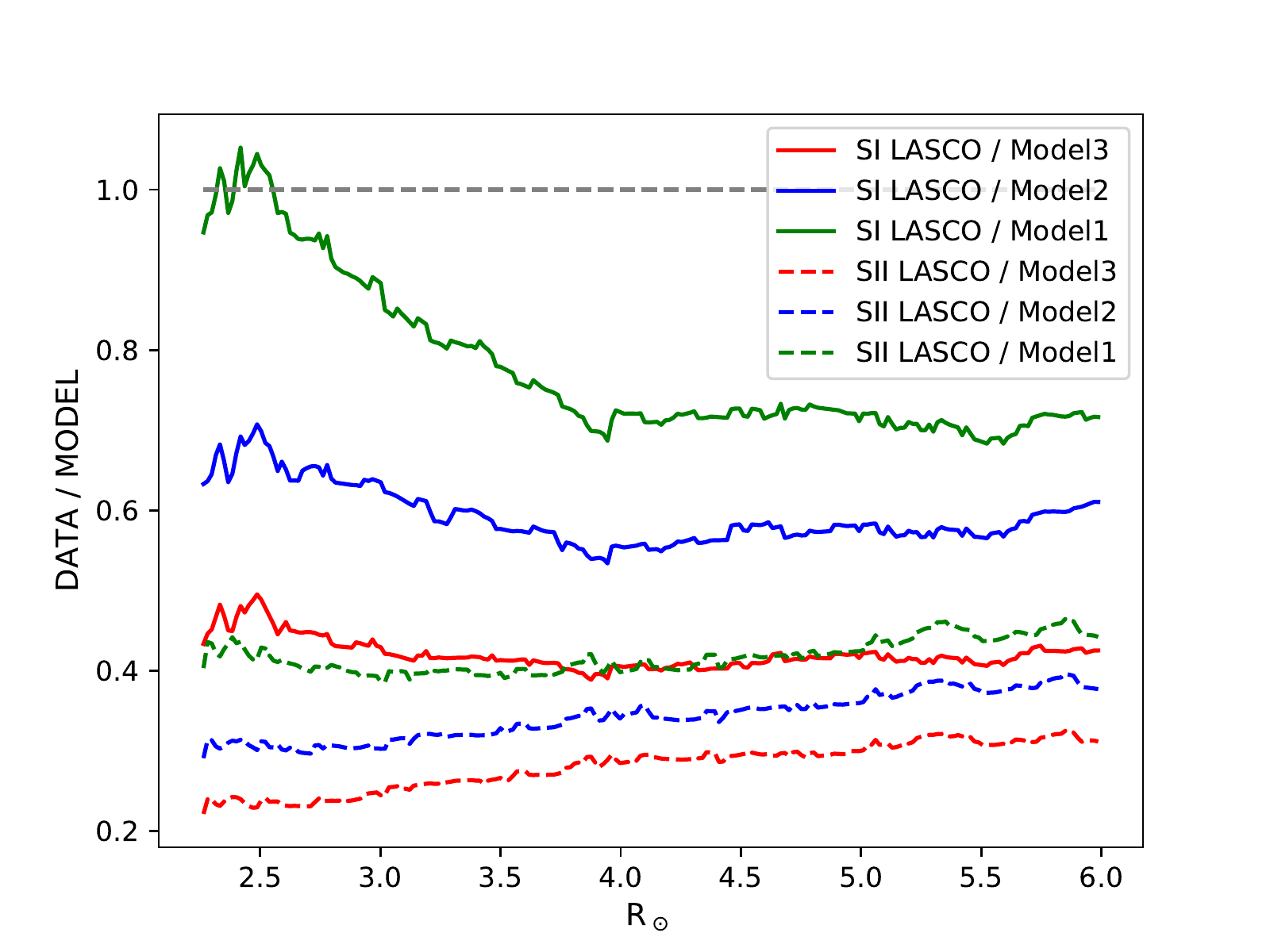}
\caption{
Left: radial averaged profile within the streamer SI from the LASCO C2 pB images of Figure \ref{fig:lasco_pa_str}. Right: ratio of LASCO to models intensities for the two streamers of  Figure \ref{fig:lasco_pa_str}.}
\label{fig:lasco_nprof6}
\end{figure*}

Figure \ref{fig:lasco_pa_str} shows the west side of the corona  $0^\circ < \text{PA} < 180^\circ$ for November 6th as seen by LASCO C2 plotted in polar coordinates. As we said earlier, the models reproduce well the two streamers, although we can notice some minor differences in the intensity and latitudinal distribution (see also Figure \ref{fig:lasco218_pa}). As expected, the denser Model 3 is the brightest, but even Model 1 has both streamers brighter than the observation (see also Figure \ref{fig:lasco_nprof6}). Additionally, the simulated streamers have similar intensity, while this is not the case in the observation. We now analyze more in detail these two streamers.

In each panel of Figure \ref{fig:lasco_pa_str} we have over plotted three black line cuts. The central one marks, for each distance, the PA corresponding to the peak of intensity (pB$_m$) within the selected streamer. This can be considered as the plane of the sky projected streamer axis. The other two are located at 0.8 ($20\%$ decrease) from the intensity probed by the central one.

As an example, we show these profiles for streamer SI in the first two panels and streamer SII in the last two.
The axis of the streamer SI within the LASCO data (first panel) is bent  below about 4 R$_\odot$. 
This property may be due to the large scale magnetic field which pushes one side of the streamer, or the result of the line of sight intensity integration of foreground and background low-lying structures. Above about 4 $R_\odot$ the streamer becomes radial. 
The second panel of the figure shows that Model 1 reproduce the behavior of the streamer axis (this is also seen in the other simulations). A similar effect is also visible in the second streamer in all cases.

Further information on the comparison between observation and modeling is obtained from Figure \ref{fig:lasco218_pa} which shows the latitudinal profiles at selected  radial distances. In the observation (black profile) an asymmetry of SI is clearly visible, and SI is more intense than SII at all distances. 
 At each distance, the models produce streamers which are too intense, and the asymmetry of SI is not strong enough compared to the observation. While Models 1 and 2 produce SI and SII similar in intensity at all distances, this is not the case for Model 3 which reproduces a more intense SI.  

To build comparable radial profiles of the intensity representative of the observed streamers, we need to take into account of this bending effect at the base of the streamers, which can be different among models and observation. Additionally, we want to minimize the effect of local small variation of the intensity. For these reasons, at each radial distance we averaged the intensity within an interval $d$PA which collects all the pixels with intensity above 0.8 pB$_m$. 
In Figure \ref{fig:lasco_pa_str} these intervals are limited by the two black profiles at the side of the projected streamer axis. Appendix \ref{app:avg} presents results on the averaged profiles using a different  method where we take into account only of a fixed number of pixels around pB$_m$.

The resulting averaged profiles with the distance are shown in Figure \ref{fig:lasco_nprof6}. In the left panel, we plot as an example the radial profiles for streamer SI. The right panel shows the ratio of the observation to each model for both streamers SI (solid lines) and SII (dashed lines). 

The first thing to be noticed is the  high intensity of the simulated streamers which, at the base, is within a factor 2.5 for SI and  within a factor 5 for SII. As we already noticed,
below about 4 R$_\odot$, the modelled SI radial dependence is not as strong as in the observation,  and this is particularly true for Model 1.
Further out, the ratios become constant.
The ratios for the second streamer SII have instead a different  radial behavior, as they remain almost constant with the distance from the Sun, suggesting that the shape and height of the observed and modelled streamers are similar, even though the intensity is much overestimated by the  simulations. To better understand if the behavior of SI below  4 R$_\odot$ is method dependent, in Appendix \ref{app:avg} we applied a different selection criteria for the pixels. We used a fixed band of 15 pixels for all the solar distances, and we concluded that this may be the case. 

Using  $d$PA intervals around the peak of intensity of the streamers as described for the first method,  we estimated the average axis position above 4 $R_\odot$ (see Table \ref{table:axes}). This choice is guided by the results previously illustrated, both concerning the bent of the streamer axis and the ratio between models and the observation. 
Table \ref{table:axes}  shows that the simulations are able to reproduce the average position of the streamers within less than 10$^\circ$ (sometimes within few degrees), which is quite satisfactory. We notice that as the density increases, the distance between the axes of the two streamers increases, as suggested by the maps in Figure \ref{fig:Expansion}.

\subsection{Results for the Coronal hole data}

\begin{figure*}[ht] 
\centering
\includegraphics[width=0.46\textwidth, trim={70 0 30 0}, clip]{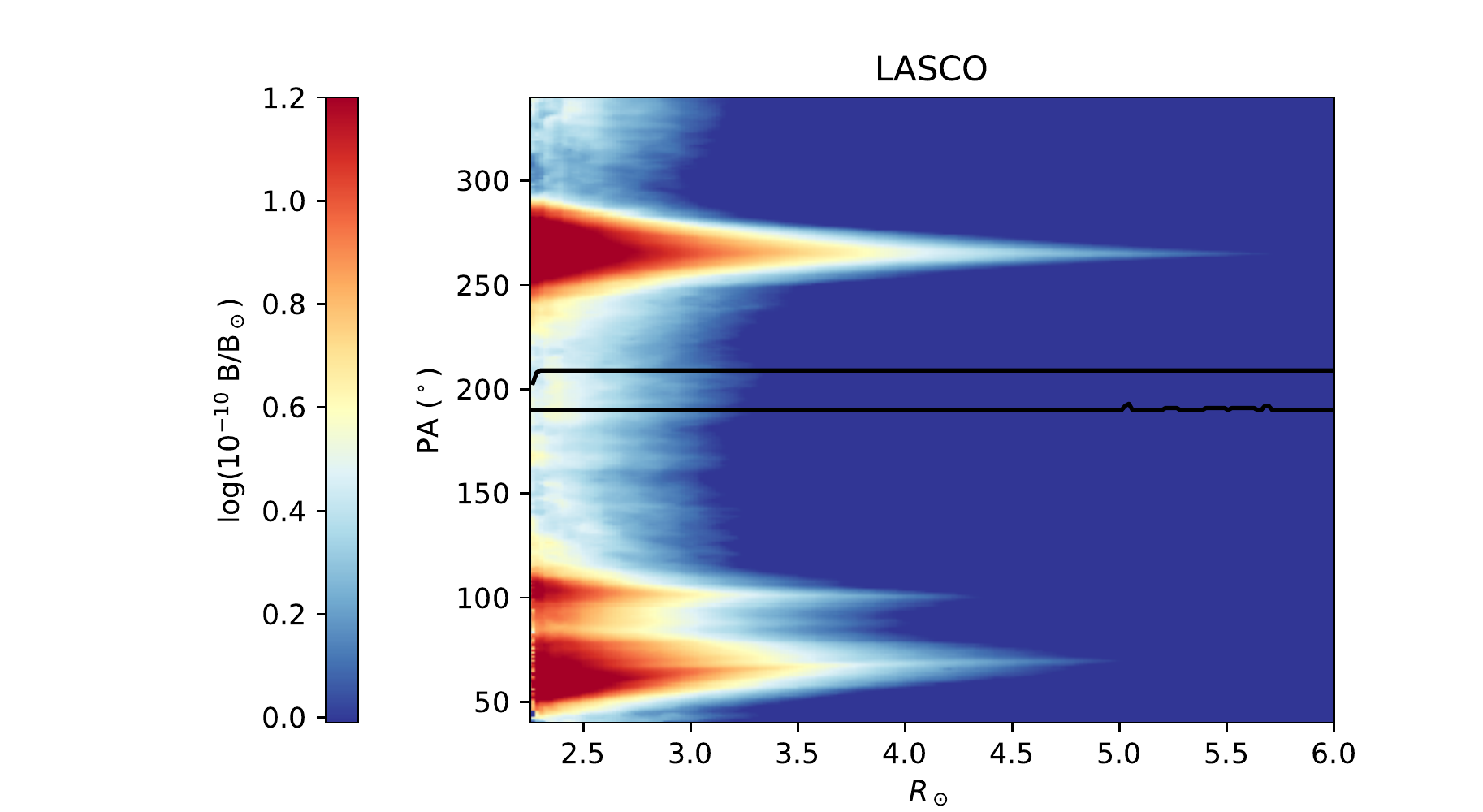}
\includegraphics[width=0.41\textwidth, trim={130 0 20 0}, clip]{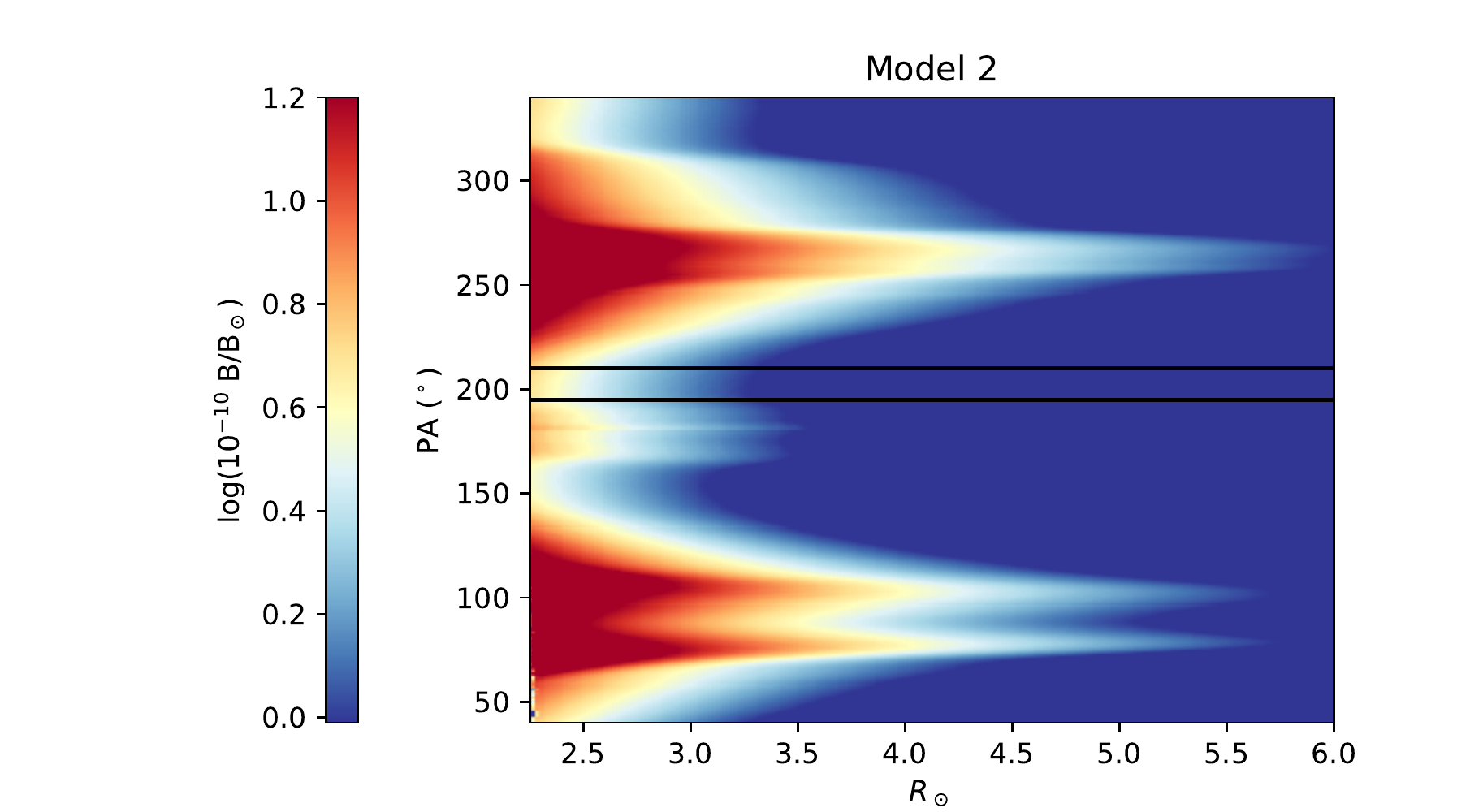}\\
\includegraphics[width=0.45\textwidth,trim={-110 0 45 0}, clip]{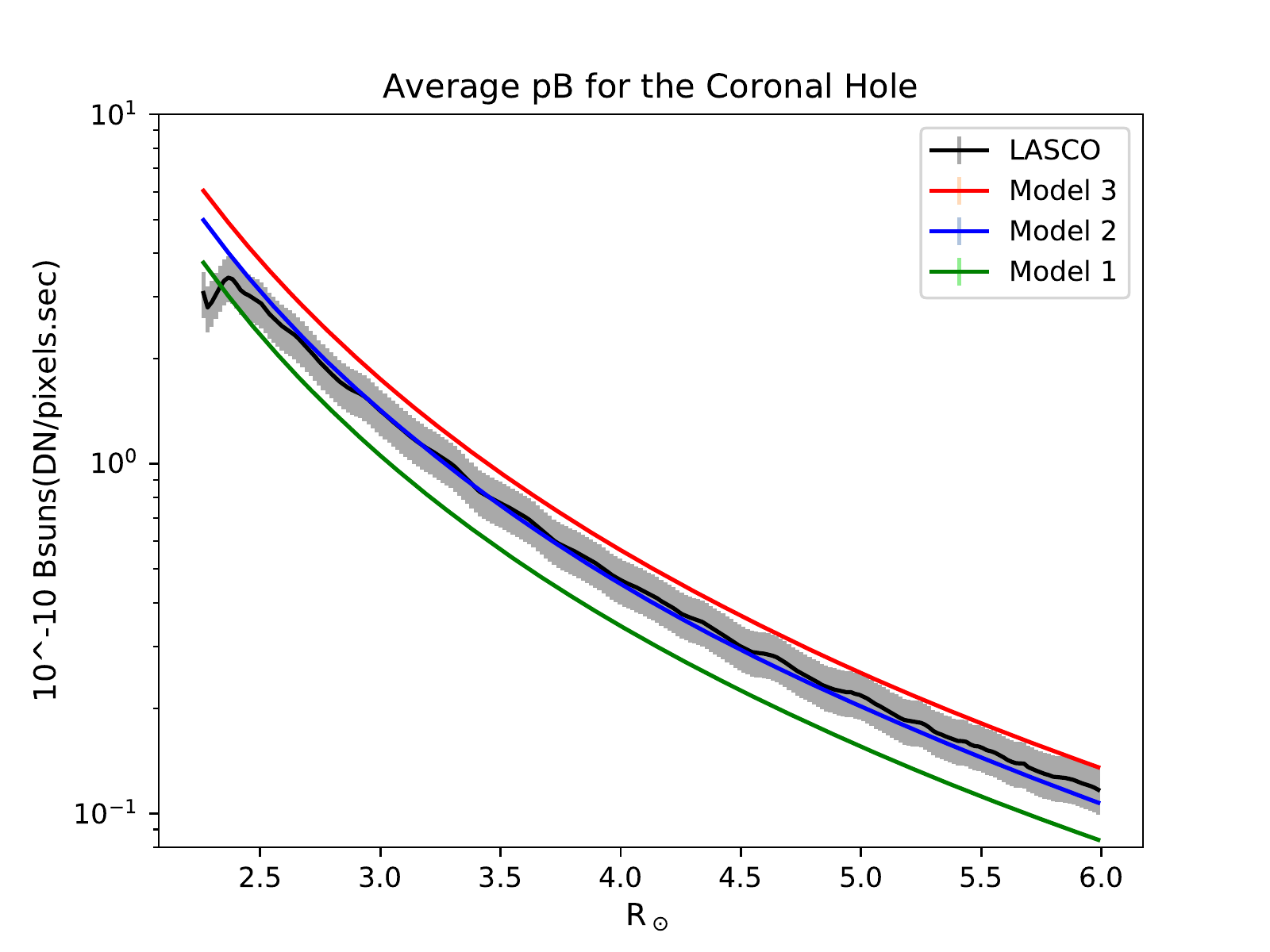}
\includegraphics[width=0.45\textwidth, trim={-50 0 -10 0}, clip]{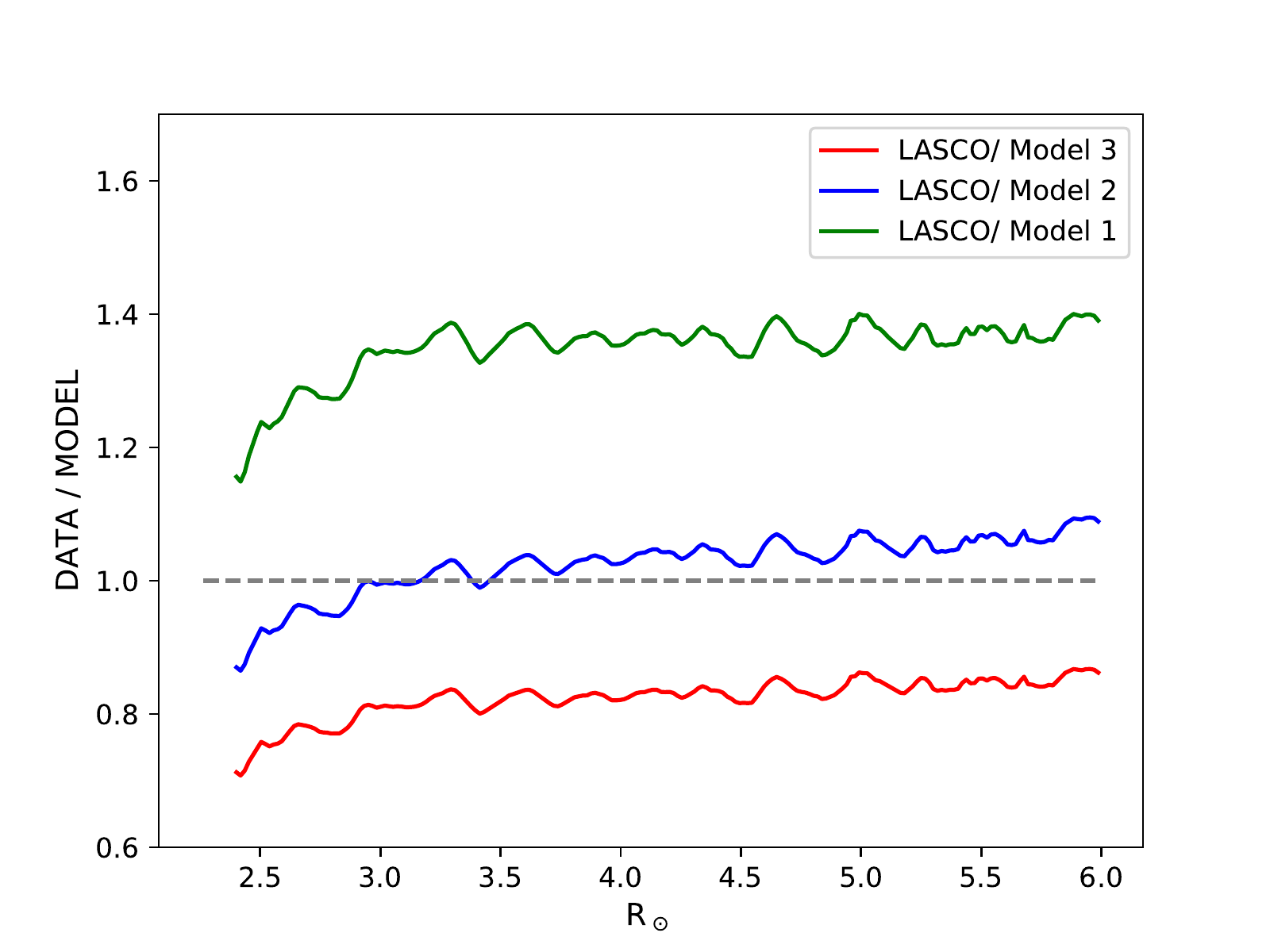}
\caption{Top: LASCO pB and Model 2  synthetic maps with the selected CH  area used to create the averaged radial profile shown in the left--bottom row. Bottom--right: ratio of the observation to models from the left panel. }
\label{fig:lasco_avg_ch}
\end{figure*}

Similar averaged radial profiles were made for a CH area as illustrated in  Figure \ref{fig:lasco_avg_ch}. We selected one of the darkest area close to the south pole and, for each solar distance, we averaged the intensities  within the black lines marked in the first two top panels (LASCO data on the left, Model 2 on the right). 
The profiles plotted in the bottom--left panel show that Model 2 completely superposes the observation, and that the Model 3 profile is also very close to it.

The ratios of the observation to the models (shown on the bottom--right panel) are about constant with the solar distance, similarly to what was obtained above 4 R$_\odot$ within the streamers. This suggests that the models reproduce correctly the pB radial dependence within open field areas.
 
We now move our attention to the 7th of November, where the results from the simulations  will also be compared to the K-Cor pB observation.

\section{pB and EUV results for the off limb corona: November 7}
\label{res:res_corona_7}

The K-Cor observation was taken the day after the PSP perihelion. We verified that the large scale corona changed, and a direct comparison with the observation taken the 6th of November was not appropriate. For this reason, we discuss separately here WL pB and EUV results for the day after. 

\begin{figure*}[ht] 
\centering
\includegraphics[width=0.47\textwidth, trim={70 0 45 0}, clip]{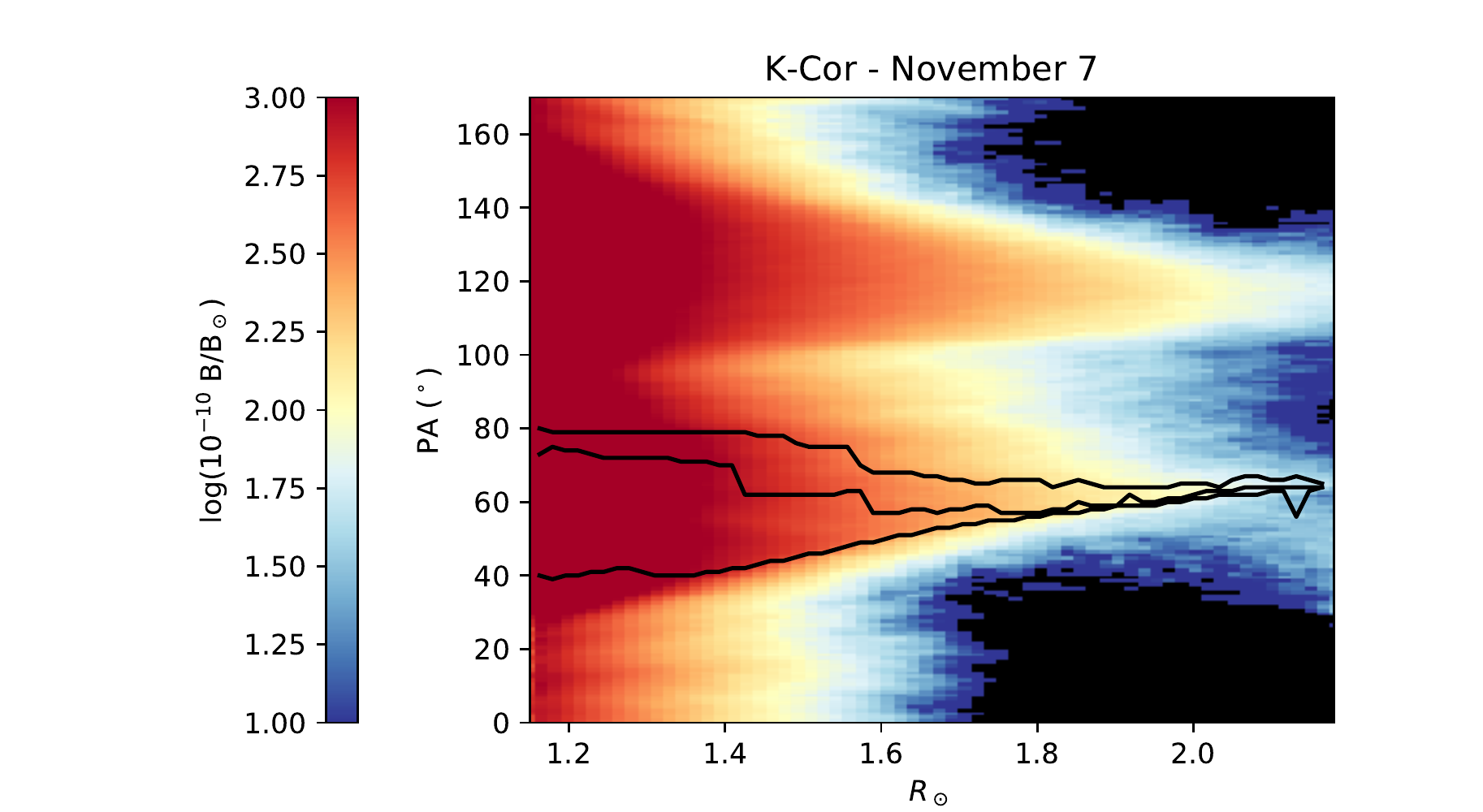}
\includegraphics[width=0.44\textwidth, trim={0 0 0 0}, clip]{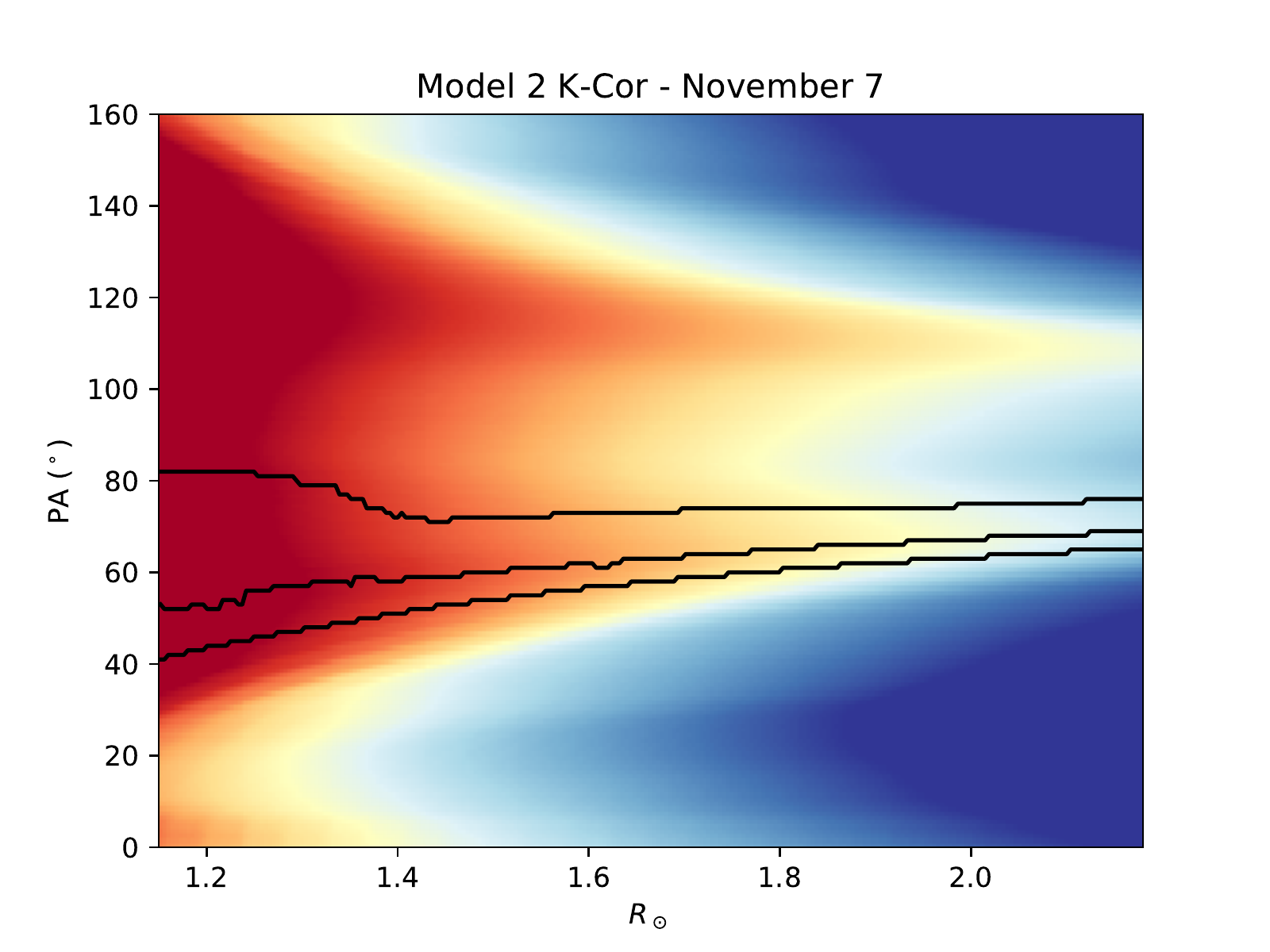}\\
\includegraphics[width=0.47\textwidth, trim={70 0 45 0}, clip]{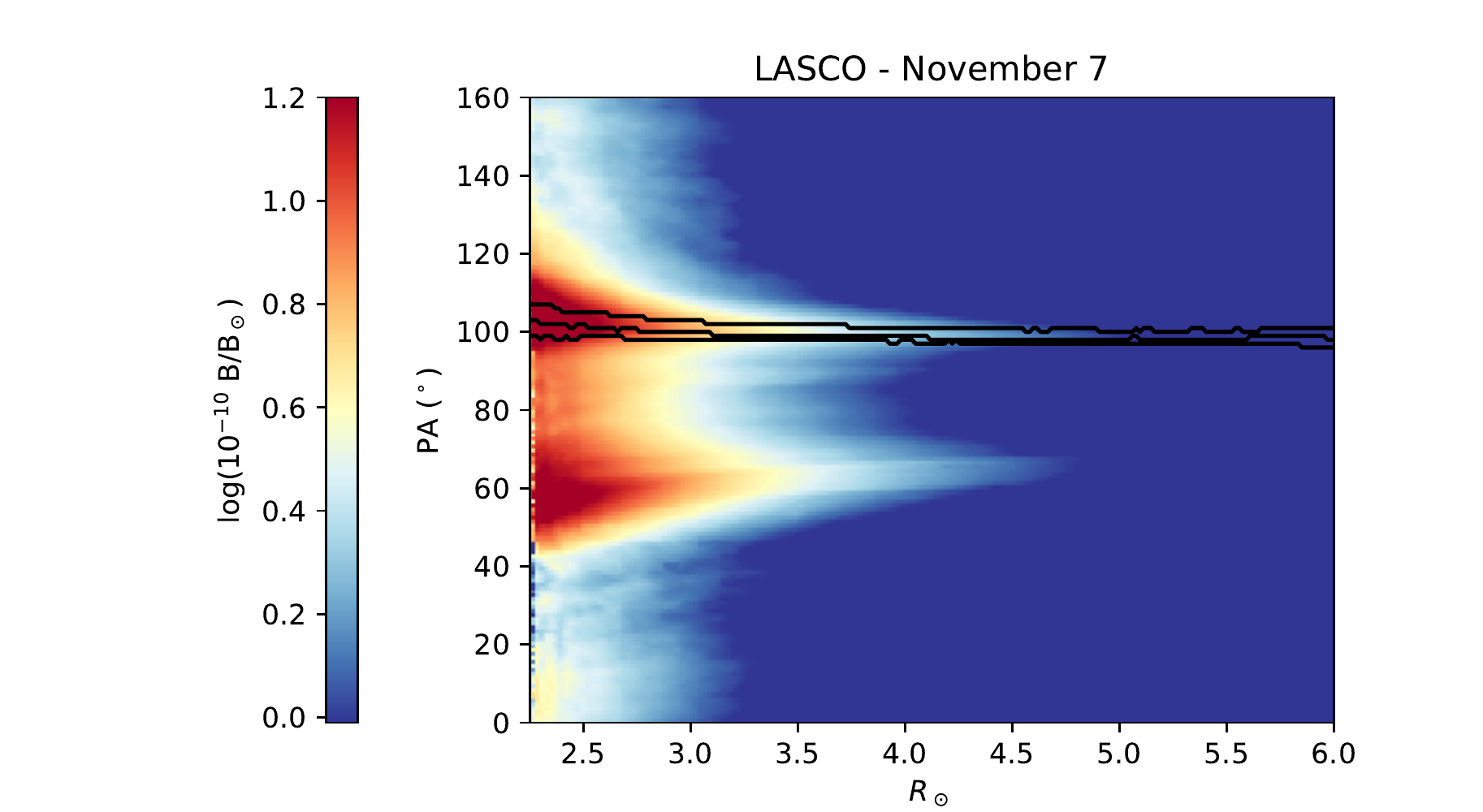}
\includegraphics[width=0.44\textwidth, trim={0 0 0 0}, clip]{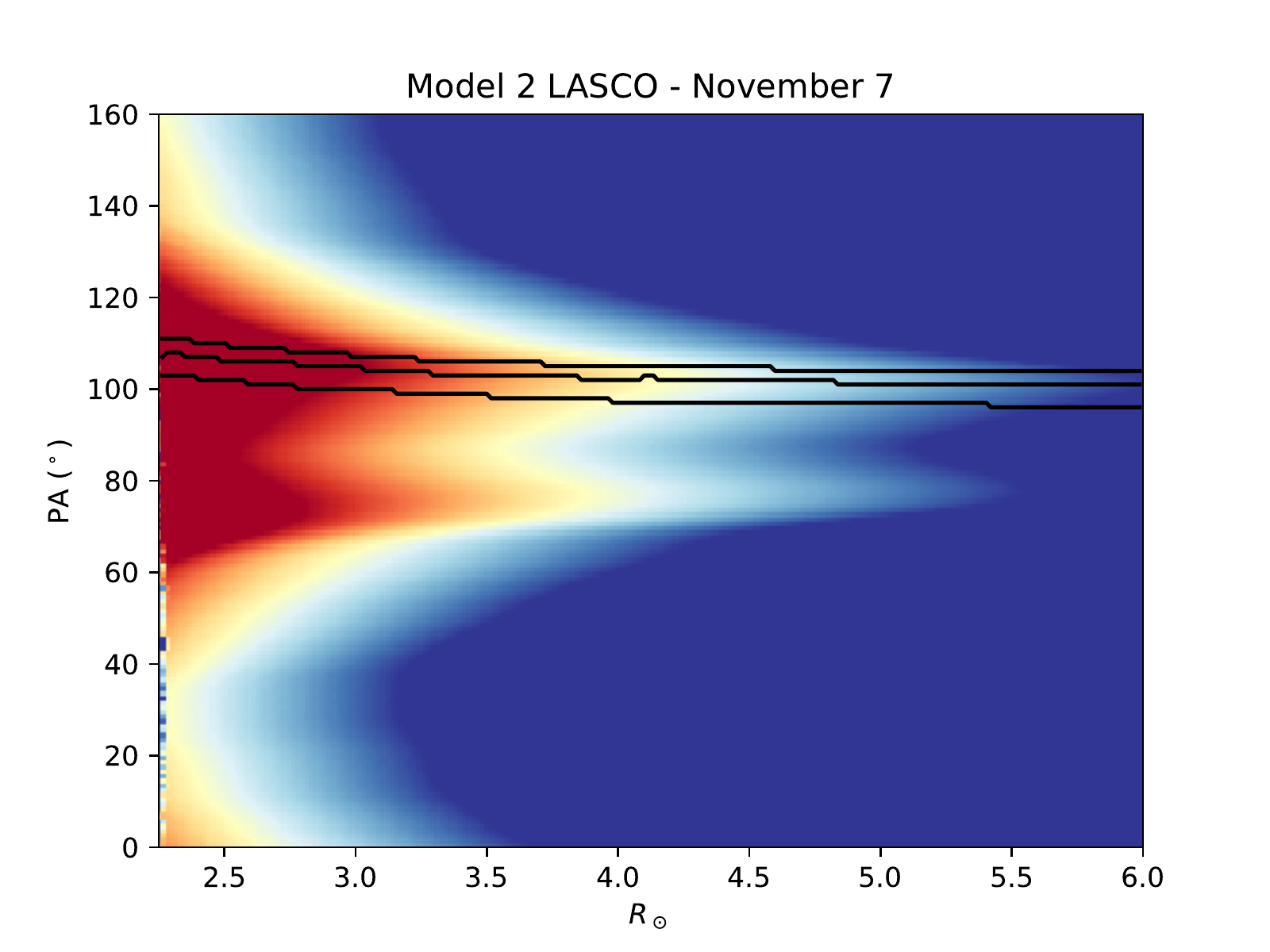}
\caption{
pB maps of the corona in polar coordinates for November 7th imaged by K-Cor (Top--left), result for Model 2 (Top--right), LASCO C2 observation (Bottom--left) and result for Model2 (Bottom--right). The color scale is the same for each row map.}
\label{fig:polar_7}
\end{figure*}

\begin{figure*}[ht] 
\centering
\includegraphics[width=0.8\textwidth]{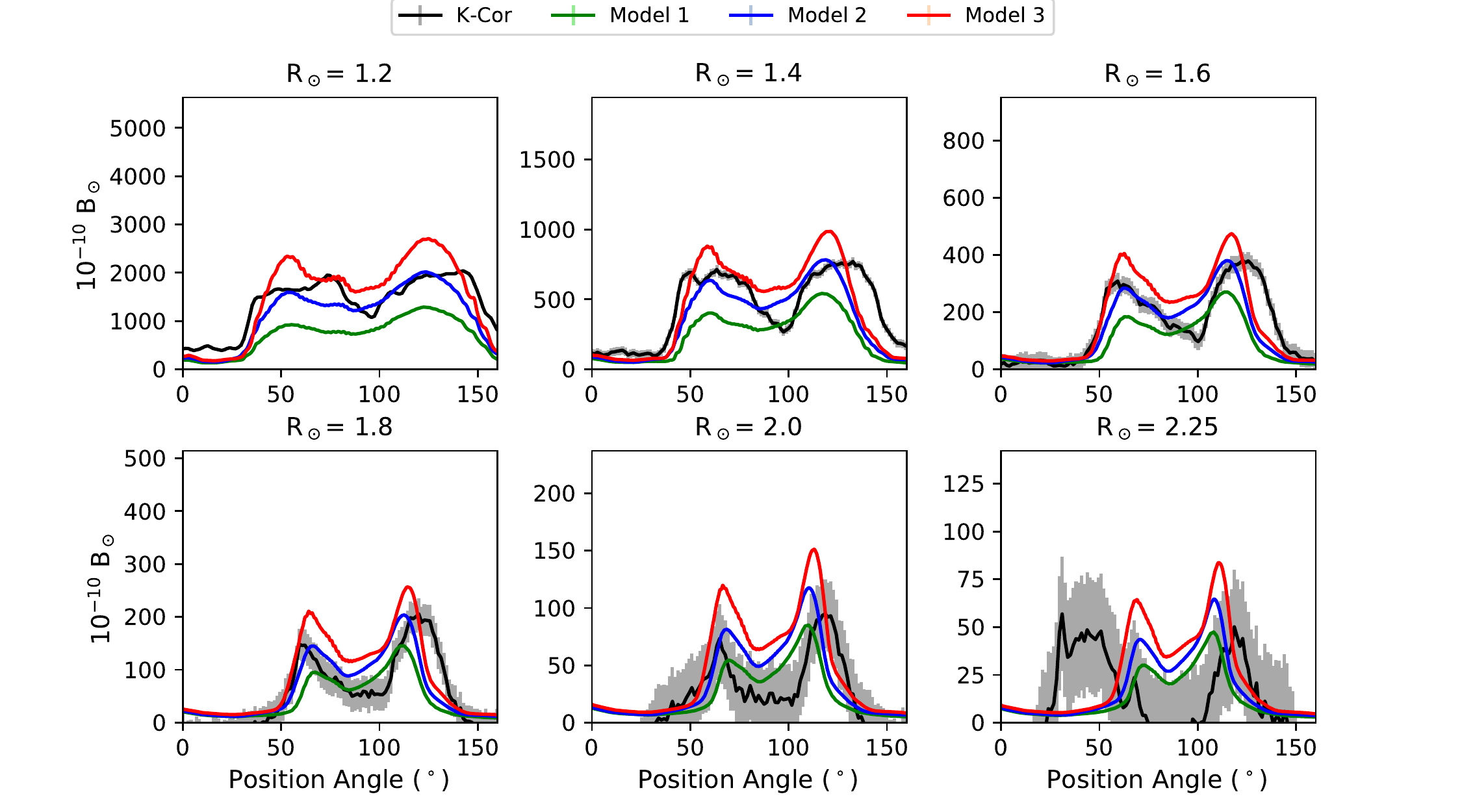}\\
\caption{
Data and simulations of K-Cor latitudinal profiles for different heights.}
\label{fig:k-Cor_pa}
\end{figure*}

Figure \ref{fig:polar_7} shows the polar maps which include the SI and SII streamers for both the pB in K-Cor and LASCO C2 data (first column) and Model 2 synthetic images (second column). The K-Cor data were binned over 4 pixels to reduce the noise.
The overplotted black profiles are those used to build the radial average intensity profiles.
The two streamers appear of similar intensity in both observations, and Model 2 well reproduces this property, contrary to the  simulation for the 6th where it was not the case. 
In Figure \ref{fig:k-Cor_pa} we show the latitudinal intensity profiles for different solar distances within the K-Cor field of view. The streamers SI and SII morphologies  are well reproduced by the models also at these lower heights, even though at R$_\odot$ = 2.25 the observation is too noisy or stray light dominated (see Figure \ref{fig:exe_m2}).
The difference in the intensity peaks of the two streamers is well replicated. Only SII appears slightly narrower in the models than in the observation, with the streamer axis closer to the equator. 
  Model 2 is the one that reproduces very well the observation (particularly SI) at all distances, and Model 3 is close to it.

\begin{figure*}[ht] 
\centering
\includegraphics[width=0.32\textwidth, trim={10 0 40 0},clip]{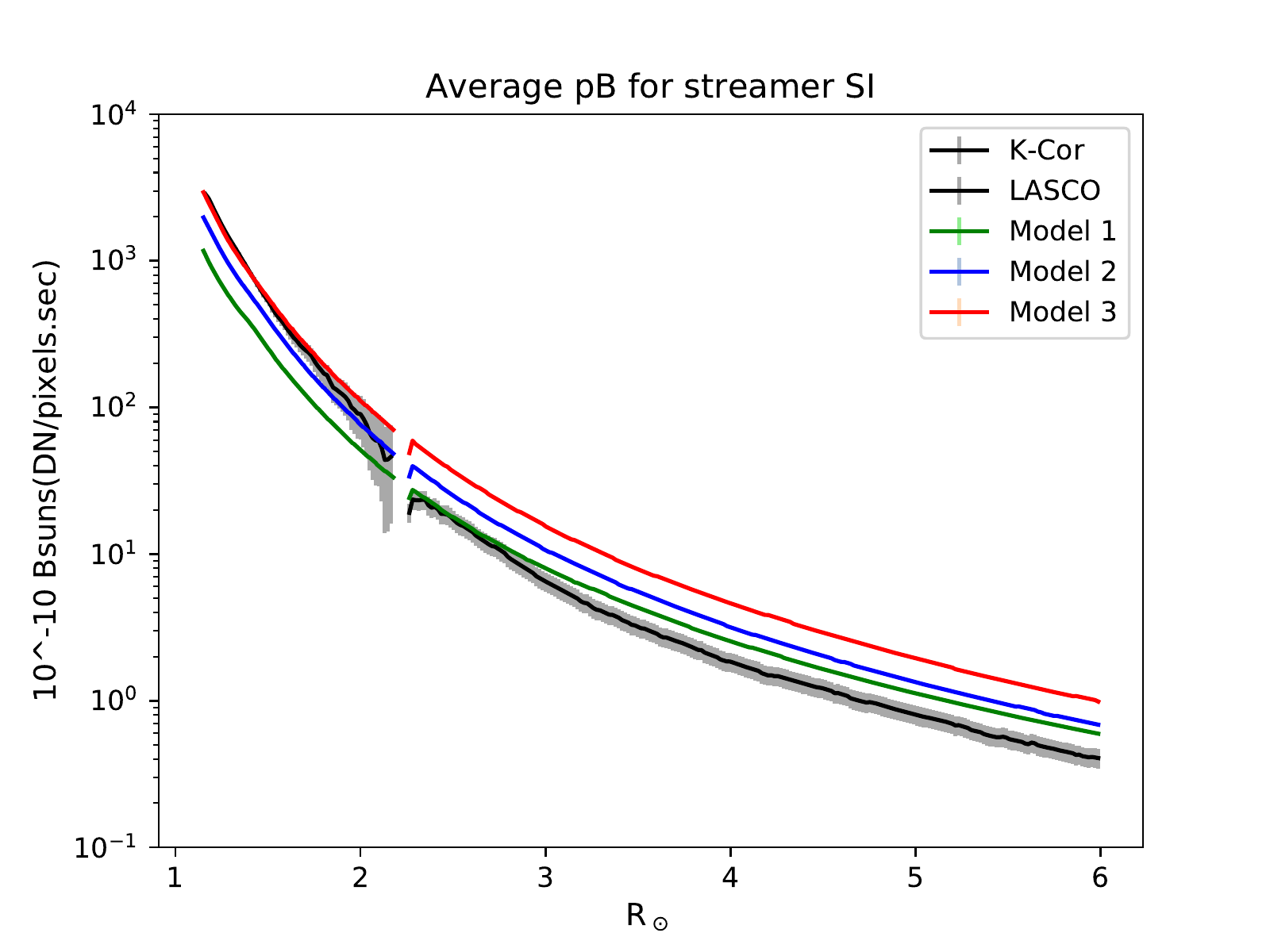}
\includegraphics[width=0.32\textwidth, trim={10 0 40 0}, clip]{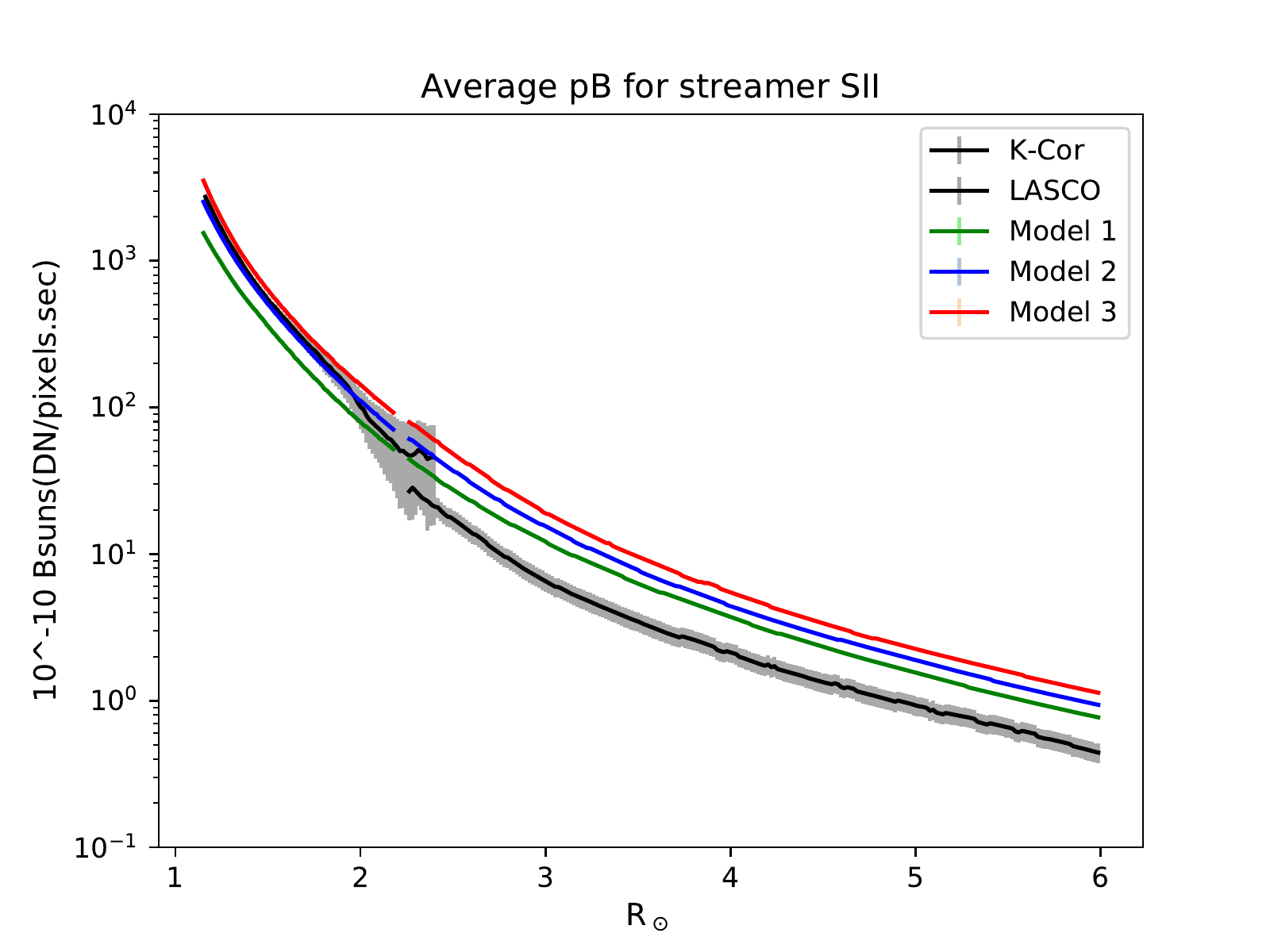}
\includegraphics[width=0.32\textwidth, trim={10 0 40 0}, clip]{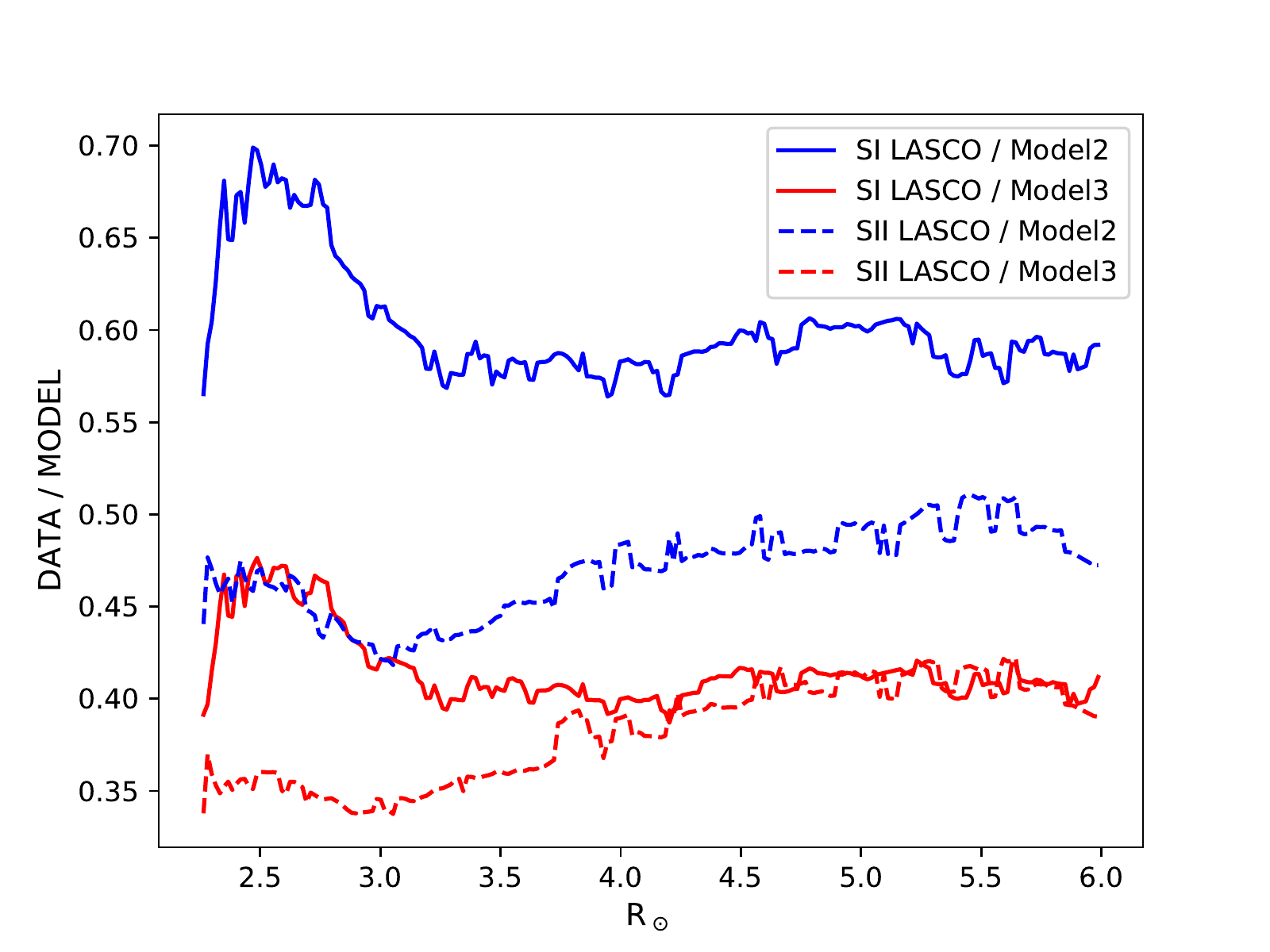}
\caption{
Left and Middle panels: K-Cor and LASCO averaged radial profiles for November 7 together with Model 2 results. Right: ratio of the LASCO observation to Model 2 (blue curves) and Model 3 (red curves).}
\label{fig:rad_7}
\end{figure*}

The first two panels of Figure \ref{fig:rad_7} report the averaged radial profiles for the two streamers. 

The K-Cor observations are very well reproduced by Model 3 for the streamer SI, while it is in between Model 2 and 3 for the second streamer.
However, the LASCO observed intensities are too low with respect to the models, similarly to what was found for November 6. The last panel of the figure quantifies this difference by reporting the ratio of the observation to Model 2 and 3 for LASCO.
For SI the ratio is about the same as for November 6, while for SII the ratio increases approaching the observational values.

Figure  \ref{fig:rad_7} shows a  possible step in intensity between the K-Cor and the LASCO observations. This can be due to the uncertainties in the data processing (as described in Section \ref{sec:data})  and inter-calibration between the two instruments.
At the same time, the plot of the ratio suggests that below about 3 R$_\odot$ the radial dependence of the observations and modeling could be improved.

Table \ref{table:axes} lists the streamers' axis position also for this day. For the K-Cor data, we made two kinds of estimation. In the first case, we selected the data within those distances where the signal looks clean from residual of contamination (case (a), $1.15 <r<2.18 ~~R_\odot$). 
The second case (case (b), $2 <r<3 ~~R_\odot$) was applied only to the simulated streamers, whose data extends  up to the LASCO distances. Doing this, we can compare the axis position within the two datasets.

For case (a), SI axis position is very well reproduced by the models within about 2$^{\circ}$, while for SII the differences are within 6$^{\circ}$. Such difference is also clearly visible in Figure \ref{fig:k-Cor_pa}. When we select only the outer part of the streamers using the case (b), we obtain that the streamers' axis get closer to the LASCO values 
 (within maximum 6$^{\circ}$), showing the continuity and consistency between the observations and modeling.

In general,  both K-Cor and LASCO axis are reproduced by the simulations within few degrees, apart from the LASCO SI case on the 7th, whose difference reaches 11$^\circ$.

\begin{table}[th]
\caption{ Results for the streamers axis position, Position Angle ($^\circ$), obtained from the averaged profiles along the radial direction. 
For LASCO the value is obtained using the data in $4<r<6$. For K-Cor (a) the range is $1.15<r<2.18$ and (b) is $2<r<3$. }
\label{table:axes}
\center
\setlength{\tabcolsep}{3pt}
\begin{tabular}{clcccc}
\hline\hline
Day & Streamer & Observation & Model 1 & Model 2 & Model 3 \\
\hline
 { 06 Nov.}&S I LASCO  & 69.9  &  77.5  & 76.2 & 73.4\\
 { 06 Nov.}&S II LASCO & 99.2 & 102.5 & 103.05 & 105.6 \\
 { 07 Nov.} &S I K-Cor  (a) & 63.5  & 63.1  & 62.3 & 61.1\\
 {  07 Nov.} & S I K-Cor (b)  &       &  72.7 & 71.0 & 69.0 \\
 {  07 Nov.} & S I LASCO   & 66.5  &  77.3   & 75.7.  &  73.4\\
 {  07 Nov.} & S II K-Cor  (a)  & 119.3 & 113.1 & 113.7 & 115.7 \\
 { 07 Nov.} & S II K-Cor (b) &          &  105.5  &  106.1 & 108.5 \\ 
 { 07 Nov.}& S II LASCO &  102  &  102.1      & 102.8 & 105.0 \\
\hline
\end{tabular}
\end{table}

\begin{figure*}[ht] 
\centering
\includegraphics[width=0.8\textwidth]{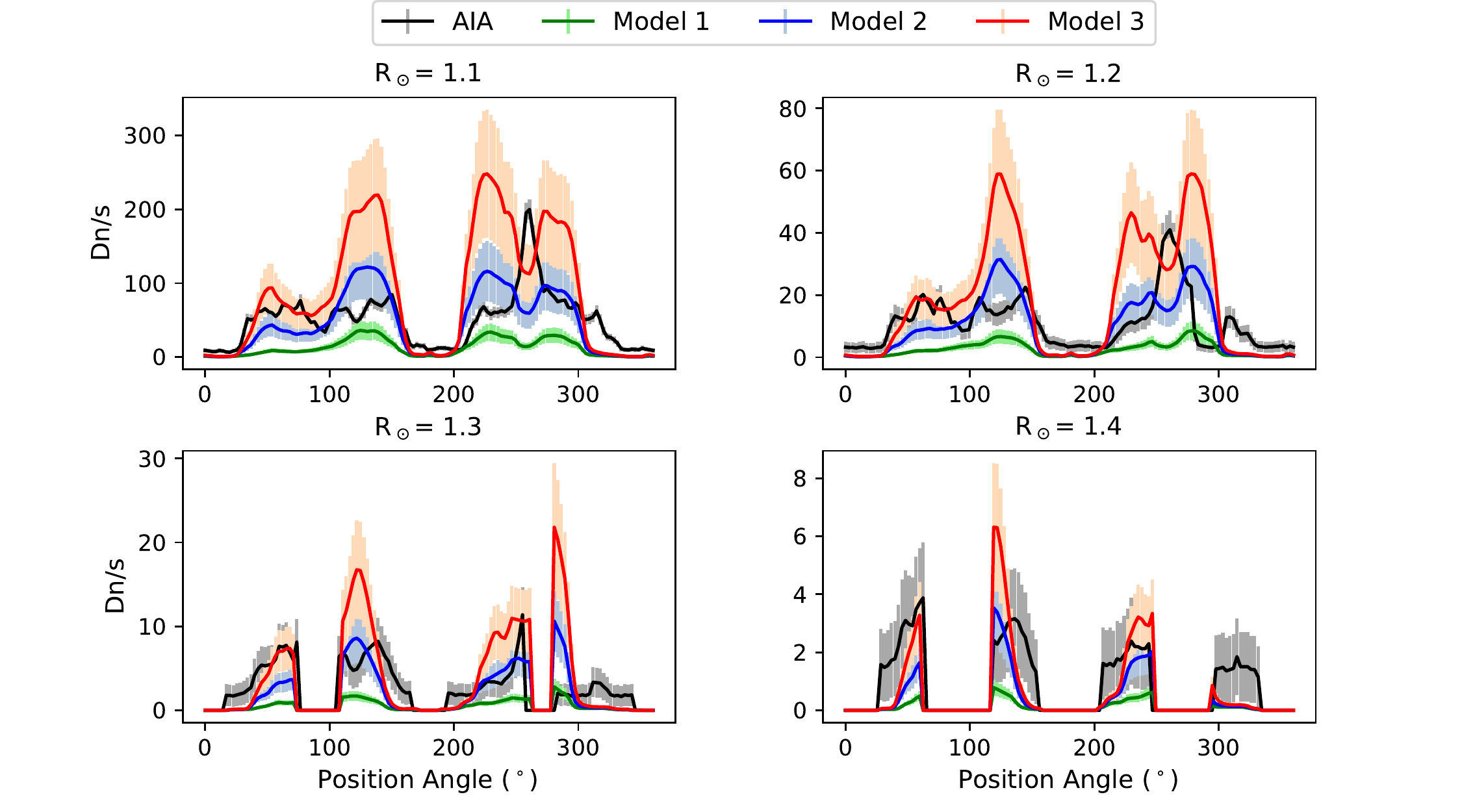}
\caption{ Observation and simulations of AIA 193\,\AA\ latitudinal intensity profiles for different off limb heights (November 7).}
\label{fig:aia_pa_7}
\end{figure*}

We conclude our validation tests on the models, with AIA 193 results, which are shown in Figure \ref{fig:aia_pa_7}. Here we present the latitudinal intensity profiles for November 7. The two streamers from the observation have similar intensity at all heights, while the simulations make SII more intense, similarly to November 6 results. This asymmetry in the peak intensity between the streamers is visible in the K-Cor observation (Figures \ref{fig:k-Cor_pa}) but only above 1.6 R$_\odot$. Nevertheless, within the errors we confirm Model 2 to be the best representative of the data. 

It is also interesting to compare the latitudinal profiles at the same distances from AIA  and K-Cor (1.2 and 1.4 R$_\odot$, Figures \ref{fig:k-Cor_pa} and \ref{fig:aia_pa_7}). In the observations, at 1.2 R$_\odot$ the two streamers have similar intensity. This property is about reproduced by the models in the WL pB profiles, 
while a strong asymmetry is seen for the EUV profiles, with S II being much more intense. A similar behavior is seen at 1.4 R$_\odot$, even though the AIA observation is quite noisy. Such a difference between the two bands may be due to a discrepancy in temperature between simulations and observations; only the EUV emission is, in fact, dependent also on the plasma temperature. 

\section{Discussion and conclusions}
\label{sec:conclusion}

In this work, we have presented a quantitative comparison between observations and synthetic coronal EUV and pB images of the solar corona, derived from the results of three simulations using the 3D MHD WindPredict-AW model. These simulations differ for the inner initial and boundary values of the base density and plasma velocity perturbations. The coronal on--disk and off--limb morphologies are well reproduced by the simulations for  all the AIA bands considered in this study.

On disk quantitative comparisons with the AIA 193 and 211 bands reveal that the intermediate Model 2 best reproduces the observations, even though the CHs are always too dark in the simulations. The equatorial CH in the observations is not well marked and some QS structuring is visible within it. This large scale modulation in the intensity is partially reproduced by the simulations (see Figure \ref{fig:aia_radial} and \ref{fig:aia_211_prof}). The model lower limit at the base of the corona of about 0.4 MK leaves out possible contributions of the cooler plasma emission visible by the AIA bands. The response function of these bands, in fact, has an extended wing at low temperature (see Figure \ref{aia_resf}). For instance, the 193 band has a secondary peak close to $\mathrm{0.3 ~~MK}$ (due mostly to \ion{O}{5} lines). This is about $20\%$ of the main peak (dominated by \ion{Fe}{11}--\ion{}{12}).  To test this possibility, we simulate the Dn/s within the bands using the CHIANTI database. We found that the contribution of such a wing in the 193 band can reach 50$\%$ of the total CH emission. Adding such a component will increase the consistency with the observations.

Additionally, we investigated the possible contribution of the scattered light to the CH measured intensity. In Appendix \ref{app:aia_PSF} we show that applying the standard PSF to the 193 channel to our simulated images,  the CHs still remain too faint with respect to the observations.
We have to note, however, that such PSF only includes the diffraction correction and not the mirror scattering. This can additionally explain the difference we found between the observation and modeling in the off limb radial profiles. In this regard, \cite{Saqri2020} estimated in the 193  band an extra stray light contribution which can reach up to 50$\%$ of the CH measured intensity, while a more variable value was found for the 211 channel which could reach higher value.  The 171 band appears to be less affected by this stray light component.

The AIA response functions depend on the plasma composition which, in our case, has been assumed to be coronal. However, the solar corona may have a different composition depending on the region of interest \citep[see for instance][]{Delzanna2018}. Tests performed in the past on the AIA bands \citep[][]{Delzanna2011} showed that such changes affect the shape of these functions mostly in the cooler wing. This is because they are dominated by the emission from a mixture of other ions than the Fe, which instead dominates the coronal bands at the maximum of their response.
Due to the present  model setup, in our simulations we have only a small part of the plasma which is at the temperatures covered by the cool wing of the AIA bands. This means that if there is a  change in the plasma composition, it will be detected only through the change of the amount of emission from the Fe ions.
The use of only the coronal composition in our simulation may certainly be a source of error for our work, but this is minimized by the dependence of on one element only. Note, that the error we assumed for the synthetic EUV intensities takes into account  the error in the atomic physics, which also includes uncertainties on the elemental composition.

The elemental composition also affects the amount of radiative losses expressed through the loss function within the energy balance equation in the model, and consequently  the amount of electron density derived. Again, this is a source of error in our validation process when we compare both the EUV and pB simulated and observed images. However, part of this is taken into account within the error we associated to our synthetic EUV intensities.

Concerning the comparison with polarized brightness measurements of the corona, we note a clear mismatch between the simulations and the observations of the east streamer. One of the reason for this mismatch could be found in the arrival of an AR at the east limb which is not fully characterized by our magnetic map. Aiming at validating the MHD model, we thus concentrated our white light analysis on the west limb. The off limb corona is very well reproduced within the K-Cor distances, that is about 2 R$_\odot$, with an excellent agreement of Model 2  and Model 3. In the LASCO field of view, the models mostly reproduce well the two streamers in terms of morphology and radial decay. More discrepancies are observed at low heights within streamer SI, with the streamer being highly asymmetric in the observations. The most important difference between the modeling and observations is the counts of the radial profiles which are too high in the models, particularly for S II. The intensity decay with the distance runs almost parallel to the observations apart from one case. The small difference goes in the direction of a too slow decay in the models. However, we also notice a possible step between the K-Cor and LASCO observations, which may explain part of this difference (see Figure \ref{fig:rad_7}).  At the junction between the two instruments the uncertainties become important and may vary with the period of the observations \citep{lamy2020}. In our case, a comparison with other published radial profiles within streamers shows similarities in the absolute pB LASCO C2 values \citep[see for instance][]{lamy2020, lamy2019}.  The off limb CH results support the picture where Models 2 and 3 are practically correct, within 20$\%$, in this area.

Combining EUV and polarized brightness results, we thus identify Model 2 as the most representative of the observed corona during the 6 and 7 November 2018. Additionally, in certain cases, Model 3 is also compatible with the observations. Model 2 and 3 display in general similar features, despite differences in the input wave energy flux (see, e.g., Figure \ref{fig:aia_radial}). Model 1 is on the contrary quite different from the other two, even though it has the same input energy flux than model 3. Analyses of typical profiles shown in Figure \ref{fig:profiles} suggests that our different setups are very sensitive to the inner boundary condition parameters, in particular the base coronal density. Lowering the base coronal density in Model 1 implies that more energy per particle is available for the heating of the corona (through $Q_w$) and the acceleration of the solar wind (through the wave pressure) over larger distances, in addition to the already larger amplitude of the Alfvén waves. Hence, the average open regions of Model 1 have a less dense and faster wind in comparison with the other models. Now, it should be stressed that setting the coronal base density, even though our scheme allows some spatial variation at the boundary depending on the outer solution, is a strong simplification of the physics of the lower atmosphere. The value of the coronal density is indeed, in reality, the result of the balance of heating, radiative losses, and thermal conduction going through the transition region. For instance, the coronal density has been shown to be roughly linearly related to the heating of a given flux tube \citep{Rosner1978} and many works have documented the physical ingredients and numerical techniques to properly render the chromosphere and transition region \citep{Lionello2001,Lionello2009,Downs2010, Johnston2020, Zhou2021}. From there, a clear path can be set for future improvements of the model. First by including the transition region. Using very high resolution 1D and 2.5D simulations with simple magnetic field, we have already tested the ability of the PLUTO code to describe the balance of thermal conduction, radiative losses and heating into sharp density gradients characteristics of the transition region \citep{Reville2018,Reville2019,Reville2021}. Transposing such resolutions to 3D requires however unrealistic computing times. Preliminary tests show that the technique described in \citet{Lionello2009} to thicken the transition region, keeping thermal conduction and radiative losses balanced as in low coronal regimes, seems compatible with tractable WindPredict-AW 3D simulations. As discussed before, this addition is necessary for a better comparison of EUV emissions in low temperature CHs. It is also essential to make the model compatible with more active phases of the solar cycle, in which active regions will represent an increasing part of the solar disk \citep[see, e.g.,][]{Mok2005, Mok2008, Mok2016}. These improvements will also help describe the small scale coronal bright points, that the current spatial resolution of the magnetograms do not include. Finally, a precise description of the low coronal regions must account rigorously for the different composition and subsequent radiative losses functions in the different layers. 

From the observational point of view, we need a more regular update of the photospheric magnetic map, particularly for the far side of Sun. This will decrease the uncertainties in the coronal magnetic field extrapolation and modeling, particularly at the east limb, and allows a stronger constraint on the lower boundary conditions of our model. With the new Solar Orbiter mission at work (as well as  the next ESA Lagrangian 5 mission) providing, among others, photospheric magnetic field \citep[through the PHI instrument,][]{Solanki2020} from different view points than Earth, this issue will be certainly partially or completely resolved. 
To reach a stronger constraint to the global corona-solar wind modeling validation, ideally we need spatially uninterrupted observations over the corona and the heliospheric distances with instruments at high photometric sensitivity \citep[such as envisaged by ASPIICs on board PROBA3,][]{Galano2018}. 

The challenge for reaching a quasi-uninterrupted radial coverage of imaging data has been taken by the Solar Solar Orbiter project through the full Sun imager of EUI \citep{Rochus2020}, the coronagraph METIS \citep{Antonucci2020} and the heliospheric Imager \citep[SOLOHI,][]{Howard2020}.
METIS will partially overlap to the full Sun EUV imager, which will allow to expand the work we have attempt here using cotempotral AIA off-limb and K-COR data. In the future the PUNCH project also envisage a very large coverage of the measurements (within about 42$^\circ$).

To have a reliable 3D MHD model, testing on multiple solar conditions using different wavebands is of primary importance. For instance, the connectivity between what is measured \emph{in-situ} and on the Sun could be much better constrained. This is one of the main goal of Solar Orbiter. Additionally, we have now regular possible connectivities configurations between the Sun, Solar Orbiter and PSP, which can be identified only with the support of a reliable coronal and heliospheric model, as the one used here.  In this regard, our aim in the near future is the co-temporal constrain of our WindPredict-AW model with both coronal and \emph{in-situ} data. This model will also be extremely useful is in support of the planning for the Solar Orbiter observations \citep[][]{Zouganelis2020, Rouillard2020}, that should be prepared weeks in advance \citep{Auchere2020}.


\begin{acknowledgements}

S.P, A.S.B, B.P thanks ESA-GSTP funding for preparing WindPredict for the  VSWMC portal.
This work was supported by the CNRS and INSU/PNST program, CNES Solar Orbiter and “Météorologie de l’espace” funds, and the ERC Synergy grant WholeSun No. 810218. V.R. acknowledges funding by the ERC SLOW{\_}\,SOURCE project (SLOW{\_}\,SOURCE - DLV-819189). Computations were carried out using CNRS IDRIS facility within the GENCI A0110410293 and A010810133 allocations and a local meso-computer founded by DIM ACAV+. The authors are grateful to A. Mignone and the PLUTO development team.

Courtesy of the Mauna Loa Solar Observatory, operated by the High Altitude Observatory, as part of the National Center for Atmospheric Research (NCAR). NCAR is supported by the National Science Foundation.

This work makes use of the LASCO-C2 legacy archive data produced by the LASCO-C2 team at the Laboratoire d’Astrophysique de Marseille and the Laboratoire Atmosphères, Milieux, Observations Spatiales, both funded by the Centre National d’Études Spatiales (CNES). LASCO was built by a consortium of the Naval Research Laboratory, USA, the Laboratoire d’Astrophysique de Marseille (formerly Laboratoire d’Astronomie Spatiale), France, the Max-Planck-Institut für Sonnensystemforschung (formerly Max Planck Institute für Aeronomie), Germany, and the School of Physics and Astronomy, University of Birmingham, UK. SOHO is a project of international cooperation between ESA and NASA.

Courtesy of NASA/SDO and the AIA, EVE, and HMI science teams.

This work utilizes data produced collaboratively between AFRL/ADAPT and NSO/NISP.

\end{acknowledgements}

\appendix

\section{Full Set of Equations}
\label{app:eq}

We recall here the full set of equations solved by the code. The MHD equations are solved in conservative form for the background flow, while the contribution of the waves' energy ($\mathcal{E} = \mathcal{E}^+ + \mathcal{E}^-$) and pressure ($p_w = \mathcal{E}/2$) is accounted for. The system can be written:

\begin{equation}
\label{MHD_1}
\frac{\partial}{\partial t} \rho + \nabla \cdot \rho \mathbf{v} = 0,
\end{equation}
\begin{equation}
\label{MHD_2}
\frac{\partial}{\partial t} \rho \mathbf{v} + \nabla \cdot (\rho \mathbf{vv}-\mathbf{BB}+\mathbf{I}p) = - \rho \nabla \Phi - 2 \rho \Omega_z{\bf e}_z\times {\bf v} - \rho \Omega_z^2 {\bf e}_z \times \left({\bf e}_z\times {\bf r} \right),
\end{equation}
\begin{equation}
\label{MHD_3}
\frac{\partial}{\partial t} (E + \mathcal{E} + \rho \Phi)  + \nabla \cdot [(E+p+\rho \Phi)\mathbf{v}-\mathbf{B}(\mathbf{v}\cdot \mathbf{B}) + \mathbf{v}_g^+ \mathcal{E}^+ + \mathbf{v}_g^- \mathcal{E}^-] = Q_h-Q_c-Q_r,
\end{equation}
\begin{equation}
\label{MHD_4}
\frac{\partial}{\partial t} \mathbf{B} + \nabla \cdot (\mathbf{vB}-\mathbf{Bv})=0,
\end{equation}
\begin{equation}
\frac{\partial \mathcal{E}^\pm}{\partial t} + \nabla \cdot \left( [\mathbf{v} \pm \mathbf{v_A}] \mathcal{E}^\pm \right) = -\frac{\mathcal{E}^\pm}{2} \nabla \cdot \mathbf{v}  - Q_w^\pm,
\label{MHD_5}
\end{equation}
where $E \equiv \rho e + \rho v^2/2 + B^2/2$ is the background flow energy, $\mathbf{B}$ is the magnetic field, $\rho$ is the mass density, $\mathbf{v}$ is the velocity field, $p = p_{\mathrm{th}}+ \mathcal{E}/2 + B^2/2$ is the total (thermal, wave and magnetic) pressure, $\mathbf{I}$ is the identity matrix and $\mathbf{v}_g^{\pm} = \mathbf{v} \pm \mathbf{v_A}$ is the group velocity of Alfvén wave packets. $\Phi=-GM_{\odot}/r$ is the gravitational potential. The equations are solved in the rotating frame with a frequency of $\Omega_z = 2.86 \times 10^{-6}$s, corresponding to the solar sidereal period of $25.38$ days. The rotation axis ${\bf e}_z$ is aligned with the North Pole of the spherical coordinates.
The Coriolis and centrifugal forces are taken into account in the momentum conservation equation (see Eq. \ref{MHD_2}). Within the PLUTO code they are split between sources and conservative updates to minimize discretization errors in spherical coordinates (see PLUTO's documentation).
 
Note that, as explained in the Erratum of \citet{Reville2020ApJS}, the wave heating term $Q_w$ does not appear explicitly in equation \ref{MHD_3}, but acts nonetheless as a source term on the fluid's energy along with $Q_h,Q_c$ and $Q_r$. The wave heating term $Q_w = Q_w^+ + Q_w^-$ is obtained using the two equations of wave energy propagation and dissipation \ref{MHD_5}, which solve the time evolution of $\mathcal{E} = \mathcal{E}^+ + \mathcal{E}^-$. These correspond to two populations of parallel and anti-parallel Alfvén waves excited from the boundary conditions. The waves' energy density can be written:
\begin{equation}
\mathcal{E}^\pm =  \rho \frac{|z^\pm|^2}{4},
\end{equation}
where the Els\"asser variables are defined as follows:
\begin{equation}
\mathbf{z}^\pm = \delta \mathbf{v} \mp \mathrm{sign}(B_r) \frac{\delta \mathbf{b}}{\sqrt{\mu_0 \rho}},
\end{equation}
so that the sign + (-), corresponds to the forward wave in a + (-) field polarity. 

Following the Kolmogorov phenomenology, each term
\begin{equation}
Q_w^\pm = \mathcal{E^\pm} \frac{|z^{\mp}|}{2 \lambda_\perp}=\rho  |z^\pm|^2 \frac{|z^\mp|}{8 \lambda_\perp},
\label{eq:kol_ph}
\end{equation}
\textbf{where $\lambda_\perp=0.0020 R_{\odot} \sqrt{B_0/|B|}$, is the turbulence correlation scale. The reference magnetic field $B_0=1$ G, and $\lambda_\perp$ increases with the distance to the Sun (and the decay of the magnetic field)}. Finally, we assume that, in open regions, there is a small reflection of forward waves giving birth to inward waves. We model this reflection through a constant reflection coefficient $\mathcal{R}=0.1$. This reflection process only appears in the dissipation terms, as we assume that the reflected component is instantly dissipated. In general, the turbulent heating can thus be written: 

\begin{equation}
Q_w^\pm = \frac{\rho}{8}  \frac{|z^\pm|^2}{\lambda_\perp} (\mathcal{R} |z^\pm| + |z^\mp|).
\label{eq:kol_ph2}
\end{equation}

We close the system using an ideal equation of state relating the internal energy and the thermal pressure, 
\begin{equation}
\rho e  = \frac{p_{\mathrm{th}}}{\gamma-1},
\end{equation}
with $\gamma = 5/3$, the ratio of specific heat for a fully ionized hydrogen gas.

\section{Comparison of magnetic maps from November 6th and 7th}
\label{app:mag_6_7}

\begin{figure}
    \centering
    \includegraphics[width=7.5in]{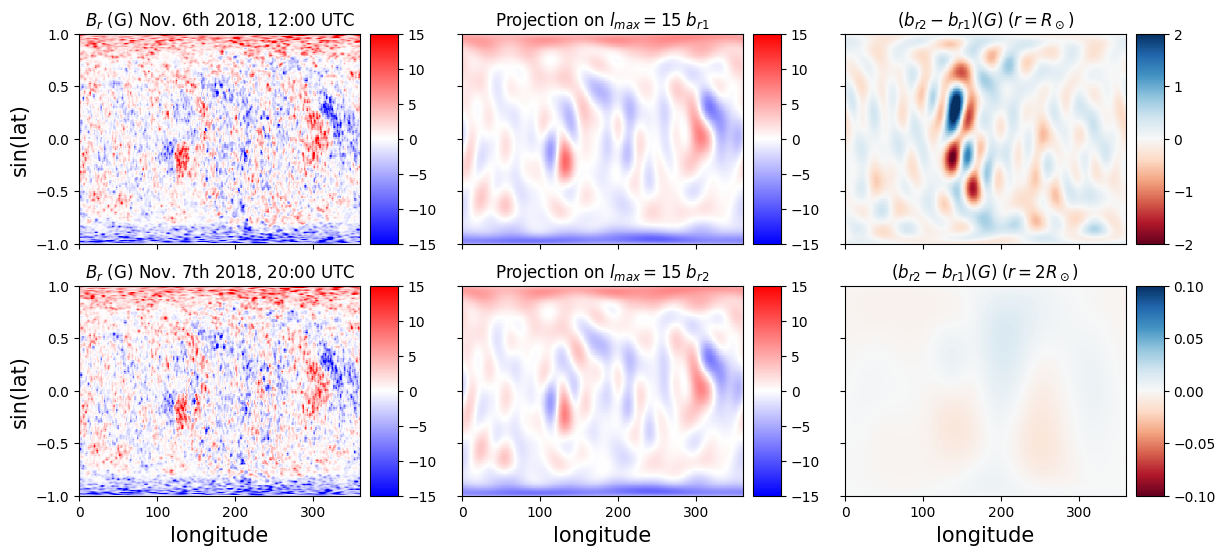}
    \caption{Comparison of the low coronal boundary conditions used in the simulations (top row on November 6th) and the magnetic field configuration on November 7th. The left column are the original ADAPT maps. The middle panel is the projection on the 15 first spherical harmonics. The bottom row shows the differences of the projected magnetic fields at the surface ($1R_{\odot}$) and at $2 R_{\odot}$, using a PFSS extrapolation.}
    \label{fig:mag_6_7}
\end{figure}

 In Figure \ref{fig:mag_6_7}, we compare the ADAPT maps of November 6th 12:00 UTC and November 7th  20:00 UTC 2018. These two magnetograms use input data from KVPT, GONG and VSM and couple them to a flux transport model to obtain an ensemble of realizations describing the complete solar surface at the desired time and date \citep{Worden2000, Hickmann2015}. The November 7th map was not used for the simulations, but since we compared our models to observations which were taken the day after the PSP perihelion, we want to demonstrate here that the change in the solar surface magnetic field during this 28 hours interval is not significant enough to justify a new simulation. The original ADAPT maps are shown in the left column. In the middle column, we show the spherical harmonics filtering up to $l_{\mathrm{max}}=15$, which is used in the initialization of the simulation. We then compare this boundary condition to the map used for the simulations and the map of November 7th 2018 20:00 UTC which is the closest time to the K-Cor data used in section \ref{res:res_corona_7}. At the surface (top right panel), we observe some differences, of the order of a few Gauss between 100 and 200 degrees of longitude. We show again the field difference in the bottom right panel, but at $1 R_{\odot}$ above the solar surface, using a PFSS model with $r_{ss} = 2R_{\odot}$. We see that very little differences remain (of the order of $0.01$ Gauss), which reinforces our confidence in the large scale structure analysis performed in this work.

\section{The effect of the AIA Point Spread Function on the disk intensity}
\label{app:aia_PSF}

\begin{figure}
\includegraphics[width=0.45\textwidth]{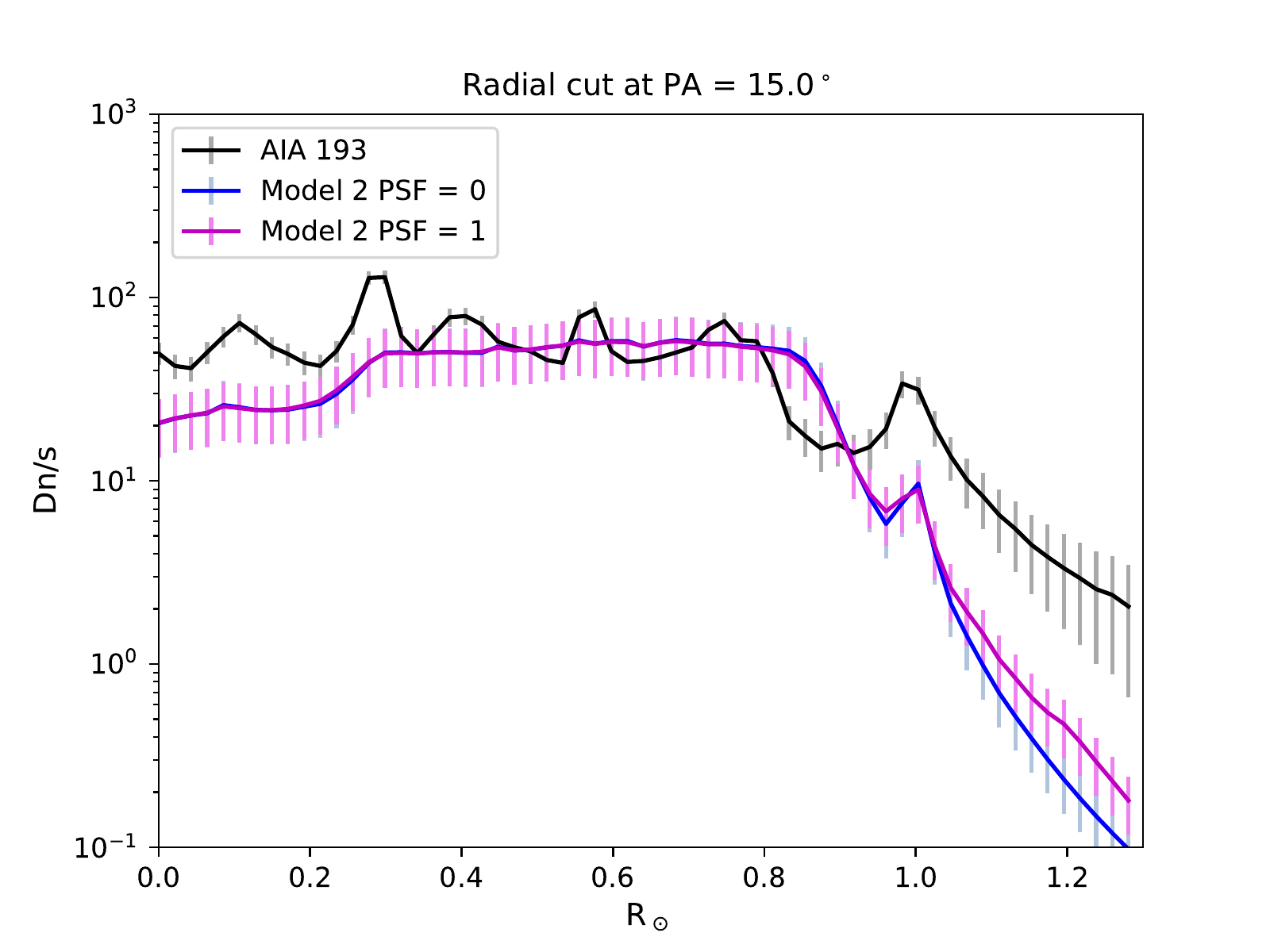}
\includegraphics[width=0.45\textwidth]{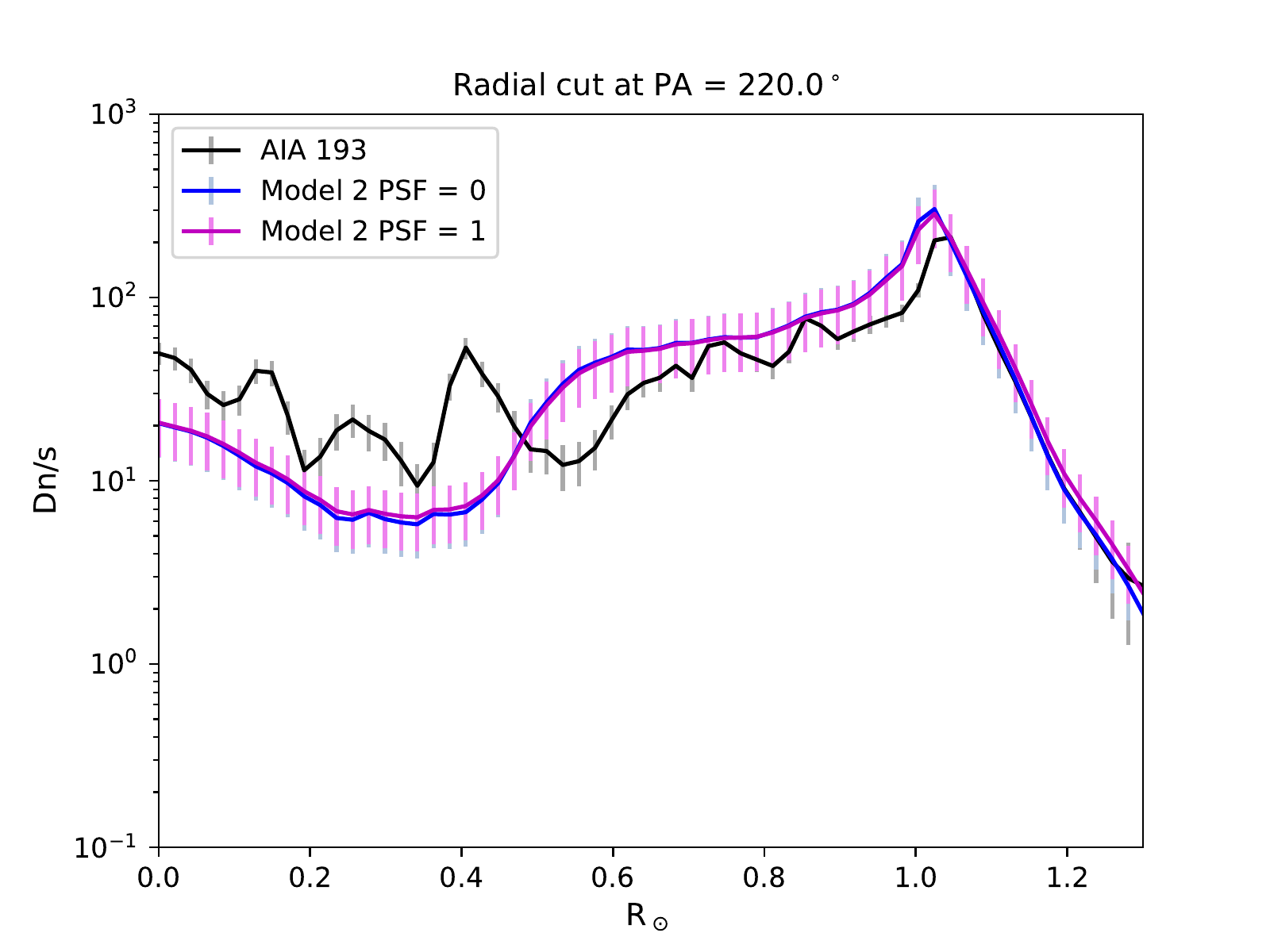}
\caption{ Radial intensity profiles for AIA 193 on November 7. The PSF main effect is visible in the CH and  off disk. }
\label{fig:aia_psf}
\end{figure}

Figure \ref{fig:aia_psf} plots two examples of AIA radial cut (15$^\circ$ and 220$^\circ$) on the solar disk for November 7. The black profile marks the observation, the blue profile is from the Model 2 intensity which was convolved with the PSF (Point Spread Function) distributed within the instrument software, and the red one shows the same model intensity without correction for the PSF. The effects of the PSF are mainly seen within weak intensity areas, such as the  CH  and the off--disk. However, such changes are only partially responsible for the higher contrast visible in the synthetic images with respect to the observations. As shown by \citet{Saqri2020} (see also Section \ref{sec:conclusion}) there may be further light diffusion effects not yet taken into account. 

\section{Transition region emission contribution to the AIA bands.}
\label{app:aia_tr}

\begin{figure*}
\centering
\includegraphics[width=0.33\textwidth, trim=5 0 0 0, clip=True]{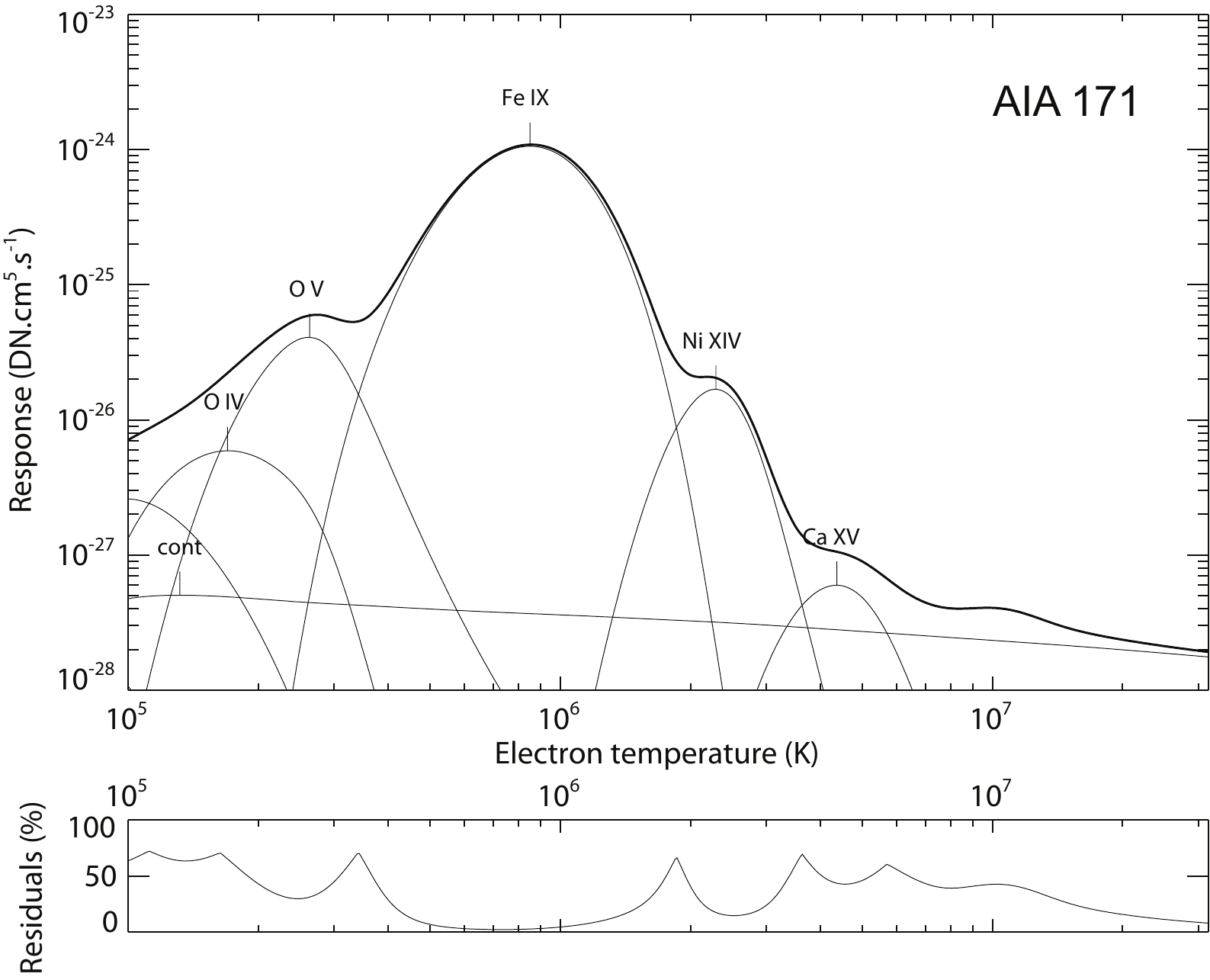}
\includegraphics[width=0.32\textwidth, trim=20 0 0 0, clip=True]{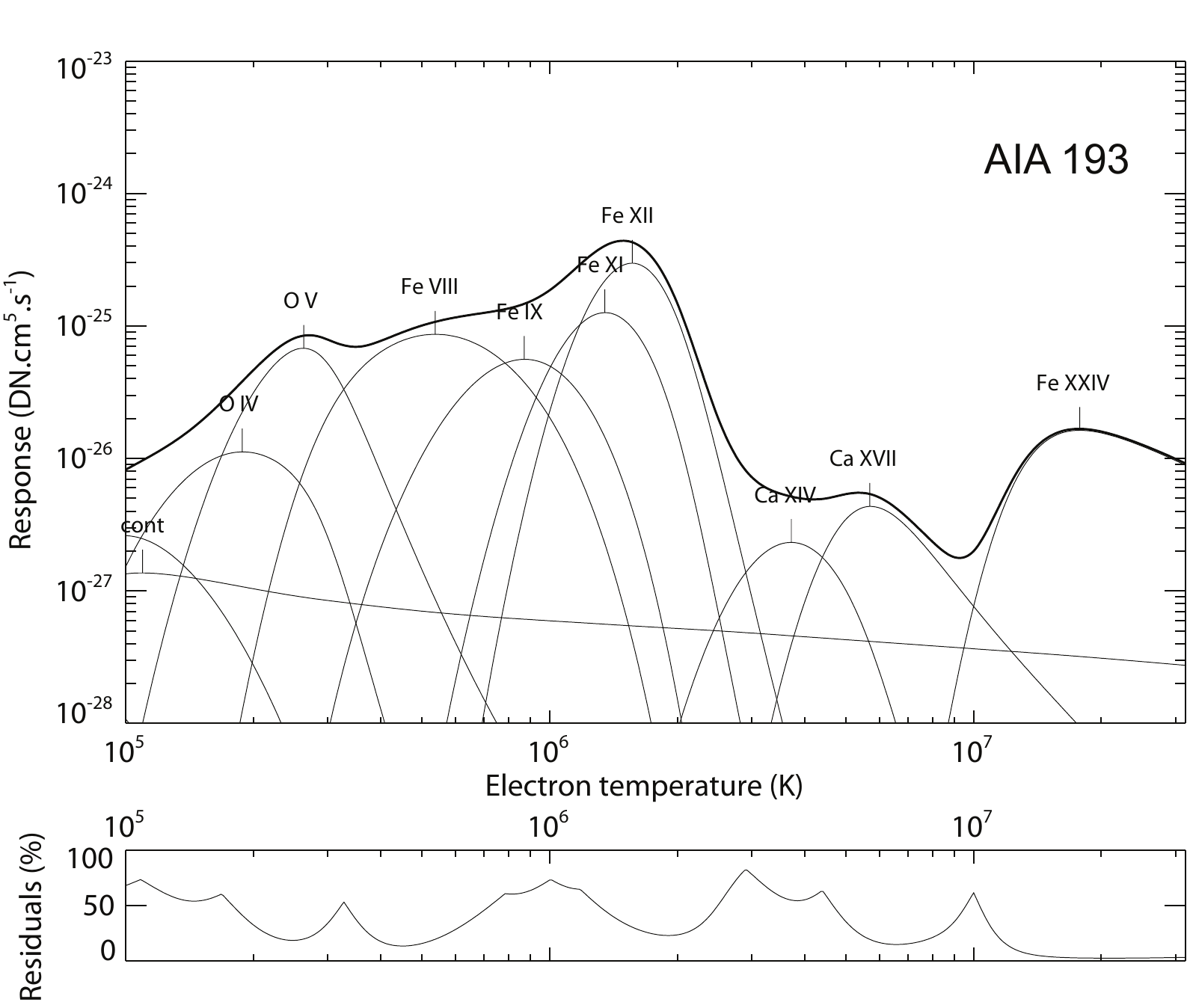}
\includegraphics[width=0.32\textwidth, trim=20 0 0 0, clip=True]{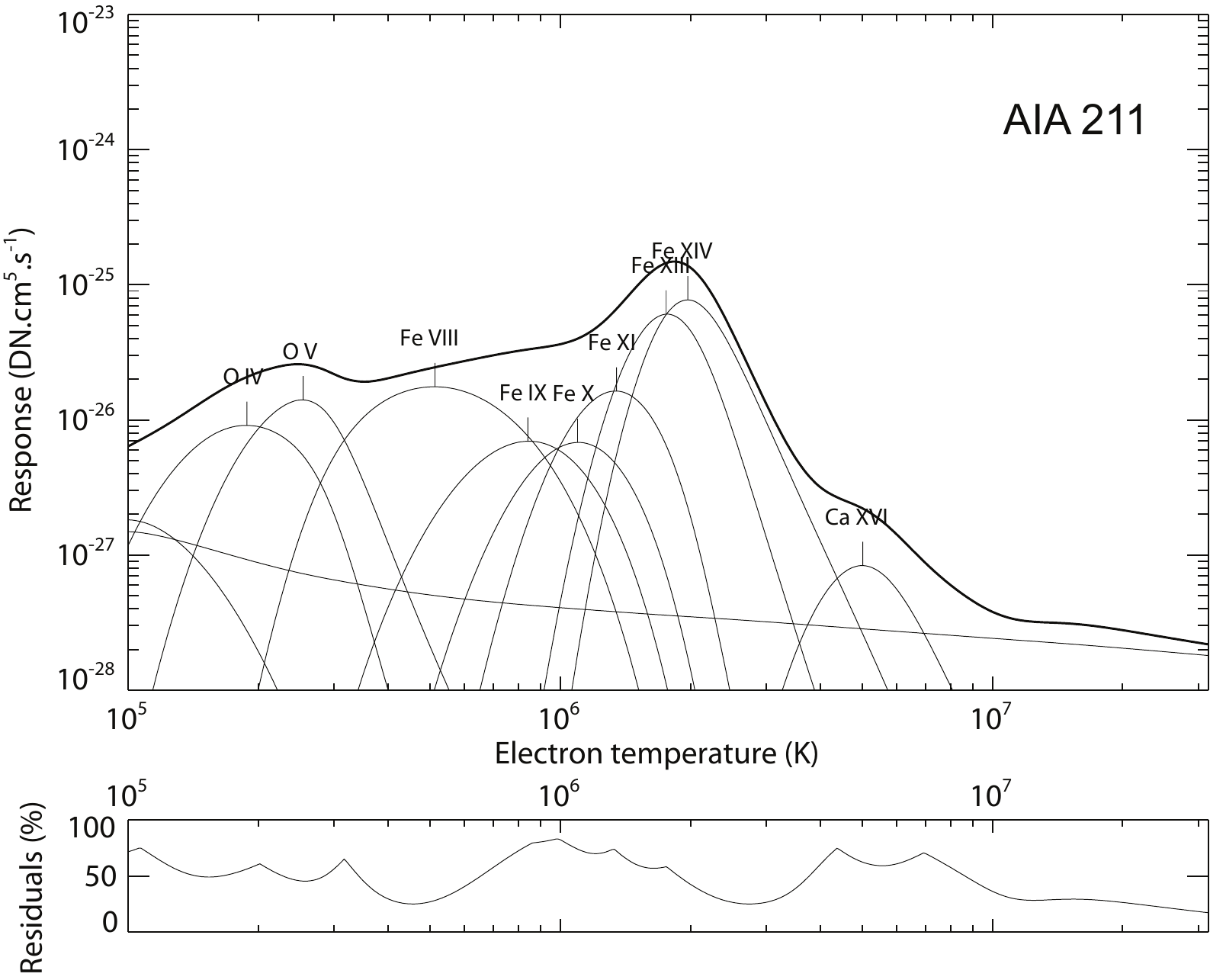}
\caption{AIA response functions (solid thick line) with the G(T, N) from the brightest lines falling in this wavelength range (thin lines). From left to right: 171\,\AA\, 193\,\AA\, and 211\,\AA\ .}
\label{aia_resf}
\end{figure*}

Figure \ref{aia_resf} shows the AIA 171 (left), 193 (center) and 211 (right) response functions, $A_f$, as a function of the temperature (solid thick line) together with the contribution functions, $G (n,T)$, of the main spectral lines from ions that are emitting within the bands. These quantities are defined in Sec. \ref{sec:mod_ima}.

The AIA 171\,\AA\, response function has the main peak at the \ion{Fe}{9}  temperature formation, the  193\,\AA\,  at the \ion{Fe}{12} temperature formation, while the 211\,\AA\,  band main peak is at the temperature of \ion{Fe}{13} - \ion{Fe}{14}. 
All curves are extended in temperature, with important contribution from transition region emission lines. Here we test, as an example, how much the QS and CH transition region emission contribute to the total intensity measured in the 193\,\AA\, band. To do this we follow the method described in \cite{parenti2012}, that is calculating the intensity in the band as function of an ad-hoc DEM, that excludes coronal plasma.  
 
The AIA intensity is calculated from Eq. \ref{eq:i} as follows: 

\begin{equation}
    I = \int DEM(T, n) ~~A_f ~~dT 
\end{equation}

\noindent where $\mathrm{DEM(n, T) = n^2 ~~dV/dT }$ \citep[see for instance][]{parenti2015}.

\begin{figure}
\centering
\includegraphics[width=0.52\textwidth]{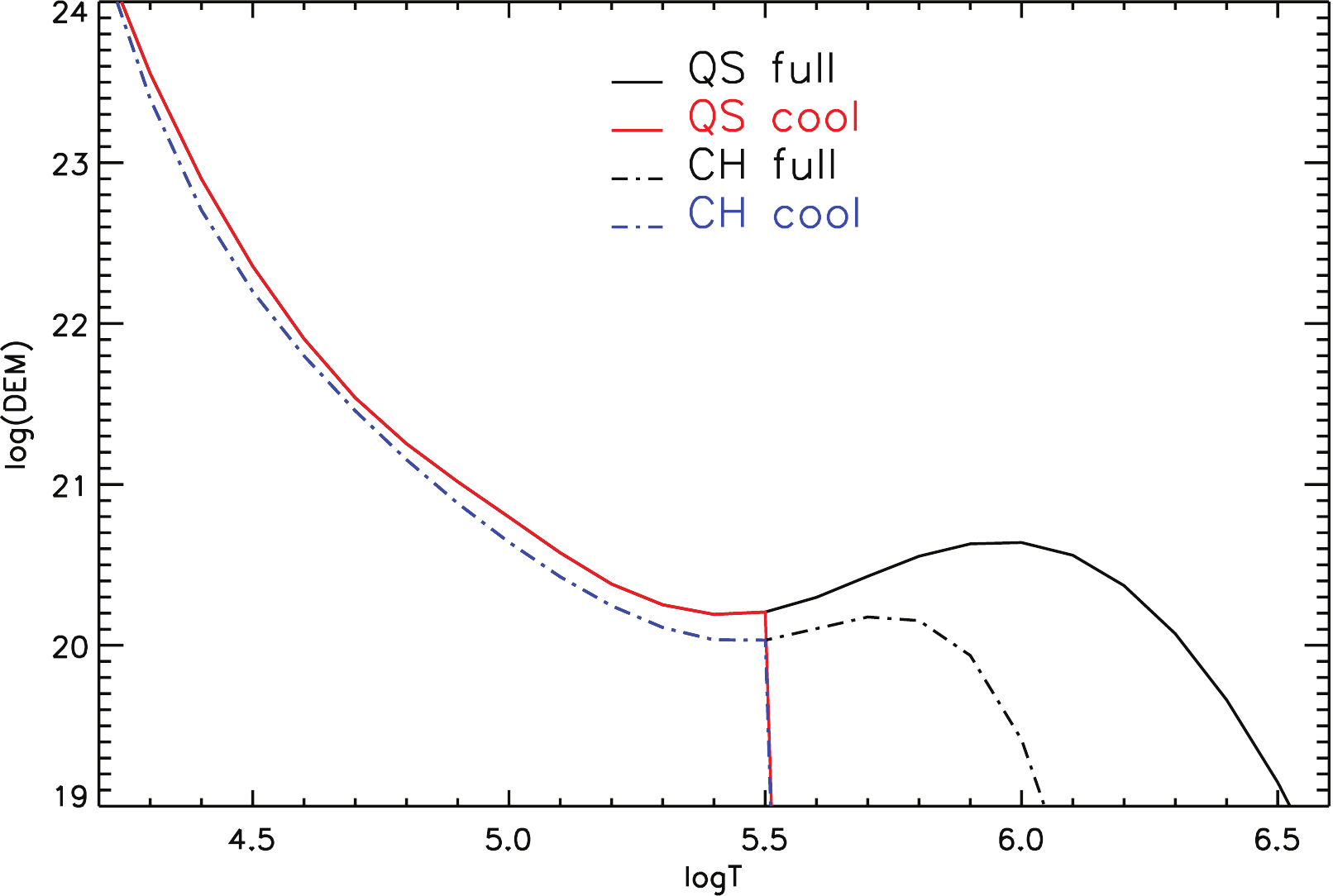}
\caption{Quiet Sun and coronal hole DEMs used for the tests described in this section.}
\label{sol_dems}
\end{figure}

Figure \ref{sol_dems}
shows the  DEMs used for our  tests. The black solid line represents the measured QS DEM obtained by \citep{parenti2007}, the black dashed line is from the CHIANTI database (obtained using \cite{vernazza78} line intensity) and represents the CH. The MHD model inner boundary provides a lower temperature limit of about 0.4 MK. We then cut these DEMs to such a limit and calculate the missing $\mathrm{Dn}$ within each AIA. These two news DEMs are the red and blue ones in the figure.

Assuming a number density of $5~10^9~\mathrm{cm^{-3}}$, we obtain 39 $\mathrm{Dn/s}$ for the full QS DEM. This is about consistent with the QS measured intensity (see Figure \ref{fig:aia_radial} black curve). When we use the ad-hoc (cooler) DEM we obtain 3.2 $\mathrm{Dn/s}$, which is small and representing less than 10$\%$ of the total. The same test on the CH gives a different result. Using the full DEM provides 2.5 $\mathrm{Dn/s}$, while we obtain 1.1 $\mathrm{Dn/s}$ for the cooler CH DEM. This latter is almost $50\%$ of the full DEM contribution and shows how important is the transition region emission to the AIA 193 band within CHs. 
 \cite{parenti2012} showed already that in absence of coronal emission this band is dominated by the \ion{O}{5} TR lines.

A final note concerns the difference between the CH intensity in the data and in this simulation which is discussed in Section \ref{sec:conclusion}.


\section{Streamer radial intensity profile variation with pixels selection criteria}
\label{app:avg}

\begin{figure*}[ht] 
\centering
\includegraphics[width=0.6\textwidth]{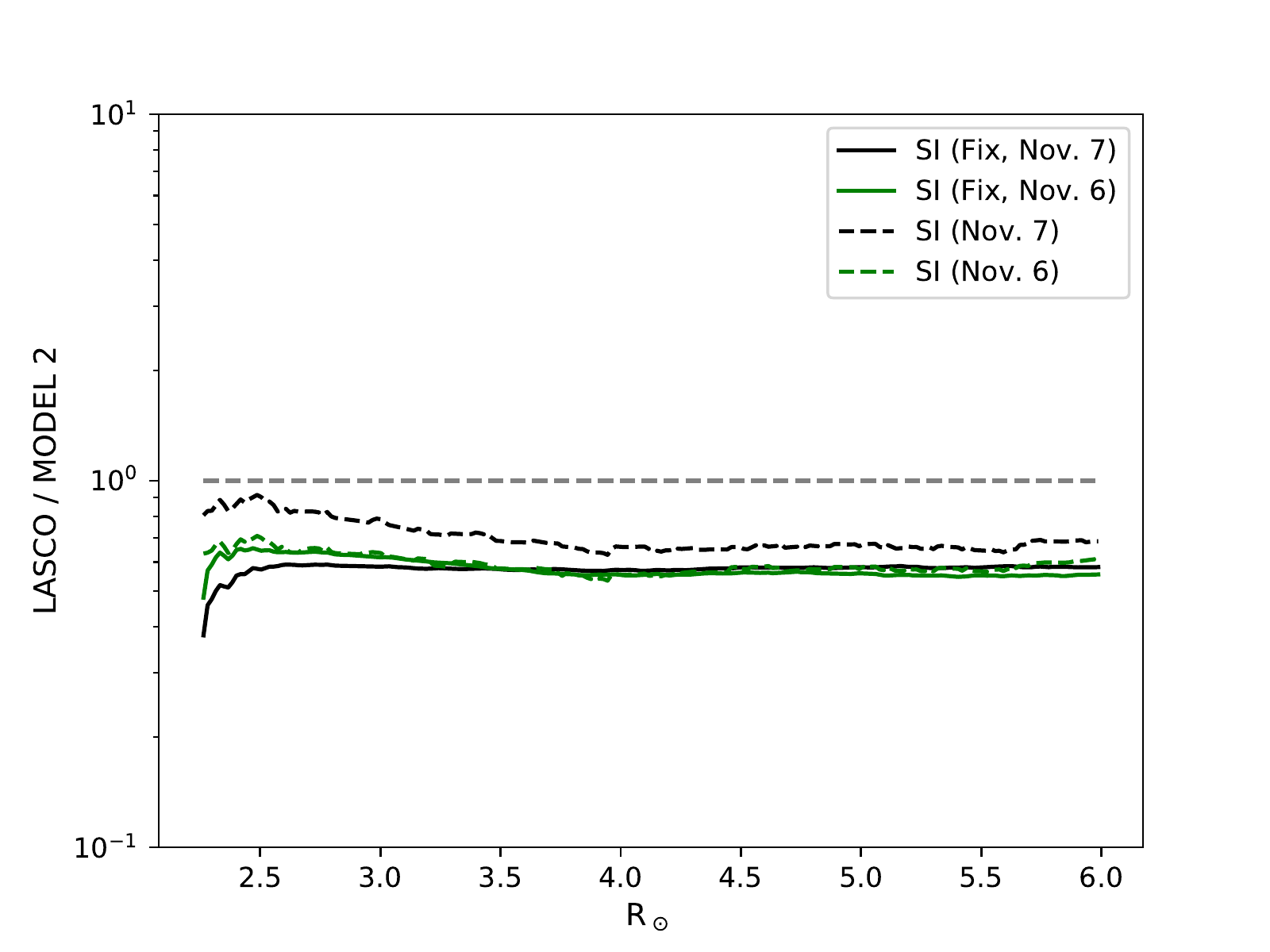}
\caption{ Ratios of LASCO C2 pB observation to Model 2 for November 6 (green) and 7 (black) as function of the solar heights. The dashed curves are obtained by selecting, at each distance, the pixels within 0.8 from the peak of intensity ($20\%$ decrease). The solid curves are obtained selecting 15 pixels around the identified streamer axis (see text). }
\label{fig:avg_test}
\end{figure*}

Figure \ref{fig:avg_test} shows the variation with the solar distance of the ratio of observational data to Model 2 for streamer SI. The dashed curves are the same of those of Figures \ref{fig:lasco_nprof6} and \ref{fig:rad_7} right plots (SI for November 6 and 7). 
The solid curves are obtained using a different criterion to select the averaged intensity profiles: from Figure \ref{fig:lasco_pa_str} first and third panels, we selected a radial stripe of 15 pixels centered in the streamer. This method does not take into account of the possible different size of the streamers, or a possible different shape. 
Figure \ref{fig:avg_test} shows that for November 7 this latest method results in a constant ratio with the solar distance. This result shows that there is some method dependence in extracting the area of interest, which may give an indication of the uncertainties in the method. For the plotted case the difference is within about $20\%$.

\bibliography{UV-WL}{}
\bibliographystyle{aasjournal}

\end{document}